\documentclass[11pt,a4paper]{article}
\pdfoutput=1

\usepackage{jheppub}
\usepackage[english]{babel}
\usepackage{relsize}
\usepackage{mathrsfs}  

\usepackage{amstext}  
\usepackage{amsfonts}
\usepackage{amsthm}
\usepackage{graphicx}
\usepackage{amsmath}
\usepackage{amsfonts}
\usepackage{mathtools}

\usepackage{array}                
\usepackage{xcolor} 

\usepackage{slashed}

\renewcommand{\refeq}[1]{\mbox{\eqref{#1}}}

\newcommand{\reffi}[1]{\mbox{Fig.~\ref{#1}}}

\newcommand{\refta}[1]{\mbox{Table~\ref{#1}}}

\newcommand{\refse}[1]{\mbox{Section~\ref{#1}}}
\newcommand{\refses}[2]{\mbox{Sections~\ref{#1}--\ref{#2}}}
\newcommand{\refapp}[1]{\mbox{Appendix~\ref{#1}}}

\newcommand{\ie}{i.e.\ }

\newcommand{\f}[2]{\frac{#1}{#2}}

\newcommand{\nosss}[1]{#1}

\newcommand{\ben}{\begin{enumerate}}
\newcommand{\een}{\end{enumerate}}
\newcommand{\bit}{\begin{itemize}}
\newcommand{\eit}{\end{itemize}}
\newcommand{\bea}{\begin{eqnarray}}
\newcommand{\eea}{\end{eqnarray}}
\newcommand{\be}{\begin{equation}}
\newcommand{\ee}{\end{equation}}
\newcommand{\ba}{\begin{align}}
\newcommand{\ea}{\end{align}}
\newcommand{\beas}{\begin{eqnarray*}}
\newcommand{\eeas}{\end{eqnarray*}}
\newcommand{\bes}{\begin{equation*}}
\newcommand{\ees}{\end{equation*}}
\newcommand{\bas}{\begin{align*}}
\newcommand{\eas}{\end{align*}}

\newcommand{\eps}{{\varepsilon}}
\newcommand{\als}{\alpha}
\newcommand{\gs}{g}
\newcommand{\ms}{\mathrm{MS}}
\newcommand{\msz}{\mathrm{MS}_0}
\newcommand{\msbar}{\overline{\mathrm{MS}}}

\newcommand{\lb}{\left(}
\newcommand{\rb}{\right)}

\newcommand{\dendim}{d}
\newcommand{\numdim}{D}
\renewcommand{\dendim}{D}
\renewcommand{\numdim}{D_{\mathrm{n}}}

\def\rcarg{\chi}

\def\inv{\mathrm{inv}}
\newcommand{\msfact}{S}
\newcommand{\scheme}{X}
\newcommand{\schemez}{\scheme_0}

\newcommand{\dscheme}{\Delta \scheme}
\newcommand{\dschemez}{\Delta \schemez}
\def\param{\theta}

\newcommand{\denbar}{\bar}

\newcommand{\Dbar}[1]{D_{\nosss{#1}}}

\newcommand{\calA}{\mathcal{A}}
\newcommand{\calC}{\mathcal{C}}
\newcommand{\calD}[1]{\mathcal{D}^{(#1)}(\bar q_{#1})}
\newcommand{\calF}{\mathcal{F}}

\newcommand{\calK}{\mathcal{K}}

\newcommand{\calN}{\mathcal{N}}
\newcommand{\calP}{\mathcal{P}}
\newcommand{\ntilde}{\tilde\calN}

\newcommand{\calR}{\mathcal{R}}
\newcommand{\calT}{\mathcal{T}}
\newcommand{\calZ}{\mathcal{Z}}
\newcommand{\dcalZ}{\delta \hat{\mathcal{Z}}}
\newcommand{\bfM}{{\textbf{M}}}

\newcommand{\bfK}{\textbf{K}}

\newcommand{\textR}{\textbf{R}}
\newcommand{\bfR}{\textbf{R}}

\newcommand{\bfT}{\textbf{T}}
\newcommand{\bfS}{\textbf{S}}
\newcommand{\bfSX}[1]{\bfS^{(#1)}_{X_{#1}}}
\newcommand{\barN}{\bar{\mathcal{N}}}

\newcommand{\calV}{\mathcal{V}}

\newcommand{\Nc}{N}
\newcommand{\dA}{d_\mathrm{A}}
\newcommand{\dF}{d_\mathrm{F}}
\newcommand{\nf}{n_{\mathrm{f}}}

\newcommand{\mf}{m_{f}}
\newcommand{\CA}{C_{\mathrm{A}}}
\newcommand{\CF}{C_{\mathrm{F}}}
\newcommand{\TF}{T_{\mathrm{F}}}
\newcommand{\TA}{T_{\mathrm{A}}}

\newcommand{\ff}{\mathrm{f\hspace{.001ex}f}}
\newcommand{\psif}{f}

\newcommand{\singlearg}[1]{
\ifx&#1&
\else
(#1)   
\fi
}

\newcommand{\doublearg}[2]{
\ifx&#2&
(#1)  
\else
(#1,#2)   
\fi
}

\newcommand{\ampindices}[5]{{#1}_{{#2,#3}}^{#4 }\singlearg{#5}}

\newcommand{\amp}[4]{{\ampindices{\calA}{#1}{#2}{#3}{#4}}}
\newcommand{\ampbar}[4]{{\ampindices{\bar\calA}{#1}{#2}{\hspace{0.6pt}#3}{#4}}}
\newcommand{\ratamp}[4]{{\ampindices{\delta \calR}{#1}{#2}{#3}{#4}}}
\newcommand{\deltaZ}[4]{{\ampindices{\delta Z}{#1}{#2}{#3}{#4}}}
\newcommand{\deltaZtilde}[4]{{\ampindices{\delta \tilde Z}{#1}{#2}{#3}{#4}}}

\newcommand{\barq}{\bar q}
\newcommand{\barqidx}[2]{{\bar q}_{#1}^{\hspace{0.9pt}#2}}
\newcommand{\tildeqidx}[2]{{\tilde q}_{#1}^{\hspace{0.6pt}#2}}

\newcommand{\gpar}{{\mathrm{gp}}}

\def\tad{\mathrm{tad}}

\def\rem{\mathrm{rem}}
\def\div{\mathrm{div}}

\def\baralpham{\bar \alpha}
\def\alpham{\alpha}

\newcommand{\Tr}{\mathrm{Tr}}

\newcommand{\srV}{{\scriptscriptstyle \mathrm V}}

\newcommand{\srp}{{\scriptscriptstyle \mathrm P}}
\newcommand{\srG}{{\scriptscriptstyle \mathrm G}}
\newcommand{\srm}{{\scriptscriptstyle \mathrm m}}

\newcommand{\rA}{\mathrm A}
\newcommand{\rB}{\mathrm B}
\newcommand{\rC}{\mathrm C}
\newcommand{\rI}{\mathrm{I}}
\newcommand{\rII}{\mathrm{II}}

\newcommand{\rR}{\mathrm R}
\newcommand{\rF}{\mathrm F}
\newcommand{\rL}{\mathrm L}
\newcommand{\rS}{\mathrm S}
\newcommand{\rT}{\mathrm T}
\newcommand{\ri}{\mathrm i}
\newcommand{\rd}{\mathrm d}
\newcommand{\ord}{\mathcal O}

\definecolor{bluemar}{rgb}{0,0,.5}
\definecolor{redmar}{rgb}{.8,0,0}
\definecolor{greenmar}{rgb}{0,.5,0}

\graphicspath{{./figures/}}

\newcommand{\diawidth}{1.8cm}

\preprint{
\begin{flushright}
PSI-PR-20-10\\
ZU-TH 24/20\\
\end{flushright}
}

\title{\boldmath Two-Loop Rational Terms in Yang--Mills Theories}

\author[a]{Jean-Nicolas Lang}
\author[a]{Stefano Pozzorini}
\author[a]{Hantian Zhang}
\author[b]{Max F. Zoller}

\affiliation[a]{Physik-Institut, Universit\"at Z\"urich, CH-8057 Z\"urich, Switzerland}
\affiliation[b]{Paul Scherrer Institut, Forschungsstrasse 111, CH-5232 Villigen PSI, Switzerland}

\emailAdd{jlang@physik.uzh.ch}
\emailAdd{pozzorin@physik.uzh.ch}
\emailAdd{hantian.zhang@physik.uzh.ch}
\emailAdd{max.zoller@psi.ch}

\abstract{Scattering amplitudes in $D$ dimensions involve particular terms that
originate from the interplay of UV poles with the $(D-4)$-dimensional parts of
loop numerators. 
Such contributions can be controlled through a finite set of
process-independent rational counterterms, which make it possible to compute
loop amplitudes with numerical tools that construct the loop numerators in
four dimensions.
Building on a recent study~\cite{Pozzorini:2020hkx} of the general
properties of two-loop rational counterterms, in this paper we investigate
their dependence on the choice of renormalisation scheme.
We identify a nontrivial form of 
scheme dependence, which originates from the 
interplay of mass and field renormalisation with the
$(D-4)$-dimensional parts of loop numerators, 
and we show that it can be controlled through a new kind of 
one-loop counterterms.
This guarantees that the two-loop rational counterterms for a given
renormalisable theory can be derived once and for all in terms of generic
renormalisation constants, which can be adapted a posteriori to any scheme.
Using this approach, we present the first calculation of the full set of
two-loop rational counterterms in Yang--Mills theories.
The results are applicable to SU(N) and U(1) gauge theories 
coupled to $\nf$ fermions with arbitrary masses.
}

\arxivnumber{}

\begin{document}
\maketitle
\flushbottom

\section{Introduction}

Dimensional regularisation~\cite{tHooft:1972tcz}
is the most widely used method to regularise the
ultraviolet (UV) and infrared (IR) singularities of scattering amplitudes in quantum-field theory.  
In this approach, loop amplitudes are computed in a continuous number
$D$ of space-time dimensions, and the 
divergences of UV and IR kind assume the form of 
$1/(D-4)$ poles. For this reason, the $(D-4)$-dimensional parts of loop momenta,
metric tensors and Dirac matrices need to be manipulated with special care.
In a computer algebra framework this is rather straightforward,
while in the context of numerical algorithms, where algebraic quantities
need to be implemented in 
an integer
 number of space-time dimensions, 
the consistent treatment of $(D-4)$-dimensional terms raises nontrivial 
technical and conceptual
problems.

In the literature a variety of methods have been proposed 
that aim at 
restricting the calculation of 
loop amplitudes to an integer number of space-time dimensions~\cite{%
,Bern:1991aq%
,Kilgore:2012tb%
,Siegel:1979wq%
,Signer:2005iu%
,Cherchiglia:2010yd%
,Pittau:2012zd%
,Page:2015zca%
,Fazio:2014xea%
,Ossola:2008xq%
,Giele:2008ve%
,Abreu:2017hqn%
,Soper:1999xk%
,Nagy:2003qn%
,Catani:2008xa%
,Sborlini:2016gbr%
,Verdugo:2020kzh%
,Becker:2010ng%
,Becker:2012bi%
,Anastasiou:2018rib%
,Capatti:2019edf%
}.
At one loop, the most widely used 
method is based on the idea of splitting
loop amplitudes into two parts according to the dimensionality $\numdim$ of
the numerators of loop integrands.
In this approach, loop amplitudes can be constructed 
by means of automated numerical algorithms in 
$\numdim=4$ dimensions,
while the remaining $(\numdim-4)$-dimensional parts contribute only 
in combination with UV poles, and can be reconstructed a posteriori 
by means of process-independent rational
counterterms~\cite{Ossola:2008xq,Draggiotis:2009yb,Garzelli:2009is,Pittau:2011qp}.
This method is a key ingredient of the most efficient and flexible NLO 
automated tools on the market~\cite{Buccioni:2019sur,Denner:2017wsf,
vanHameren:2009dr,Hirschi:2011pa}, and its extension to two loops 
is a natural strategy towards NNLO automation.

As a first step in this direction, recently it was shown that 
renormalised two-loop amplitudes in dimensional
regularisation can be computed in terms of 
quantities in $\numdim=4$ dimensions and rational counterterms~\cite{Pozzorini:2020hkx}.
The relevant relation is encoded in the general formula
\bea 
\label{eq:mfor} 
{\textbf{R}}\, \ampbar{2}{\Gamma}{}{}   
&=&  \amp{2}{\Gamma}{}{} + 
\sum  \limits_{\gamma} \lb \deltaZ{1}{\gamma}{}{} +\deltaZtilde{1}{\gamma}{}{} + \ratamp{1}{\gamma}{}{} \rb \cdot \amp{1}{\Gamma/\gamma}{}{}
\,+\, 
\deltaZ{2}{\Gamma}{}{} + \ratamp{2}{\Gamma}{}{}
\,,
\label{eq:masterformula2intro}
\eea
where ${\textbf{R}}\,\bar \calA_{\Gamma,2}$ is the renormalised
amplitude of a two-loop vertex function\footnote{
Here and throughout the paper 
by vertex, or vertex function, we mean
any $N$-point function with $N\ge 2$ external lines.
}
or a single two-loop diagram
$\Gamma$ in $D$ dimensions.
The corresponding one-loop subdiagrams and their complements are labelled
$\gamma$ and $\Gamma/\gamma$, respectively, while
$\calA_{2,\Gamma}$ and $\calA_{1,\Gamma/\gamma}$ 
denote the unrenormalised amplitudes of 
$\Gamma$ and $\Gamma/\gamma$
in $\numdim=4$ dimensions. 
The above formula features a similar structure as the 
well-known \textR-operation~\cite{Bogoliubov:1957gp,hepp1966,Zimmermann1969,Kennedy_ROp},
\ie it involves the unrenormalised two-loop amplitude
of $\Gamma$
in combination with 
one-loop counterterms associated with 
the UV divergences of the subdiagrams $\gamma$,
as well as two-loop counterterms
associated with the remaining local two-loop
divergence.
For the subtraction of UV divergences in $\numdim=4$ dimensions, the 
standard counterterms $\delta Z_{1,\gamma}$ and $\delta Z_{2,\Gamma}$ 
are supplemented by additional one-loop 
counterterms $\delta \tilde Z_{1,\gamma}$.
Such extra UV counterterms are required only 
for quadratically divergent selfenergy subdiagrams 
$\gamma$ and are proportional to $\tilde q^2/\eps$,
where $\tilde q$ is the $(D-4)$-dimensional part 
of the loop momentum that flows through 
$\Gamma/\gamma$.
The role of the remaining counterterms $\delta \calR_1$ and 
$\delta \calR_2$ in~\refeq{eq:mfor}
is to reconstruct all 
parts of the renormalised two-loop 
amplitude that originate from the
interplay of UV divergences with the $(D-4)$-dimensional 
terms in the loop numerator.
The parts stemming from 
one-loop subdivergences are reconstructed 
through the well-known one-loop rational counterterms 
$\delta \calR_1$, while the two-loop rational counterterms
$\delta \calR_2$ account for the parts stemming from
the remaining local two-loop divergence.

The relation~\refeq{eq:mfor} holds for any process in any renormalisable
theory, and,
similarly as for the usual UV counterterms,
also $\delta \tilde Z_1$, $\delta \calR_1$ and $\delta
\calR_2$ 
are process-independent local counterterms
that depend only on the theoretical model. 
Thus, once the $\delta \tilde Z_1$,  $\delta \calR_1$ and $\delta
\calR_2$ counterterms are available,
their implementation amounts to 
a straightforward extension of the Feynman rules.
A general method to derive the two-loop counterterms $\delta \calR_2$ 
in any renormalisable model was presented 
in~\cite{Pozzorini:2020hkx}.
Technically, this procedure needs to be 
applied to all one-particle irreducible
(1PI)
vertex functions that involve a 
global UV divergence, and the
required one- and two-loop integrals
can be simplified using expansions 
that give rise to tadpole integrals
with a single mass scale.

So far, the study of two-loop rational terms was restricted to effects of UV
origin, assuming that IR divergences are either absent or 
are subtracted in a way that does not interfere with the rational terms.
In fact, at one loop IR divergences do not give rise to any rational 
term~\cite{Bredenstein:2008zb}.
However, the implications of IR divergences at two loops
remain to be investigated.
Another limitation of the original study of~\cite{Pozzorini:2020hkx} 
lies in the fact that the
derivation of the master formula~\refeq{eq:mfor} 
and the available $\delta \calR_2$ counterterms for QED~\cite{Pozzorini:2020hkx}
are based on specific renormalisation schemes, namely the 
$\ms$ or $\msbar$ schemes.


In this paper---after a review 
in~Sections~\ref{sec:oneloop}--\ref{se:irredtwoloop}
of the previous work on $\delta \calR_2$ terms~\cite{Pozzorini:2020hkx}---the study of two-loop rational terms 
is extended in two new directions. 
First, in \refse{se:schemetrans} we demonstrate that the 
master formula~\refeq{eq:mfor} 
is valid for arbitrary renormalisation schemes, and we 
present a general analysis of the scheme dependence of 
$\delta \calR_2$ counterterms of UV origin.
To this end, we consider a finite multiplicative renormalisation of 
couplings, masses and fields, and we study its interplay with the 
projection of loop numerators to $\numdim=4$ dimensions.
As we will show, these two operations do not commute at two loops, 
but the effect of their commutator can be encoded in 
a new set of scheme- and process-independent one-loop 
counterterms $\delta \hat\calK_1$.
This allows us to derive  
the general formulas~\refeq{eq:RtransfG}--\refeq{eq:straKdec}, which describe
the scheme dependence of $\delta \calR_2$ counterterms 
as the result of the multiplicative renormalisation of the known 
$\delta \calR_1$ counterterms 
plus a nontrivial part that can be written as a
combination of one-loop renormalisation constants
and $\delta \hat\calK_1$ counterterms.
In this way,
the $\delta \calR_2$ counterterms for a given theoretical model can be
derived, 
once and for all, in the form of a linear combination of generic one-loop renormalisation 
constants, which can be adapted a posteriori to any desired renormalisation scheme.

The second main novelty of this 
paper
is the first calculation of the full set of $\delta \calR_2$ counterterms 
in Yang--Mills theories.
As detailed in~\refse{sec:qcdres},  
the relevant calculations are carried out in a generic renormalisation scheme 
and for a generic gauge group,
while the results are presented in a form that is applicable both to 
SU(N) and U(1) gauge theories 
coupled to massless or massive fermions.
The various tadpole expansions that have been used to 
compute and validate the required loop integrals 
are documented in detail in~\refapp{se:tadexp}. There 
we present the expansion techniques that have already been used in~\cite{Pozzorini:2020hkx}, 
as well as a new optimised approach.
Finally, for convenience of the reader, 
in \refapp{app:msbarRCs} we have collected all relevant UV 
renormalisation constants for the case of the $\msbar$ scheme.

\section{Rational terms at one loop} \label{sec:oneloop}

In this section we introduce the conventions 
used throughout this paper and we briefly review 
the properties of one-loop rational terms following~\cite{Pozzorini:2020hkx}.

\subsection{Notation and conventions}
\label{se:notation}

For the regularisation of UV divergences 
we use the 't~Hooft--Veltman scheme~\cite{tHooft:1972tcz}, 
where external states are four-dimensional, while 
loop momenta as well as the metric tensors and Dirac
matrices inside the loops live in 
$\dendim=4-2\eps$
dimensions.
For the decomposition of these objects into 
four-dimensional parts and
$(D-4)$-dimensional remnants we use the notation\footnote{For more details see~\cite{Pozzorini:2020hkx}.}
\bea
\label{eq:ddimnotG}
\denbar q^\mu &=& q^\mu+\tilde q^{\tilde\mu}\,,
\qquad
\denbar \gamma^\mu = \gamma^\mu+\tilde \gamma^{\tilde\mu}\,,
\qquad
\denbar g^{\denbar \mu\denbar \nu} = g^{\mu\nu}+\tilde g^{\tilde\mu\tilde
\nu}\,,
\eea
where the bar and the tilde are used to mark, respectively,
the $D$-dimensional and $(D-4)$-dimensional parts.
To keep track of the dimensionality of loop numerators
we use the parameter $\numdim$, which can assume the values 
$D$ or $4$.
The case $\numdim=D$  corresponds to standard
calculations in dimensional regularisation,
while in $\numdim=4$ all loop numerators are projected 
to four dimensions keeping loop denominators in $\dendim$
dimensions.

For the integration measure in loop-momentum space we use the
shorthand
\bea
\int\!\rd\barq & = & \mu_0^{2\eps} \int \f{\rd^{^D}\! \bar
q}{(2\pi)^{^D}}\,, 
\label{eq:intmeasure}
\eea
where $\mu_0$ is the scale of dimensional regularisation.
For the renormalisation scale we use the symbol $\mu_\rR$ and,
at variance with~\cite{Pozzorini:2020hkx}, 
in this paper $\mu_0$ and $\mu_\rR$ 
are treated as independent scales.

In the 't~Hooft--Veltman scheme the renormalisation of UV divergences and
the discussion of rational term of UV origin
can be restricted to amputated 
1PI
vertex functions. 
For more details see~\cite{Pozzorini:2020hkx}.

\subsection{One-loop amplitudes with four-dimensional external momenta}
\label{se:oneloopratterms}

Let us consider the amplitude of a 1PI 
one-loop diagram $\Gamma$,
\bea 
\label{eq:rtoneloopA}
\ampbar{1}{\Gamma}{}{} &=& 
\int\!\rd\barq_1\, \f{{\barN}(\bar{q}_1)}{\Dbar{0}(\barq_1)\cdots
\Dbar{N-1}(\barq_1)}\,,
\eea
with denominators
\bea
\label{eq:rtoneloopB}
\Dbar{j}(\barq_1)&=& (\barq_1+p_j)^2-m_j^2\,,
\eea
where $p_j$ are combinations of four-dimensional external momenta.
In $\numdim=\dendim$ dimensions, the numerator $\bar \calN(\barq_1)$ can be split into
\bea
\label{eq:rtoneloopD}
\bar \calN(\barq_1)&=& \calN(q_1) + \tilde \calN(\barq_1)\,,
\eea
where 
$\calN(q_1)$
is the four-dimensional part,
obtained by projecting the metric tensor, Dirac matrices
and the loop momentum to four dimensions.
The remnant  part $\tilde \calN(\barq_1)$ is of $\ord(\eps, \tilde q_1)$
and will be referred to as the $(\dendim-4)$-dimensional part of the numerator.
As discussed in the following, its contribution 
can be controlled through a finite set of process-independent rational 
counterterms.

In view of the analysis of rational terms beyond one loop, it is 
convenient to discuss rational counterterms
at the level of renormalised amplitudes. 
In the minimal subtraction ($\ms$) scheme, for 
renormalised one-loop amplitudes we use the notation
\bea
\label{eq:oneloopdivE}
\bfR\, \ampbar{1}{\Gamma}{}{}
&=&
(1-\bfK)\, \ampbar{1}{\Gamma}{}{}
\,=\,
\ampbar{1}{\Gamma}{}{}
+\deltaZ{1}{\Gamma}{}{}\,,
\eea
where $\bfK$ is an operator that extracts the UV divergence 
according to the $\ms$ prescription, \ie
in the form of pure $1/\eps$ poles,
and $\deltaZ{1}{\Gamma}{}{}$ is the corresponding 
counterterm. 
Here and in the following, $\bfR$ and $\bfK$
should be understood as linear operators.
Thus~\refeq{eq:oneloopdivE} 
is applicable both when $\Gamma$ is a single Feynman diagram or
a set of diagrams, in which case the result is equivalent to
the sum of the contributions of individual diagrams.
This linearity property holds for all
renormalisation identities in this paper.

At one loop, the renormalised amplitude 
can be constructed from quantities with $\numdim=4$
by means of the identity
\bea 
\label{eq:masterformula1}
\textbf{R}\,\ampbar{1}{\Gamma}{}{} &=& 
\amp{1}{\Gamma}{}{} + \deltaZ{1}{\Gamma}{}{} + \ratamp{1}{\Gamma}{}{}
\,,
\eea
where $\amp{1}{\Gamma}{}{}$ denotes the amplitude in $\numdim=4$ dimensions,
\bea 
\label{eq:rtoneloopH}
\calA_{1,\Gamma} &=& 
\int\!\rd\barq_1\, \f{\calN(q_1)}{\Dbar{0}(\barq_1)\cdots
\Dbar{N-1}(\barq_1)}\,,
\eea
which can be computed with numerical tools
that handle the numerator in four dimensions, while retaining the full
$\dendim$-dependence of the loop momentum in the denominator.
The UV divergence of $\calA_{1,\Gamma}$ is cancelled by the same
$\delta Z_{1,\Gamma}$
counterterm as in~\refeq{eq:oneloopdivE}, and 
the $\delta \calR_{1,\Gamma}$ counterterm embodies the contribution 
of the $\ntilde$-part of the numerator.
At one loop, such $\delta \calR_{1,\Gamma}$ counterterms
originate only from the interplay of $\ntilde$
with poles of UV type~\cite{Bredenstein:2008zb}.
Thus, similarly as for UV counterterms,
they can be derived once and for all 
for the set of UV divergent 1PI vertex 
functions~\cite{Ossola:2008xq,Draggiotis:2009yb,Garzelli:2009is,Pittau:2011qp},
where they take the form 
of homogeneous polynomials of degree $X$ in the external momenta $\{p_k\}$ 
and internal masses $\{m_k\}$, with $X$ being the degree of UV divergence
of the vertex at hand.

In this paper we focus on the rational terms that originate from $\ntilde$, 
for which we use the symbols $\delta\calR_L$ at $L$ loops.
Such terms will be referred to as $\ntilde$ rational terms or
simply rational terms.\footnote{Note that, in the literature on one-loop 
rational terms, the $\ntilde$-terms of type $\delta \calR_1$ are usually
labelled $R_2$, while the label $R_1$
is used for one-loop rational terms stemming from the $(\dendim-4)$-dimensional
part of loop denominators.  
The latter kind of rational terms cannot
be described by local counterterms,
but can be controlled
in a process-independent way through appropriate reduction
algorithms in four dimensions (see
e.g.~\cite{delAguila:2004nf,Ossola:2006us})
and will not be discussed in this paper.}

\subsection{One-loop amplitudes with $D$-dimensional external momenta}
\label{se:ddimoneloop}

\begin{figure}
\begin{center}
\includegraphics[height=0.15\textheight]{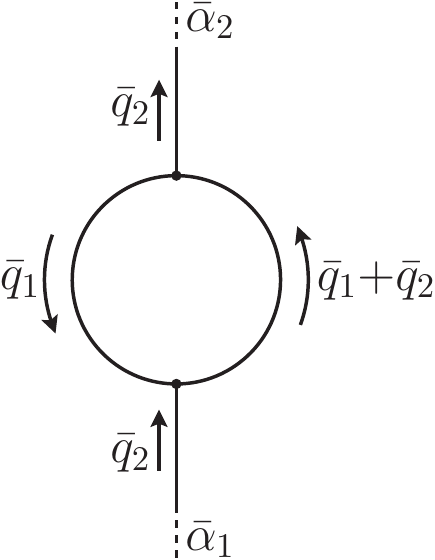}
\hspace{20mm}
\includegraphics[height=0.15\textheight]{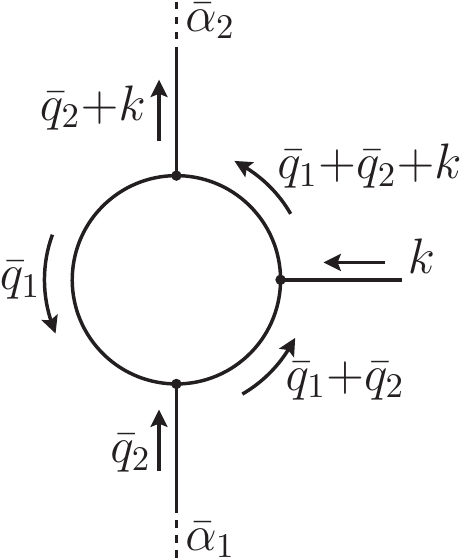}
\end{center}
\caption{Examples of UV divergent one-loop subtopologies.
The loop momentum $\barq_1$ circulates inside the subdiagram, while
the two external lines that are going to be embedded in 
a two-loop diagram depend on the $D$-dimensional loop momentum $\barq_2$
and carry the Lorentz/Dirac indices $\bar\alpha_1,\bar\alpha_2$.}
\label{fig:oneloopsubdiag}
\end{figure}

An identity of type~\refeq{eq:masterformula1} is needed also for the
one-loop subdiagrams of two-loop diagrams.
As depicted in~\reffi{fig:oneloopsubdiag}, 
this kind of one-loop (sub)diagrams involve
an internal loop momentum $\bar q_1$ and an external
loop momentum $\barq_2$. Thus the relation~\refeq{eq:masterformula1} 
needs to be extended to the case 
of $D$-dimensional external kinematics.

For the renormalised amplitude of a generic 
one-loop subdiagram in $\numdim=D$ we have
\bea
\label{eq:subdivN}
\bfR\,\ampbar{1}{\gamma}{\baralpham}{\barq_2}
&=&
(1-\bfK)\,\ampbar{1}{\gamma}{\baralpham}{\barq_2}
\,=\,
\ampbar{1}{\gamma}{\baralpham}{\barq_2}
+
\deltaZ{1}{\gamma}{\baralpham}{\barq_2}\,,
\eea
where we explicitly indicate the dependence on the 
$D$-dimensional external loop momentum $\barq_2$
and the multi-index $\baralpham = (\bar\alpha_1, \bar\alpha_2)$, which
embodies the two Lorentz/Dirac indices associated with the two $\barq_2$-dependent external
lines (see~\reffi{fig:oneloopsubdiag}). 
Since in $\numdim=D$ the momentum $\barq_2$ 
has the same dimensionality in the loop numerator and
denominator, the renormalised amplitudes~\refeq{eq:subdivN}
and~\refeq{eq:oneloopdivE} have the same form, 
and the corresponding UV counterterms
are related through the simple replacements $q_2\to \barq_2$ 
and $\alpham\to \bar \alpham$.

The extension of the identity~\refeq{eq:masterformula1} is more subtle, 
and the generalised formula for $D$-dimensional external kinematics 
reads~\cite{Pozzorini:2020hkx}
\bea
\label{eq:masterf1ddim}
\bfR\,\ampbar{1}{\gamma}{\baralpham}{\barq_2}
&=& 
\amp{1}{\gamma}{\alpham}{q_2}
+
\deltaZ{1}{\gamma}{\alpham}{q_2}
+
\deltaZtilde{1}{\gamma}{\alpham}{\tilde q_2}
+
\ratamp{1}{\gamma}{\alpham}{q_2}
+\ord(\eps,\tilde q_2 )\,.
\eea
Here the amplitude in $\numdim=4$ dimensions on the rhs is defined as
\bea 
\label{eq:fdq2baramp}
\amp{1}{\gamma}{\alpham}{q_2}
&=& 
\int\!\rd\barq_1\, \f{\calN^{\alpham}(q_1,q_2)}{\Dbar{0}(\barq_1,\barq_2)\cdots
\Dbar{N-1}(\barq_1,\barq_2)}\,,
\eea
where all parts of the loop numerator, including
$\alpham$ and $q_2$, are projected to four dimensions, while
$\barq_2$ is kept in $\dendim$
dimensions in the loop denominator.
The counterterms $\delta Z_{1,\gamma}$ and
$\delta \calR_{1,\gamma}$ on the rhs of~\refeq{eq:masterf1ddim}
are equivalent to the ones in~\refeq{eq:masterformula1}, 
while $\delta \tilde Z_{1,\gamma}$ is a new UV counterterm 
that cannot be obtained from~\refeq{eq:masterformula1}
via naive  $q_2\to \barq_2$ continuation.
The $\delta \tilde Z_{1,\gamma}$ 
counterterm 
is required in order to cancel the UV divergence of 
the amplitude in $\numdim=4$ dimensions~\refeq{eq:fdq2baramp},
\bea
\label{eq:4dimsubdiagB2}
\bfK\, \amp{1}{\gamma}{\alpham}{q_2}
&=& 
- \deltaZ{1}{\gamma}{\alpham}{q_2}
- \deltaZtilde{1}{\gamma}{\alpham}{\tilde q_2}\,.
\eea
In renormalisable theories $\delta \tilde Z_{1,\gamma}$ 
is required only for
quadratically divergent selfenergies, and its general form~\cite{Pozzorini:2020hkx}
is
\bea
\label{eq:4dimsubdiagE8}
\deltaZtilde{1}{\gamma}{\alpham}{\tilde q_2}
&=&
v^\alpham \frac{\tildeqidx{2}{2}}{\eps}\,,
\eea
where $v^\alpha$ is independent of $q_2$.
The origin of this $\ord(\tildeqidx{2}{2}/\eps)$ counterterm
lies in the fact that $\bar q_2$ is kept in $D$ dimensions in 
the loop denominator while it is projected to 
four dimensions in the numerator.

\section{Rational terms at two loops}
\label{se:irredtwoloop}

In this section we review the general analysis of two-loop rational 
terms presented in~\cite{Pozzorini:2020hkx}.

\begin{figure}[t]
\begin{center}
\includegraphics[width=0.3\textwidth]{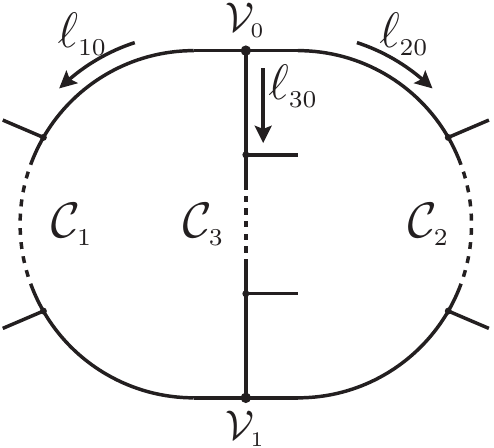}
\end{center}
\caption{A generic irreducible two-loop diagram consists of 
two vertices, $\calV_0$, $\calV_1$, that connect 
three chains, $\calC_1$, $\calC_2$, $\calC_3$, which 
contain, respectively, all propagators that depend on the loop momenta $q_1$, $q_2$,
$q_3=-q_1-q_2$.
For the sum of the loop momenta $\ell_{i0} = q_i+p_{i0}$, which 
flow out of $\calV_0$, momentum conservation requires 
$\sum_i \ell_{i0}= \sum_i p_{i0} = k_{\mathrm{ext}}$, 
where $k_{\mathrm{ext}}$ is the 
external momentum that flows into 
$\calV_0$. 
In practice, if $\calV_0$ is a triple vertices (as in the picture)
then $k_{\mathrm{ext}}=0$ and all $p_{i0}$ can be set equal to zero, 
while quartic vertices requires at least one
non-vanishing $p_{i0}$.
} 
\label{fig:twoloop_irred}
\end{figure}

\subsection{Notation for two-loop diagrams and subdiagrams}
\label{se:twoloopnot}

Two-loop amplitudes involve reducible and irreducible two-loop diagrams.
The former can be factorised into one-loop parts, which 
generate, upon renormalisation, only one-loop rational terms~\cite{Pozzorini:2020hkx}.
Thus genuine two-loop rational terms originate only from 
irreducible diagrams.
A generic irreducible two-loop 
diagram (see~\reffi{fig:twoloop_irred})
consists of three chains, $\calC_{1},\calC_{2},\calC_{3}$, 
that are connected to each
other by two vertices, $\calV_0,\calV_1$.
Each chain $\calC_{i}$ includes a certain number $N_i$ of
propagators
that depend on the loop
momentum $q_i$ and 
$N_i-1$ vertices.
The loop momenta are related to each other by 
$\bar q_1+\barq_2+\barq_3=0$.
The two-loop integral associated with a generic two-loop 
diagram $\Gamma$ has the form
\bea
\label{eq:twoloopnotA}
\ampbar{2}{\Gamma}{}{} 
&=&
\int\rd\barq_1
\int \rd\barq_2\,
\frac{
\bar\calN(\barq_1,\barq_2,\barq_3)}
{\calD{1}\,\calD{2}\,\mathcal{D}^{(3)}(\barq_3)}\bigg|_{\barq_3\,=\,-\barq_1-\barq_2}\,,
\eea 
where 
each chain $\calC_i$ contributes through the corresponding set of loop
denominators,
\bea
\label{eq:twoloopnotB}
\calD{i}&=&
D^{(i)}_0(\barq_i)\cdots
D^{(i)}_{N_i-1}(\barq_i)\,,
\eea
with 
\bea
\label{eq:twoloopnotB2}
D^{(i)}_a(\barq_i) \,=\, 
\bar \ell_{ia}^{\,2}-m_{ia}^2\,,
\qquad\mbox{and}\qquad
\bar \ell_{ia}\,=\,\barq_i+p_{ia}\,.
\eea
The form of the loop numerator is
\bea
\label{eq:twoloopnumA}
\bar\calN(\barq_1,\barq_2,\barq_3)
&=&
\bar\Gamma^{\bar\alpha_1\bar\alpha_2\bar\alpha_3}(\barq_1,\barq_2,\barq_3)\,
\bar\calN^{(1)}_{\bar\alpha_1}(\barq_1)\,
\bar\calN^{(2)}_{\bar\alpha_2}(\barq_2)\,
\bar\calN^{(3)}_{\bar\alpha_3}(\barq_3)\,,
\eea 
where the parts $\bar\calN^{(i)}_{\bar\alpha_i}(\barq_i)$ associated 
to each chain $\calC_i$ are connected 
through the 
multi-indices
$\bar \alpha_i\equiv (\bar \alpha_{i1},\bar \alpha_{i2})$
to the tensor $\bar \Gamma^{\bar\alpha_1\bar \alpha_2\bar \alpha_3}$,
which embodies the two vertices $\calV_0$ and $\calV_1$.

Irreducible two-loop diagrams $\Gamma$ involve 
three one-loop subdiagrams $\gamma_i$, which 
result from $\Gamma$ by 
truncating the chain $\calC_i$.
More precisely, 
each partition $i | jk$ of $123$ defines a
subdiagram $\gamma_i$
that contains the chains
$\calC_j$ and $\calC_k$. Its amplitude 
reads
\bea
\label{eq:subdiagnotA}
\ampbar{1}{\gamma_i}{\bar\alpha_i}{\barq_i} &=&
\int\rd\barq_j\,
\frac{
\bar\Gamma^{\bar\alpha_1\bar\alpha_2\bar\alpha_3}(\barq_1,\barq_2,\barq_3)\,
\bar\calN^{(j)}_{\bar\alpha_j}(\barq_j)\,
\bar\calN^{(k)}_{\bar\alpha_k}(\barq_k)}
{\calD{j}\,\calD{k}}\Bigg|_{\barq_k\,=\,-\barq_i-\barq_j}\,,
\eea
where $\barq_i$ plays the role of external momentum,
and $\bar\alpha_i$ connects $\gamma_i$ to its complement 
$\Gamma/\gamma_i$, which contains the chain $\calC_i$,
and is derived from $\Gamma$ by 
shrinking $\gamma_i$ to a vertex.

Similarly as in~\refeq{eq:rtoneloopD},
the two-loop
numerator can be split into
four-dimensional and $(\dendim-4)$-dimensional parts as
\bea
\label{eq:rttwoloopDD}
\bar \calN(\barq_1, \barq_2, \barq_3)&=& \calN(q_1, q_2, q_3) + \tilde \calN(\barq_1, \bar
q_2, \bar q_3)\,.
\eea
As discussed below, explicit two-loop calculations can be restricted to the
four-dimensional $\calN$ contribution, while
all $\ntilde$-terms can be reconstructed by means of 
rational counterterms.

\subsection{UV poles and rational parts at two loops}
\label{se:polestructure}

In general, two-loop amplitudes involve subdivergences and
additional local two-loop divergences.
These two kinds of divergences can be subtracted by means of the so-called 
$\bfR$-operation~\cite{Bogoliubov:1957gp,hepp1966,Zimmermann1969,Kennedy_ROp}.
For a single two-loop diagram or a full two-loop vertex function
$\Gamma$, 
the subtracted amplitude has the form
\bea 
\label{eq:twoloopren} 
{\textbf{R}}\, \ampbar{2}{\Gamma}{}{}   
&=&  \ampbar{2}{\Gamma}{}{} + 
\sum_\gamma \deltaZ{1}{\gamma}{}{} \cdot \ampbar{1}{\Gamma/\gamma}{}{}
+ \deltaZ{2}{\Gamma}{}{}\,,
\eea
where $\ampbar{2}{\Gamma}{}{}$ is the 
unrenormalised two-loop amplitude in $\dendim$ dimensions.
The second term on the rhs subtracts all relevant subdivergences.
When $\Gamma$ is a single two-loop diagram the sum 
involves the three one-loop subdiagrams  $\gamma = \gamma_1,\gamma_2,
\gamma_3$ of $\Gamma$. The corresponding UV divergences are subtracted by the  counterterms 
\bea
\deltaZ{1}{\gamma_i}{}{}
&=&
- \bfK \,
\ampbar{1}{\gamma_i}{}{}\,,
\eea
and their insertion into the complementary one-loop diagrams
$\Gamma/\gamma_i$ read
\bea
\label{eq:subdiagnotB}
\deltaZ{1}{\gamma_i}{}{}\cdot
\ampbar{1}{\Gamma/\gamma_i}{}{}
&=&
\int\rd\barq_i\,
\deltaZ{1}{\gamma_i}{\bar\alpha_i}{\barq_i}\,
\frac{\bar\calN^{(i)}_{\bar\alpha_i}(\barq_i)}{\calD{i}}
\,.
\eea
The counterterm $\deltaZ{2}{\Gamma}{}{}$ in~\refeq{eq:twoloopren} 
subtracts the local two-loop divergence that is left after subtraction of
the subdivergences.
If $\Gamma$ is 
a full two-loop vertex function, the
identity~\refeq {eq:twoloopren} 
can be applied at the level of the individual two-loop 
diagrams that contribute to $\Gamma$
by handling $\bfR$ as a linear operator.
Alternatively, \refeq{eq:twoloopren}
can be directly applied to the full vertex $\Gamma$. 
In this case, 
$\delta Z_{2,\Gamma}$ corresponds  to the full UV counterterm 
for $\Gamma$,
and $\delta Z_{1,\gamma}$ 
are the complete one-loop counterterms
for the vertices $\gamma$
that can be inserted into 
the one-loop vertex function $\calA_{1,\Gamma}$,
while 
$\deltaZ{1}{\gamma}{}{} \cdot \ampbar{1}{\Gamma/\gamma}{}{}$
embodies
all possible insertions of a certain counterterm
$\delta Z_{1,\gamma}$ into the various one-loop diagrams
that contribute to $\calA_{1,\Gamma}$.

The presence of a UV divergence in a 
subdiagram $\gamma_i$ can be identified by means of the 
degree of (sub)divergence
\bea
\label{eq:Xijdef}
X(\gamma_i) &=& X_{jk}(\Gamma) 
\,=\, 4+ U_j(\Gamma) +U_k(\Gamma) +
\sum_{a=0}^1 Y_a(\Gamma)\,,
\eea
where $U_m(\Gamma)$ denotes the maximum power in $q_m$ along the chain 
$\calC_m$, and  $Y_a(\Gamma)$ is the generic power in $q$ of the vertex
$\calV_a$.
Subdiagrams with $X(\gamma_i)\ge 0$ are UV divergent.
The remaining local two-loop divergences
can be identified by means of the
global degree of divergence 
\be
\label{eq:Xdef}
X(\Gamma) \,=\, 8+\sum_{i=1}^3 U_i(\Gamma) +\sum_{a=0}^1 Y_a(\Gamma)\,,
\ee
which corresponds to the total loop-momentum power of 
the full two-loop diagram.
Globally divergent diagrams, \ie diagrams with 
$X(\Gamma)\ge 0$, involve local divergences.

\begin{figure}[t]
\begin{center}
\bea
\hspace{-5mm}
&&\bfR\,
\left[\;
\vcenter{\hbox{\includegraphics[width=\diawidth]{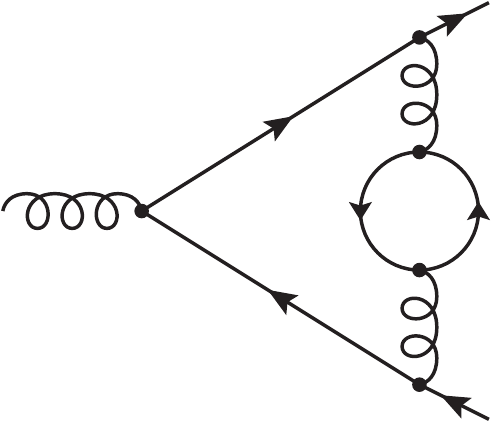}}}\;\;\right]_{\numdim\,=\,\dendim}
\hspace{-3mm}
\,=\, \nonumber\\[5mm]
\hspace{-5mm}&&=\,
\left[\;
\vcenter{\hbox{\includegraphics[width=\diawidth]{QCDvtxIILoop}}}  
\;\;+\;
\vcenter{\hbox{\includegraphics[width=\diawidth]{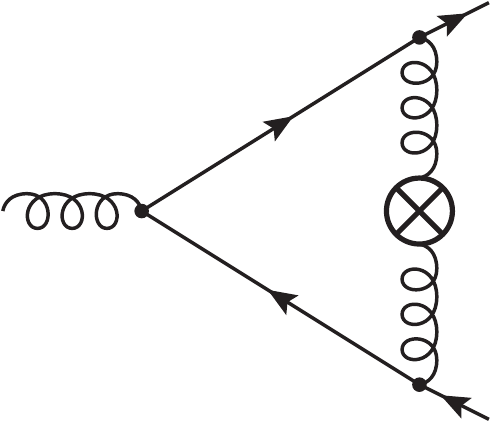}}}{
\left(\deltaZ{1}{\gamma_i}{}{}+ \deltaZtilde{1}{\gamma_i}{}{} 
+\ratamp{1}{\gamma_i}{}{}\right)
} 
\;+\;
\vcenter{\hbox{\includegraphics[width=\diawidth]{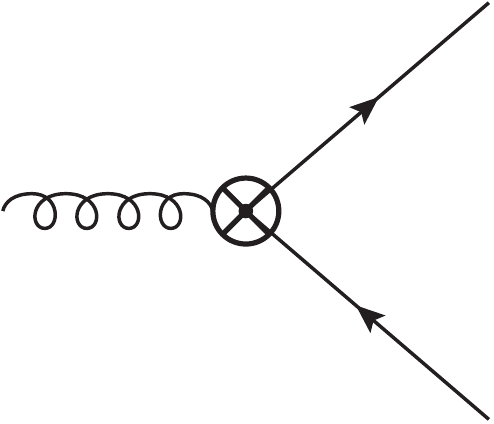}}} \hspace{-5mm}
\left(\deltaZ{2}{\Gamma}{}{} 
+\ratamp{2}{\Gamma}{}{}\right)
\;\right]_{\numdim\,=\,4}
\nonumber
\eea
\end{center}
\caption{Graphical representation of the master
formula~\refeq{eq:masterformula2} 
for the case of a globally divergent two-loop QCD diagram with a single subdivergence.
} 
\label{fig:twoloopratQED}
\end{figure}

As demonstrated in~\cite{Pozzorini:2020hkx},
the renormalised two-loop amplitude~\refeq{eq:twoloopren} 
in $\numdim=\dendim$ dimensions can be expressed in terms of 
amplitudes 
in $\numdim=4$ dimensions plus appropriate rational counterterms.
The corresponding master formula 
reads
\bea 
\label{eq:masterformula2} 
{\textbf{R}}\, \ampbar{2}{\Gamma}{}{}   
&=&  \amp{2}{\Gamma}{}{} + 
\sum  \limits_{\gamma} \lb \deltaZ{1}{\gamma}{}{} +\deltaZtilde{1}{\gamma}{}{} + \ratamp{1}{\gamma}{}{} \rb \cdot \amp{1}{\Gamma/\gamma}{}{}
\,+\, 
\deltaZ{2}{\Gamma}{}{} + \ratamp{2}{\Gamma}{}{}
\,,
\eea
and is illustrated in \reffi{fig:twoloopratQED}.
The first term on the rhs is
the unrenormalised two-loop amplitude in $\numdim=4$ dimensions, 
which corresponds to $\bar\calA_{2,\Gamma}$ with the
$\ntilde$-part of the numerator~\refeq{eq:rttwoloopDD} set to zero.
The second term contains all required 
one-loop counterterms---see~\refeq{eq:masterf1ddim}---for the cancellation of the
UV poles of the subdiagrams $\gamma$
and for the reconstruction of
the associated rational parts.
As for the remaining two-loop counterterms, $\delta Z_{2,\Gamma}$
is the same UV counterterm 
as in~\refeq{eq:twoloopren}, 
while the rational counterterm $\ratamp{2}{\Gamma}{}{}$
reconstructs all remaining contributions of order $\eps^{-1}$ and $\eps^0$ 
that originate form the interplay of the
$\ntilde$-part of the numerator 
with local UV divergences.

As demonstrated in~\cite{Pozzorini:2020hkx}, 
the $\ratamp{2}{\Gamma}{}{}$ terms are process-independent 
local counterterms, and can be computed once and for all in terms of 
tadpole integrals.

\subsection{Sketch of the proof}

In the following we review the 
key aspects of the 
proof of the master formula~\refeq{eq:masterformula2}
and we outline how to compute $\delta \calR_2$ terms
from tadpole integrals with one mass scale.

In renormalisable theories, 
two-loop diagrams $\Gamma$ with $X(\Gamma)<0$ 
involve at most one divergent subdiagram $\gamma$, for which, 
according 
to~\refeq{eq:masterf1ddim}, 
\bea
\label{eq:proofsketchA0}
\bar\calA_{1,\gamma}+
\delta  Z_{1,\gamma}
&=&
\calA_{1,\gamma}
+
\delta  Z_{1,\gamma}
+\delta {\tilde Z}_{1,\gamma}
+\delta \calR_{1,\gamma}
+\ord(\eps,\tilde q_2^2)\,,
\eea
where $\barq_2$ is the loop momentum that circulates
through the complementary part $\Gamma/\gamma$ of the two-loop diagram. Using this identity one can 
show 
that, up to negligible $\ord(\eps)$ terms,
\bea
\label{eq:proofsketchB}
\bar\calA_{2,\Gamma}+
\delta  Z_{1,\gamma}
\cdot
\bar\calA_{1,\Gamma/\gamma}
&=&
\calA_{2,\Gamma}
+
\left(\delta  Z_{1,\gamma}
+\delta {\tilde Z}_{1,\gamma}
+\delta \calR_{1,\gamma}
\right)\cdot
\calA_{1,\Gamma/\gamma}\,,
\eea
which is equivalent to the master formula~\refeq{eq:masterformula2} with
\bea
\label{eq:proofsketchC}
\deltaZ{2}{\Gamma}{}{}\,=\,0
\quad \mbox{and}\quad
\ratamp{2}{\Gamma}{}{}\,=\,0
\qquad \mbox{for}\quad
X(\Gamma)<0\,.
\eea
This means that two-loop rational terms $\delta \calR_{2,\Gamma}$ occur only in 
the presence of a local divergence. 
Therefore they can be determined, once and for
all,
by inverting the
master formula~\refeq{eq:masterformula2}, \ie by computing
\bea 
\label{eq:proofsketchD}
\ratamp{2}{\Gamma}{}{}
&=&
\ampbar{2}{\Gamma}{}{}  
-\amp{2}{\Gamma}{}{} + 
\sum  \limits_{\gamma} \deltaZ{1}{\gamma}{}{}
\cdot \ampbar{1}{\Gamma/\gamma}{}{} 
-
\sum  \limits_{\gamma} \lb \deltaZ{1}{\gamma}{}{}
+\deltaZtilde{1}{\gamma}{}{} + \ratamp{1}{\gamma}{}{} \rb \cdot
\amp{1}{\Gamma/\gamma}{}{}\qquad
\eea
for all 1PI amputated vertex functions $\Gamma$ with  $X(\Gamma)\ge 0$.
The fact 
that
$\delta \calR_{2,\Gamma}$ terms originate only from local UV
divergences makes it possible to express~\refeq{eq:proofsketchD}
in terms of tadpole integrals~\cite{Pozzorini:2020hkx}.
Technically, this 
is achieved by decomposing
two-loop diagrams into two parts, 
\bea
\label{eq:proofsketchE}
\ampbar{2}{\Gamma}{}{} &=& 
\ampbar{2}{\Gamma_\tad}{}{} 
+
\ampbar{2}{\Gamma_\rem}{}{}\,,
\eea
where $\Gamma_\tad$ is constructed through a systematic
expansion that embodies the local divergence of $\Gamma$ in the
form of tadpole integrals. 
Here we focus on the general properties of such tadpole
expansions, while their 
explicit form and various possible 
optimisations are discussed in detail in~\refapp{se:tadexp}.
By construction, the remnant of the expansion in~\refeq{eq:proofsketchE}
is free from local divergences, 
i.e.\footnote{This identity should hold in $\numdim=D$ dimensions, \ie both for the 
four-dimensional and the $(\numdim-4)$-dimensional parts of the 
loop numerator, where terms of $\ord(\tilde q_i)$ and $\ord(\eps q_i)$
should be counted on the same footing as $\ord(q_i)$.
The same holds also for~\refeq{eq:proofsketchI}.
}
\bea
\label{eq:proofsketchF}
X(\Gamma_\rem) &<& 0\,.
\eea
Therefore, according to~\refeq{eq:proofsketchC}
the $\delta\calR_{2}$ terms of $\Gamma$ arise only from its
$\Gamma_\tad$ part and can be calculated with the 
formula
\bea 
\label{eq:proofsketchG}
\ratamp{2}{\Gamma}{}{}
&=&
\ampbar{2}{\Gamma_\tad}{}{}  
-\amp{2}{\Gamma_\tad}{}{} 
+
\sum  \limits_{\gamma} \deltaZ{1}{\gamma_\tad}{}{}
\cdot \ampbar{1}{\Gamma_\tad/\gamma_\tad}{}{} 
\nonumber\\&&{}
-
\sum  \limits_{\gamma} \lb \deltaZ{1}{\gamma_\tad}{}{}
+\deltaZtilde{1}{\gamma_\tad}{}{} + \ratamp{1}{\gamma_\tad}{}{} \rb \cdot
\amp{1}{\Gamma_\tad/\gamma_\tad}{}{}\,.
\eea
Here $\gamma_\tad$ are the various subdiagrams of $\Gamma_\tad$, 
and $\deltaZ{1}{\gamma_\tad}{}{}$, $\deltaZtilde{1}{\gamma_\tad}{}{}$ and 
$\ratamp{1}{\gamma_\tad}{}{}$ 
are the one-loop counterterms associated with the corresponding
subdivergences. The latter do not need to coincide with the  
subdivergences of the original diagram $\Gamma$.
However, as discussed in~\refapp{se:tadexp}, 
the tadpole decomposition~\refeq{eq:proofsketchE} can be implemented in such a 
way that $\Gamma_\tad$ and $\Gamma$ have the same subdivergences.
More precisely, one can require that the difference between the
corresponding subdiagrams,
\bea
\label{eq:proofsketchH}
\bar\calA_{1,\delta\gamma_\tad}
&=&
\bar\calA_{1,\gamma}
-
\bar\calA_{1,\gamma_\tad}\,,
\eea
is free from UV divergences, \ie
\bea
\label{eq:proofsketchI}
X(\delta \gamma_\tad)&<&0\,.
\eea
If this condition is fulfilled, then 
all subdiagrams of $\Gamma_\tad$ and $\Gamma$
have identical UV poles and rational parts, \ie
\bea
\label{eq:proofsketchJ}
\delta Z_{1,{\gamma_\tad}} \,=\,  \delta Z_{1,{\gamma}}\,,
\qquad
\delta \tilde Z_{1,{\gamma_\tad}} \,=\,  \delta \tilde Z_{1,{\gamma}}\,,
\qquad
\delta \calR_{1,{\gamma_\tad}} \,=\,  \delta \calR_{1,{\gamma}}\,.
\eea
The above considerations apply to single 
two-loop diagrams $\Gamma$ with related subdiagrams $\gamma$, 
but can be directly extended---as discussed after~\refeq{eq:subdiagnotB}---to 
the case where $\Gamma$ and $\gamma$ are full vertex functions.

Based on the formula~\refeq{eq:proofsketchG}
and the 
general properties of the tadpole expansion 
(see \refapp{se:naivetadpoledec})
one can show~\cite{Pozzorini:2020hkx} that 
the $\ratamp{2}{\Gamma}{}{}$ counterterms associated with 
1PI vertex functions
take the form of homogeneous 
polynomials of degree $X(\Gamma)$
in the external momenta  $\{p_{ia}\}$ 
and internal masses $\{m_{ia}\}$.

\section{Renormalisation scheme transformations}
\label{se:schemetrans}

The goal of this section is to generalise the master formulas
\refeq{eq:masterformula1} and~\refeq{eq:masterformula2}, which have been 
derived in the $\ms$ scheme 
and its $\msbar$ variant, 
to any renormalisation scheme. 
As we will see, the form of the master formulas 
is independent of the renormalisation scheme,
\ie we will demonstrate that the renormalised one- and two-loop amplitudes 
in a  generic scheme $\scheme$ fulfil the relations
\bea
\label{eq:genmform1}
\bfR^{(\scheme)} 
\bar \calA_{1,\Gamma} 
&=&
\calA_{1,\Gamma}
\,+\,
\delta  Z^{(\scheme)}_{1,\Gamma}+\delta
\calR^{(\scheme)}_{1,\Gamma}
\,,
\\[3mm]
\label{eq:genmform2}
\bfR^{(\scheme)} 
\bar \calA_{2,\Gamma}
&=&
\calA_{2,\Gamma}
+\sum_\gamma \left(\delta  Z^{(\scheme)}_{1,\gamma}
+\delta {\tilde Z}^{(\scheme)}_{1,\gamma}
+\delta \calR^{(\scheme)}_{1,\gamma}
\right)\cdot
\calA_{1,\Gamma/\gamma}
\,+\,
\delta  Z^{(\scheme)}_{2,\Gamma}+\delta
\calR^{(\scheme)}_{2,\Gamma}
\,,\quad
\eea
where $\delta Z_{k}^{(\scheme)}$
are the UV counterterms in the scheme $\scheme$. As we will show, the
remaining \mbox{$k$-loop} counterterms $\delta \tilde Z^{(\scheme)}_{k}$
and rational terms $\delta \calR^{(\scheme)}_{k}$ are related to the corresponding objects 
in the $\msbar$ scheme
through transformations that involve only 
lower-order counterterms and rational terms. 
Actually, apart from a trivial scale dependence,
$\delta \calR_1$ and $\delta \tilde Z_1$ are 
scheme independent. As for the
$\delta \calR_2$ terms, it turns out---see~\refeq{eq:RtransfG} 
and~\refeq{eq:straK1c}---that 
their transformation 
 amounts to a finite renormalisation of the corresponding 
$\delta \calR_1$ terms 
plus a nontrivial 
contribution 
that 
can be 
expressed---see~\refeq{eq:straKdec} and~\refeq{eq:strarec1}---as 
a combination of one-loop renormalisation constants 
and auxiliary one-loop counterterms.

Note that, for consistency with the original formulation~\cite{Pozzorini:2020hkx},
the generalised master formulas~\refeq{eq:genmform1}--\refeq{eq:genmform2} 
are expressed in the language of the $\bfR$-operation, where
one-loop subdivergences are subtracted via insertion of 
UV counterterms $\delta  Z^{(\scheme)}_{1,\gamma}$ at the level of 
subdiagrams. 
This calls for an operational definition of 
such $\delta  Z^{(\scheme)}_{1,\gamma}$ insertions
in $\numdim=4$ dimensions in a generic scheme $\scheme$.
Motivated by the practical goal of expressing renormalised amplitudes in terms of 
loop integrals with four-dimensional numerators, we adopt the definition
\bea
\label{eq:subdivprescX}
\sum_\gamma \delta  Z^{(\scheme)}_{1,\gamma}
\cdot
\calA_{1,\Gamma/\gamma}
&=&
\left[\sum_\gamma \delta  Z^{(\scheme)}_{1,\gamma}
\cdot
\bar\calA_{1,\Gamma/\gamma}\right]_{\numdim = 4}\,,
\eea
\ie the $\delta  Z^{(\scheme)}_{1,\gamma}$ insertions on the rhs of
\refeq{eq:genmform2} should be understood as a standard UV subtraction in 
$\numdim=D$ dimensions 
with a {\it subsequent} projection to $\numdim=4$.
Actually, in a generic renormalisation scheme
the subtraction of UV divergences is controlled 
through the multiplicative renormalisation of parameters and fields
at the level of the Lagrangian. 
Thus, contrary to the case of the minimal subtraction scheme, 
the prescription~\refeq{eq:subdivprescX}
cannot be applied at the level of individual two-loop 
diagrams and their subdiagrams, but should be understood at the level of
the full sets of (sub)diagrams that are associated with the renormalisation
of a certain parameter or field.
In practice~\refeq{eq:subdivprescX} can be implemented,
as detailed in~\refeq{eq:straK1d}--\refeq{eq:straK1g},
via multiplicative renormalisation 
of the amplitude
$\bar\calA_{1,\Gamma}$ 
of a full $n$-point vertex function $\Gamma$
with a posteriori projection to 
$\numdim=4$.

In order to derive the above mentioned properties,
in \refses{se:multren}{se:msmsbarconv} we first
introduce an appropriate scheme-transformations
formalism for amplitudes in $\numdim=\dendim$ dimensions.
The connection to amplitudes in $\numdim=4$ dimensions is established 
in \refses{se:schdep4d}{se:nontrschdep4d}, and 
the main results for the scheme dependence of rational terms 
are presented in~\refeq{eq:RtransfG}--\refeq{eq:straK1c}
and in~\refse{se:fullfschemedep}.

\subsection{Multiplicative renormalisation}
\label{se:multren}
Let us consider a generic renormalisable theory with a certain set of
fields $\{\varphi_j\}$ and a set of parameters $\{\param_i\}=\{\alpha,
\lambda,m\}$.
For simplicity we restrict ourselves to a single coupling constant $\alpha$, 
a gauge-fixing parameter $\lambda$, and a mass
parameter $m$, but the formalism introduced in the following is 
applicable to any number of parameters.

To define a generic renormalisation scheme
we adopt the multiplicative renormalisation approach,
where the cancellation of UV divergences is controlled through
the identities
\bea
\label{eq:straA}
\varphi_{j,0} &=& \lb \calZ^{(\scheme)}_{\varphi_j} \rb^{1/2}
 \,\varphi_{i,\scheme} \,,
\qquad
\param_{i,0} \,=\, \calZ^{(\scheme)}_{\param_i}\,\param_{i,\scheme}\,
\qquad\mbox{for}\qquad \param_i=\alpha,\lambda,m\,,
\quad
\eea
where $\varphi_{j,0}$
and $\param_{i,0}$
denote the scheme-independent bare fields and parameters,
while $\varphi_{j,\scheme}$ and $\param_{i,\scheme}$
are their renormalised counterparts.
The label $\scheme$ corresponds to a generic renormalisation scheme, 
which may be the $\msbar$ scheme, the on-shell scheme, or any other scheme.
For the perturbative expansion of the renormalisation constants
we use the notation
\bea
\label{eq:straB}
\calZ^{(\scheme)}_{\rcarg}
\,=\, 1+ \sum_{k=1}^{\infty}\delta\calZ^{(\scheme)}_{k,\rcarg}
\qquad\mbox{for}\qquad
\rcarg\,=\,\alpha, \lambda, m, \varphi_j\,,
\eea
where the index $k$ is the order in the coupling constant,
\ie $\delta\calZ^{(\scheme)}_{k,\rcarg}\propto \alpha_\scheme^k$.
Note that
\refeq{eq:straA}--\refeq{eq:straB}
imply that
\bea
\label{eq:straB1}
\frac{\alpha_0}{\alpha_\scheme}
&=& 1+ \ord(\alpha_\scheme)\,.
\eea

The renormalisation formula for 
a scattering amplitude reads
\newcommand{\lndev}[1]{#1\frac{\partial}{\partial #1}}
\newcommand{\lndevsq}[2]{#1 #2 \frac{\partial^2}{\partial #1\partial#2}}
\bea
\label{eq:straC0}
\bfR^{(\scheme)} \bar \calA_\Gamma\left(\{\param_{i,\scheme}\}\right)
&=& 
\bigg[\prod_{j}
\Big(\calZ^{(\scheme)}_{\varphi_j}
\Big)^{1/2}\bigg]\,
\bar \calA_\Gamma\left(\{\param_{i,0}\}\right)\,,
\eea
where the index $j$ runs over all external legs, and
the scattering amplitude $\calA_\Gamma$ 
includes terms of any order in the
coupling constant.
The combined effect of field and 
parameter renormalisation in~\refeq{eq:straC0} can be cast in the form
\bea
\label{eq:straC3}
\bfR^{(\scheme)} \bar \calA_\Gamma \left(\{\param_{i,\scheme}\}\right)
&=&
\sum_{k=0}^\infty D^{(\scheme)}_{k} 
\,
\bar \calA_\Gamma \left(\{\param_{i,\scheme}\}\right)\,,
\eea
where $D^{(\scheme)}_{k}$ are operators of order
$\alpha_\scheme^k$ that contain combinations of 
renormalisation constants and derivatives with respect to the 
corresponding parameters, $\param_i=\alpha, \lambda, m$.
Up to second order they read
\bea
\label{eq:frc3}
D^{(\scheme)}_{0}
&=& 1\,,
\\[2mm]
\label{eq:frc3a}
D^{(\scheme)}_{1}
&=& 
\sum_{i} \delta\calZ^{(\scheme)}_{1,\param_i}\;\lndev{\param_i}
+
\sum_{j}\frac{1}{2}\delta\calZ^{(\scheme)}_{1,\varphi_j} \,,
\\[1mm]
\label{eq:frc3b}
D^{(\scheme)}_{2}
&=&
\sum_i \delta\calZ^{(\scheme)}_{2,\param_i}\;\lndev{\param_i}
+\frac{1}{2}
\sum_{i,k}
\delta\calZ^{(\scheme)}_{1,\param_i}
\delta\calZ^{(\scheme)}_{1,\param_k}
\,\lndevsq{\param_i}{\param_j}
+
\frac{1}{2}\sum_{i}\sum_{j} 
\delta\calZ^{(\scheme)}_{1,\varphi_j}
\delta\calZ^{(\scheme)}_{1,\param_i}\;\lndev{\param_i}
\nonumber\\
&&
{}+
\frac{1}{4}\sum_{i}\sum_{j< i}
\delta\calZ^{(\scheme)}_{1,\varphi_i}
\delta\calZ^{(\scheme)}_{1,\varphi_j} 
+
\sum_{j}\left[\frac{1}{2}\delta\calZ^{(\scheme)}_{2,\varphi_j} 
-\frac{1}{8}\left(\delta\calZ^{(\scheme)}_{1,\varphi_j}\right)^2 
\right]
\,.
\eea

Let us now consider the interplay of~\refeq{eq:straC3}
with the perturbative expansion of the unrenormalised 
scattering amplitude,
\bea
\label{eq:straB2}
\bar\calA_{\Gamma} \left(\{\param_{i,\scheme}\}\right)
&=&
\sum_{k=0}^{\infty}
\bar\calA_{k,\Gamma}\left(\{\param_{i,\scheme}\}\right)\,,
\eea
where $\bar\calA_{k,\Gamma}$  denotes the $k$-loop contribution,
and we assume that 
$\bar \calA_{k,\Gamma}\propto\alpha_\scheme^{p+k}$.
Combining~\refeq{eq:straC3} with~\refeq{eq:straB2} we can write the 
$n$-loop contribution to the 
renormalised amplitude as
\bea
\label{eq:straD0}
\bfR^{(\scheme)} \ampbar{n}{\Gamma}{}{} 
&=& 
\sum_{m=0}^n
D^{(\scheme)}_{n-m}\, \ampbar{m}{\Gamma}{}{}\,,
\eea
or, more explicitly, up to two loops
\bea
\label{eq:straD1}
\bfR^{(\scheme)} \ampbar{0}{\Gamma}{}{} 
&=& 
\amp{0}{\Gamma}{}{}
\\[1mm]
\bfR^{(\scheme)} \ampbar{1}{\Gamma}{}{} 
&=& 
\ampbar{1}{\Gamma}{}{}
+
D^{(\scheme)}_{1} \amp{0}{\Gamma}{}{}\,,
\\[1mm]
\label{eq:straD2}
\bfR^{(\scheme)} \ampbar{2}{\Gamma}{}{}
&=& 
\ampbar{2}{\Gamma}{}{}
+D^{(\scheme)}_{1} \ampbar{1}{\Gamma}{}{}
+D^{(\scheme)}_{2} \amp{0}{\Gamma}{}{}\,.
\eea
Note that here we assume that the tree amplitude is free from
$(D-4)$-dimensional parts, \ie $\ampbar{0}{\Gamma}{}{}=\amp{0}{\Gamma}{}{}$.

\subsection{Scale dependence and scheme transformations}
\label{se:msmsbarconv}

The subtraction of UV singularities is associated with a
renormalisation scale $\mu_\rR$, which enters the
renormalisation formulas through a dimensionless ratio of the form
\bea
\label{eq:msstI}
t_\scheme &=&
\frac{\msfact_\scheme \mu_0^2}{\mu_\rR^2}\,, 
\eea
where $\mu_0$ and $\mu_\rR$ are, respectively, the 't~Hooft scale 
of dimensional regularisation and the renormalisation scale.
Note that $\mu_\rR$ can be introduced 
``by hand'' as part of the technical prescription for the 
renormalisation of $\alpha$, like in the $\ms$ and $\msbar$ schemes,
or it can arise from a physical renormalisation condition, 
like in on-shell
schemes, in which case $\mu_\rR$ is typically a physical mass or energy
scale.
The term $\msfact_\scheme$ in~\refeq{eq:msstI} 
is a scheme-dependent factor, and its 
explicit values in the $\ms$ and $\msbar$ schemes are indicated 
in~\refeq{eq:msrescfact} and~\refeq{eq:msstG}.

The scale dependence can be implemented in two different ways,
which can be regarded as two equivalent formulations
of a scale-dependent renormalisation scheme. These two approaches 
will be referred to as scheme $X_0$ and scheme $X$.
The first approach corresponds to renormalisation identities of the form 
\bea
\label{eq:genrsA}
\varphi_{j,0} &=& \lb \calZ^{(\schemez)}_{\varphi_j} \rb^{1/2}
\,\varphi_{i,\schemez}\,,
\quad
\alpha_{0} \,=\,
t_\scheme^{-\eps}\,
\calZ^{(\schemez)}_{\alpha}\,\alpha_{\schemez}(\mu_\rR^2)\,,
\quad
\param_{i,0} \,=\,
\calZ^{(\schemez)}_{\param_i}\,\param_{i,\schemez}(\mu_\rR^2)\,,
\quad
\eea
for $\param_i=\lambda,m$.
Here the scale dependence is entirely controlled through the 
factor $t_\scheme^{-\eps}$ in the renormalisation of the coupling constant, 
while the renormalisation constants $\calZ^{(\schemez)}_{\rcarg}$
are free from any explicit scale dependence.
Their perturbative expansion has the same form as~\refeq{eq:straB}, \ie
\bea
\label{eq:genrsB}
\calZ^{(\schemez)}_{\rcarg}
\,=\,
1+\sum_{k=1}^\infty 
\delta\calZ^{(\schemez)}_{k,\rcarg} 
\qquad\mbox{for}\qquad
\rcarg\,=\,\alpha, \lambda, m, \varphi_j\,.
\eea

At variance with~\refeq{eq:straB1}, the
above renormalisation identities imply
\bea
\label{eq:genrsB2}
\frac{\alpha_0}{\alpha_{\schemez}}
&=& t_\scheme^{-\eps} + \ord(\alpha_\scheme)\,,
\eea
and are thus inconsistent with the formalism of~\refse{se:multren}.
This issue can be circumvented by 
converting the above renormalisation identities
into the form~\refeq{eq:straA}--\refeq{eq:straB}. This can be achieved 
by rescaling 
the coupling constant as
\bea
\label{eq:genrsB3}
\alpha_{\schemez}(\mu_\rR^2) &=&
t_\scheme^\eps\alpha_\scheme(\mu_\rR^2)\,,
\eea
while keeping all other parameters and fields unchanged.
This finite renormalisation%
\footnote{To be more explicit, such finite renormalisation 
is defined through~\refeq{eq:genrsB3} in combination with 
\bea
\varphi_{j,\schemez}\,=\,\varphi_{j,\scheme}\,,
\qquad
\param_{i,\schemez}(\mu_\rR^2)\,=\,
\param_{i,\scheme}(\mu_\rR^2)
\qquad\mbox{for}\quad \param_i=\lambda,m\,,
\eea
and the renormalisation constants in the schemes $\schemez$ and $\scheme$
are related via 
\bea
\calZ^{(\scheme)}_\chi \,=\,
\calZ^{(\schemez)}_\chi\Big|_{\alpha\,=\,t_\scheme^\eps\alpha_\scheme(\mu_\rR^2)}
\qquad\mbox{for}\quad \chi=\alpha,\lambda,m,\varphi_j\,.
\eea
} 
turns the scheme $\schemez$ into the equivalent scheme $\scheme$, and the 
resulting renormalisation identities read
\bea
\label{eq:genrsC}
\varphi_{j,0} &=& \lb \calZ^{(\scheme)}_{\varphi_j} \rb^{1/2} \,\varphi_{i,\scheme}\,,
\qquad
\param_{i,0} \,=\,
\calZ^{(\scheme)}_{\param_i}\,\param_{i,\scheme}(\mu_\rR^2)\,
\qquad\mbox{for}\qquad \param_i=\alpha,\lambda,m\,,
\qquad
\eea
where 
\bea
\label{eq:genrsD1}
\calZ^{(\scheme)}_{\rcarg}
\,=\,
1+\sum_{k=1}^\infty 
\delta\calZ^{(\scheme)}_{k,\rcarg}
\qquad\mbox{for}\qquad
\rcarg\,=\,\alpha, \lambda, m, \varphi_j\,,
\eea
with
\bea
\label{eq:genrsD2}
\delta\calZ^{(\scheme)}_{k,\rcarg} \,=\,
\left(t_\scheme^{\eps}\right)^k\,
\delta\calZ^{(\schemez)}_{k,\rcarg}
\Big|_{\alpha\,=\,\alpha_\scheme(\mu_\rR^2)}
\,.
\eea
In this way, all scale-dependent factors are 
reabsorbed into the coupling factors associated with 
the renormalisation constants.

The coupling constants corresponding to the schemes $\scheme$ and $\schemez$, 
defined through~\refeq{eq:genrsA}--\refeq{eq:genrsB} 
and~\refeq{eq:genrsC}--\refeq{eq:genrsD2},
differ by a factor $t_\scheme^\eps=1+\ord(\eps)$. 
Thus these two schemes are not identical. Nevertheless they are equivalent
since finite quantities in the scheme $\schemez$ and $\scheme$
differ only by irrelevant terms of $\ord(\eps)$.
In particular, at the level of renormalised amplitudes the schemes
$\scheme$ and $\schemez$ yield equivalent results.
In the following sections we will treat scale-dependent terms 
as in~\refeq{eq:genrsC}--\refeq{eq:genrsD2},
which guarantees the consistency with the formalism of~\refse{se:multren}
and makes it possible to use the renormalisation identities
\refeq{eq:straC3}--\refeq{eq:straD2}.

Note that the scale-independent parts $\delta\calZ^{(\schemez)}_{k,\rcarg}$ 
of~\refeq{eq:genrsD2} 
can involve terms of arbitrary high order in 
$\eps$,
and the $n$-loop renormalised amplitudes~\refeq{eq:straD0}
receive finite contributions from all 
$\eps$-suppressed terms of $\delta\calZ^{(\schemez)}_{k,\rcarg}$  up to order 
$\eps^{n-k}$.

\subsubsection*{Minimal subtraction schemes}

The $\ms$ and $\msbar$ schemes are simple realisations of the above
scale-dependent renormalisation prescriptions. In the case of the
$\ms$ scheme, the rescaling factor in~\refeq{eq:msstI} 
is simply
\bea
\label{eq:msrescfact}
\msfact_{\ms} = 1\,.
\eea
Thus for the rescaled renormalisation constants 
\refeq{eq:genrsD2} we have
\bea
\label{eq:genrsE}
\delta\calZ^{(\ms)}_{k,\rcarg} 
\,=\,
\left(\frac{\mu_0^2}{\mu_\rR^2}\right)^{k\eps}\,
\delta\calZ^{(\msz)}_{k,\rcarg}
\qquad\mbox{for}\qquad
\rcarg\,=\,\alpha, \lambda, m, \varphi_j\,,
\eea
while their scale-independent parts 
involve only pure $1/\eps$ poles, \ie
\bea
\label{eq:genresF}
\delta\calZ^{(\msz)}_{k,\rcarg}
&=&
\left[\alpha_\ms(\mu_\rR^2)\right]^k
\sum_{m=1}^k \,
\frac{b^{(m)}_{k,\rcarg}}{\eps^m}\,.
\eea
The difference between the $\ms$ and $\msbar$ schemes lies only in the 
rescaling parameter $\msfact_{\msbar}$, 
which is defined through \cite{Kataev:1988sq}
\bea
\label{eq:msstG}
\left(\msfact_{\msbar}\right)^\eps
&=& 
(4\pi)^\eps\Gamma(1+\eps)=\left(4\pi
e^{-\gamma_{\mathrm{E}}}\right)^\eps+\ord(\eps^2)\,.
\eea
Otherwise, the scale-independent parts of the renormalisation constants 
are equivalent.
Thus
\bea
\label{eq:genrsG}
\delta\calZ^{(\msbar)}_{k,\rcarg} 
\,=\,
\left(\frac{\msfact_{\msbar}\,\mu_0^2}{\mu_\rR^2}\right)^{k\eps}\,
\delta\calZ^{(\msz)}_{k,\rcarg}
\qquad\mbox{for}\qquad
\rcarg\,=\,\alpha, \lambda, m, \varphi_j\,.
\eea
Here it is implicitly understood that $\alpha_{\ms}(\mu_\rR^2)$ is replaced by 
$\alpha_{\msbar}(\mu_\rR^2)$ in $\delta\calZ^{(\msz)}_{k,\rcarg}$.

For later convenience, we introduce also a generalised minimal subtraction
scheme that we label $\ms_\scheme$
and is defined through the renormalisation constants
\bea
\label{eq:genrsF}
\delta\calZ^{(\ms_\scheme)}_{k,\rcarg}
&=&
\left(t_\scheme^{\eps}\right)^k\,\delta\calZ^{(\msz)}_{k,\rcarg}
\,=\,
\left(\frac{\msfact_{\scheme}\,\mu_0^2}{\mu_\rR^2}\right)^{k\eps}\,
\,\delta\calZ^{(\msz)}_{k,\rcarg}
\,,
\eea
where the rescaling factor $\msfact_\scheme$ is a freely adjustable
parameter. The relation between the coupling constants in the
schemes $\ms_{\scheme}$ and $\ms$ 
can be easily derived from the scale-independence of $\alpha_0$
and reads
\bea
\label{eq:msxcoupling}
\alpha_{\ms_{\scheme}}(\mu_\rR^2) \,=\, 
\alpha_{\ms}\left(\mu_\rR^2\,\msfact^{-1}_{\scheme}
\right)\,.
\eea
Let us analyse the scale dependence of renormalised amplitudes in the 
$\ms_{\scheme}$ scheme,
\bea
\label{eq:msxramp}
\bfR^{(\ms_{\scheme})} \ampbar{n}{\Gamma}{}{} (\mu^2_\rR)
&=& 
\sum_{m=0}^n
D^{(\ms_{\scheme})}_{n-m}\, \ampbar{m}{\Gamma}{}{}\,.
\eea
The rhs depends on $\mu_\rR$ and $\msfact_\scheme$ 
via the renormalisation operators
\bea
D^{(\ms_{\scheme})}_{n-m}
&=&
\left(t_\scheme^{\eps}\right)^{n-m} 
D^{(\ms_0)}_{n-m}\,.
\eea
Here the 
part $\left(t_\scheme^{-\eps}\right)^{m}$ of the scale-dependent
factor 
can be reabsorbed into the
unrenormalised amplitude 
$\bar\calA_{m,\Gamma}$ 
by using 
\bea
\left(t_\scheme^{-\eps}\right)^{m}
\ampbar{m}{\Gamma}{}{}
\,=\,
\ampbar{m}{\Gamma}{}{}
\Big|_{\mu^2_0\,=\,\mu^2_\rR\msfact_\scheme^{-1}}\,,
\eea
which follows from the fact that the dependence of
$\bar\calA_{m,\Gamma}$ on $\mu_0$ amounts to an overall factor 
$(\mu_0^{2\eps})^m$ stemming from the integration measure
\refeq{eq:intmeasure}.
This yields
\bea
\label{eq:mszrampres}
\bfR^{(\ms_{\scheme})} \ampbar{n}{\Gamma}{}{} (\mu^2_\rR)
&=& 
\left(t_\scheme^{\eps}\right)^{n}
\sum_{m=0}^n
D^{(\ms_0)}_{n-m}\, \ampbar{m}{\Gamma}{}{}
\Big|_{\mu^2_0\,=\,\mu^2_\rR\msfact_\scheme^{-1}}\,,
\eea
where, due to the finiteness of the renormalised amplitude, the
overall factor $(t_\scheme^{\eps})^n$ generates only irrelevant $\ord(\eps)$
contributions, while the rest of the scale dependence is entirely 
controlled through the prescription $\mu^2_0=\mu_\rR^2\msfact_\scheme^{-1}$
for unrenormalised amplitudes.
For the special case of the $\ms$ scheme the above identity reads
\bea
\label{eq:minsubrampres}
\bfR^{(\ms)} \ampbar{n}{\Gamma}{}{} (\mu^2_\rR)
&=& 
\left(t_{\ms}^{\eps}\right)^{n}
\sum_{m=0}^n
D^{(\ms_0)}_{n-m}\, \ampbar{m}{\Gamma}{}{}
\Big|_{\mu^2_0\,=\,\mu^2_\rR}\,,
\eea
where $t_{\ms} = \mu_0^2/\mu_\rR^2$.
Comparing the above equations we see that 
the renormalised amplitudes
in the $\ms_{\scheme}$ and $\ms$ schemes
are connected through the same 
rescaling as in~\refeq{eq:msxcoupling}.
More precisely,
\bea
\label{eq:msmsxrelation}
\bfR^{(\ms_{\scheme})} \ampbar{n}{\Gamma}{}{} (\mu^2_\rR)
&=& 
\left(\msfact_\scheme^{\eps}\right)^n
\bfR^{(\ms)} \ampbar{n}{\Gamma}{}{} (\mu^2_\rR\msfact_\scheme^{-1})
\,=\,
\bfR^{(\ms)}
\ampbar{n}{\Gamma}{}{} (\mu^2_\rR\msfact_\scheme^{-1})
+\ord(\eps)\,.\quad
\eea
A similar relation holds also for the 
dependence on the rescaling factor $\msfact_\scheme$ in a
generic scheme $\scheme$.

\subsubsection*{Scheme transformations}

Let us now discuss generic transformations that connect two multiplicative
renormalisation schemes of the form~\refeq{eq:genrsC}--\refeq{eq:genrsD2}.
Specifically we consider the transformation that connects 
the schemes $\scheme$ and $\ms_\scheme$ defined in~\refeq{eq:genrsD2}
and~\refeq{eq:genrsF}. 
Such transformations can be formulated at the 
level of the renormalisation constants as
\bea
\label{eq:genstrA}
\calZ^{(\scheme)}_{\rcarg} &=&
\calZ^{(\dscheme)}_{\rcarg}\,
\calZ^{(\ms_\scheme)}_{\rcarg}
\qquad\mbox{for}\qquad
\rcarg\,=\,\alpha, \lambda, m, \varphi_j\,,
\eea
where the renormalisation constant 
\bea
\label{eq:genstrB}
\calZ^{(\dscheme)}_{\rcarg}
\,=\,
1+\sum_{k=1}^\infty 
\delta\calZ^{(\dscheme)}_{k,\rcarg}\,
\eea
is free from
$1/\eps$ poles. Its $k$-loop parts
can be expressed as
\bea
\label{eq:genrsG}
\calZ^{(\dscheme)}_{k,\rcarg} 
&=&
\left(t_\scheme^{\eps}\right)^k\,\calZ^{(\dschemez)}_{k,\rcarg}
\,,
\eea
where $\calZ^{(\dschemez)}_{k,\rcarg}$ denotes the scale-independent part.
Note that the scale factors $t^\eps_\scheme$  that enter the
renormalisation constants $\calZ^{(\scheme)}_{\rcarg}$, 
$\calZ^{(\dscheme)}_{\rcarg}$  and $\calZ^{(\ms_\scheme)}_{\rcarg}$ 
in~\refeq{eq:genstrA}
need to be identical. 
This is mandatory in order to match all logarithms of $t_\scheme$ that arise 
from terms of type $\left(\alpha_\scheme\, t_\scheme^{\eps}\right)^n \eps^{-m}$.

Based on the factorisation of the renormalisation constants
in~\refeq{eq:genstrA}, the renormalisation in the scheme 
$\scheme$ can be regarded as a two-step procedure
consisting of a  subtraction of UV poles in the $\ms_\scheme$ scheme
and a subsequent multiplicative renormalisation 
with the finite renormalisation constants~\refeq{eq:genstrB}.
More explicitly, the renormalised amplitudes~\refeq{eq:straD0} 
in the scheme $\scheme$ can be obtained 
starting form $\ms_\scheme$ renormalised amplitude through
\bea
\label{eq:twostepren}
\bfR^{(\scheme)} \bar \calA_\Gamma\left(\{\param_{i,\scheme}\}\right)
&=& 
\bigg[\prod_{j}
\Big(\calZ^{(\dscheme)}_{\varphi_j}
\Big)^{1/2}\bigg]\,
\bfR^{(\ms_\scheme)} 
\bar \calA_\Gamma\left(\{\param_{i,\ms_\scheme}\}\right)\,,
\eea
which is equivalent to
\bea
\label{eq:finstraD0}
\bfR^{(\scheme)} \ampbar{n}{\Gamma}{}{} 
&=& 
\sum_{m=0}^n
D^{(\dscheme)}_{n-m}\,
\bfR^{(\ms_\scheme)} 
\ampbar{m}{\Gamma}{}{}\,,
\eea
where it is understood that 
$\param_i=\param_{i,\scheme}(\mu_\rR^2)$
for all objects on the rhs,
and the operators $D^{(\dscheme)}_{k}$
are defined as in~\refeq{eq:straC3}--\refeq{eq:frc3b}
but with renormalisation constants 
$\delta \calZ_k^{(\scheme)}$ replaced by
$\delta \calZ_k^{(\dscheme)}$.
Note that in~\refeq{eq:finstraD0}
the 
parameter derivatives $\partial/\partial_{\theta_i}$ 
contained in $D^{(\dscheme)}_{k}$
act on all building blocks of $\bfR^{(\ms_\scheme)} 
\ampbar{m}{\Gamma}{}{}$, \ie both on amplitudes and renormalisation constants.
At one and two loops~\refeq{eq:finstraD0} reads
\bea
\label{eq:straE1}
\bfR^{(\scheme)} \ampbar{1}{\Gamma}{}{} 
&=& 
\bfR^{(\ms_\scheme)} 
\ampbar{1}{\Gamma}{}{}
+
D^{(\dscheme)}_{1} \amp{0}{\Gamma}{}{}\,,
\\[1mm]
\label{eq:straE2}
\bfR^{(\scheme)} \ampbar{2}{\Gamma}{}{}
&=& 
\bfR^{(\ms_\scheme)} 
\ampbar{2}{\Gamma}{}{}
+D^{(\dscheme)}_{1}\,\bfR^{(\ms_\scheme)}  \ampbar{1}{\Gamma}{}{}
+D^{(\dscheme)}_{2}\, \amp{0}{\Gamma}{}{}\,.
\eea
In the following sections, 
these identities will be used as a starting point to derive the master 
formulas~\refeq{eq:genmform1} and~\refeq{eq:genmform2}
as well as the rules to transform rational terms from the 
minimal subtraction scheme to a generic scheme $\scheme$.
To this end, we will also make use of the relation
\bea
\label{eq:finstraD2}
D^{(\scheme)}_{k}\, 
&=& 
\sum_{j=0}^k
D^{(\dscheme)}_{k-j}\, 
D^{(\ms_\scheme)}_{j}\,,
\eea
which can be derived by applying~\refeq{eq:straD0} on both sides 
of~\refeq{eq:finstraD0}.
At one and two loops, \refeq{eq:finstraD2} reads
\bea
\label{eq:finstraD3}
D^{(\scheme)}_{1}
&=& 
D^{(\dscheme)}_{1}+
D^{(\ms_\scheme)}_{1}\,,
\\[1mm]
\label{eq:finstraD4}
D^{(\scheme)}_{2}
&=& 
D^{(\dscheme)}_{2}+
D^{(\dscheme)}_{1}
D^{(\ms_\scheme)}_{1}+
D^{(\ms_\scheme)}_{2}\,.
\eea

\subsection{Renormalisation formulas in $\numdim=4$ in a generic scheme}
\label{se:schdep4d}

In this section we establish the generalised master formulas
\refeq{eq:genmform1}--\refeq{eq:genmform2} 
by making use of finite multiplicative renormalisations.
To this end, we first transform the original master formulas
\refeq{eq:masterformula1} and~\refeq{eq:masterformula2}
to the $\ms_\scheme$ scheme defined in~\refeq{eq:genrsF}, and we then 
extend them to the generic scheme $\scheme$
by means of~\refeq{eq:straE1}--\refeq{eq:straE2}.

In~\cite{Pozzorini:2020hkx} 
the analysis of rational terms 
was restricted to the $\ms$ and $\msbar$ schemes, and
the regularisation scale $\mu_0$ has been identified 
with the renormalisation scale $\mu_\rR$.
In order to generalise the results of~\cite{Pozzorini:2020hkx} 
to the $\ms_\scheme$ scheme with arbitrary scale 
$\mu_0$, let us start from the 
master formulas~\refeq{eq:masterformula1} and~\refeq{eq:masterformula2}
in the $\ms$ scheme with $\mu_0=\mu_\rR$,
\bea
\label{eq:4dmsstA0}
\bfR^{(\ms)} 
\bar \calA_1 
&=&
\calA_1
\Big|_{\mu_0^2=\mu_\rR^2}  
+
\left(\delta Z^{(\msz)}_{1,\Gamma}+\delta
\calR^{(\msz)}_{1,\Gamma}\right)\,,
\\[3mm]
\label{eq:4dmsstA1}
\bfR^{(\ms)} 
\bar \calA_2 
&=&
\calA_2
\Big|_{\mu_0^2=\mu_\rR^2}   
+\sum_\gamma \left(\delta Z^{(\msz)}_{1,\gamma}+\delta \tilde Z^{(\msz)}_{1,\gamma}
+\delta \calR^{(\msz)}_{1,\gamma}
\right)\cdot
\calA_{1,\Gamma/\gamma}
\Big|_{\mu_0^2=\mu_\rR^2}  
\nonumber\\
&&{}+
\left(\delta Z^{(\msz)}_{2,\Gamma}+\delta
\calR^{(\msz)}_{2,\Gamma}\right)\,.
\eea
Here all 
counterterms
$\delta Z_k^{(\msz)}$ and $\delta \tilde Z_k^{(\msz)}$
involve only poles or order $\eps^{-1},\dots, \eps^{-k}$,
and the rational counterterms $\delta \calR_k^{(\msz)}$
involve only finite terms and poles of order $\eps^{0},\dots, \eps^{-k+1}$.
In particular,  for $\mu_0=\mu_\rR$  
all $\ms$ counterterms, including $\delta \calR_k^{(\msz)}$,
do not depend on any scale~\cite{Pozzorini:2020hkx}. Note also that on the
lhs of~\refeq{eq:4dmsstA0}--\refeq{eq:4dmsstA1}
we do not indicate the special choice
$\mu_0=\mu_\rR$ since renormalised amplitudes are independent of $\mu_0$.

Based on~\refeq{eq:msmsxrelation} 
the above relations can be easily generalised 
to the $\ms_{\scheme}$ scheme by setting
$\mu_0^2=\mu_\rR^2\msfact_\scheme^{-1}$ on the rhs.
Moreover, along similar lines as
in~\refeq{eq:msxramp}--\refeq{eq:mszrampres},
the scale dependence can be reabsorbed into 
the counterterms by using 
\bea
\label{eq:4dmsstB}
\calA_{m}\Big|_{\mu_0^2=\mu_\rR^2\msfact_\scheme^{-1}} \,=\,
\left(t_\scheme^{-\eps}\right)^{m}
\calA_{m}\,,
\eea
and multiplying 
$\bfR^{(\ms_\scheme)} 
\bar \calA_n$
by an 
overall factor 
$\left(t_\scheme^{\eps}\right)^{n}$.
In this way, discarding irrelevant $\ord(\eps)$ terms,
one arrives at
\bea
\label{eq:4dmsstC1}
\bfR^{(\ms_\scheme)} 
\bar \calA_1 
&=&
\calA_1
\,+\,
Z^{(\ms_\scheme)}_{1,\Gamma}+\delta
\calR^{(\ms_\scheme)}_{1,\Gamma}
\,,
\\[3mm]
\label{eq:4dmsstC2}
\bfR^{(\ms_\scheme)} 
\bar \calA_2 
&=&
\calA_2
+\sum_\gamma \left(\delta  Z^{(\ms_\scheme)}_{1,\gamma}
+\delta {\tilde Z}^{(\ms_\scheme)}_{1,\gamma}
+\delta \calR^{(\ms_\scheme)}_{1,\gamma}
\right)\cdot
\calA_{1,\Gamma/\gamma}
\nonumber\\[1mm]&&{}
+
\delta  Z^{(\ms_\scheme)}_{2,\Gamma}+\delta
\calR^{(\ms_\scheme)}_{2,\Gamma}
\,,
\eea
where all scale-dependent factors are absorbed 
into the counterterms and rational terms
\bea
\delta {Y}^{(\ms_\scheme)}_{k,\Gamma} 
&=&
\left(t_\scheme^{\eps}\right)^{k}
\delta Y^{(\msz)}_{k,\Gamma} 
\qquad\mbox{for}\qquad
\delta Y \, =\, 
\delta Z,\, 
\delta \tilde Z,\, 
\delta\calR\,.
\eea
These relations can be 
directly converted to the $\ms$ or the $\msbar$ schemes by simply
replacing $t^{\eps}_\scheme$
by 
$t^{\eps}_{\ms} =
\left(\mu_0^2/\mu_\rR^2\right)^\eps$
or
$t^{\eps}_{\msbar} =
\left(\msfact_{\msbar}\,\mu_0^2/\mu_\rR^2\right)^\eps$, respectively.

The explicit calculations of rational terms in~\cite{Pozzorini:2020hkx} 
and in the present paper have been carried out in the
$\msbar$ scheme using the 
special scale choice $\mu_0^2=\mu_\rR^2\msfact^{-1}_{\msbar}$, 
where $t_{\msbar}=1$, such that all counterterms are free from scale
factors,
and loop integrals are free from 
logarithms of $\msfact_{\msbar}$.

In the following we derive the master formulas~\refeq{eq:genmform1}--\refeq{eq:genmform2}
as well as scheme transformation identities for 
the
rational terms 
by combining~\refeq{eq:4dmsstC1}--\refeq{eq:4dmsstC2}
with the scheme transformations~\refeq{eq:straE1}--\refeq{eq:straE2}.
In this context we use the trivial identity
\bea
\label{eq:RtransfA}
\delta Z^{(\scheme)}_{k,\Gamma}
&=& 
D^{(\scheme)}_{k} \amp{0}{\Gamma}{}{}\,,
\eea
which relates counterterms in the language of the $\bfR$-operation
to the multiplicative renormalisation formalism.

The one-loop master formula~\refeq{eq:genmform1} is 
obtained by applying~\refeq{eq:4dmsstC1} on the rhs of
\refeq{eq:straE1}. This results into
\bea
\label{eq:RtransfB}
\bfR^{(\scheme)} \ampbar{1}{\Gamma}{}{} 
&=& 
\calA_{1,\Gamma}
+
\left(\delta  Z^{(\ms_\scheme)}_{1,\Gamma}+\delta  \calR^{(\ms_\scheme)}_{1,\Gamma}\right)
+
D^{(\dscheme)}_{1} \amp{0}{\Gamma}{}{}\,,
\eea
which is equivalent to~\refeq{eq:genmform1}.
Equating the two identities, and
using 
\refeq{eq:RtransfA} together with \refeq{eq:finstraD3},
one arrives at the following scheme-transformation 
formula for rational terms,
\bea
\label{eq:RtransfC}
\calR^{(\scheme)}_{1,\Gamma}
&=&
\calR^{(\ms_\scheme)}_{1,\Gamma}
\,=\,
t_\scheme^{\eps}\,\calR^{(\msz)}_{1,\Gamma}\,.
\eea
This means that, apart from the trivial  $t_\scheme^{\eps}$ scale factor,
one-loop rational terms are scheme independent.
As we will see, this property holds also for the
$\delta \tilde Z_1$ terms that appear on the rhs of~\refeq{eq:genmform2}, \ie
\bea
\label{eq:RtransfD}
\delta \tilde Z^{(\scheme)}_{1,\Gamma}
&=&
\delta \tilde Z^{(\ms_\scheme)}_{1,\Gamma}
\,=\,
t_\scheme^{\eps}\,\delta \tilde Z^{(\msz)}_{1,\Gamma}\,.
\eea

The two-loop master formula~\refeq{eq:genmform2} 
can be derived by applying~\refeq{eq:4dmsstC1} and~\refeq{eq:4dmsstC2} 
on the rhs of~\refeq{eq:straE2}. This yields
\bea
\label{eq:RtransfE}
\bfR^{(\scheme)} \ampbar{2}{\Gamma}{}{}
&=& 
\calA_{2,\Gamma}
+\sum_\gamma \left(\delta  Z^{(\ms_\scheme)}_{1,\gamma}
+\delta {\tilde Z}^{(\ms_\scheme)}_{1,\gamma}
+\delta \calR^{(\ms_\scheme)}_{1,\gamma}
\right)\cdot
\calA_{1,\Gamma/\gamma}
+
\left(\delta  Z^{(\ms_\scheme)}_{2,\Gamma}+\delta
\calR^{(\ms_\scheme)}_{2,\Gamma}\right)
\nonumber\\
&&{}+D^{(\dscheme)}_{1}\,
\left(\calA_{1,\Gamma}+
\delta  Z^{(\ms_\scheme)}_{1,\Gamma}+\delta  \calR^{(\ms_\scheme)}_{1,\Gamma}\right)
+D^{(\dscheme)}_{2}\, \amp{0}{\Gamma}{}{}\,,
\eea
which is equivalent to~\refeq{eq:genmform2}.
Equating the two relations,
and using~\refeq{eq:RtransfA}
with~\refeq{eq:finstraD4}
and~\refeq{eq:RtransfC}, 
yields~\refeq{eq:RtransfD} 
together with the following 
identity between rational 
terms in the $\ms_\scheme$ and $\scheme$ schemes,
\bea
\label{eq:RtransfF}
\sum_\gamma \delta  Z^{(\scheme)}_{1,\gamma}
\cdot\calA_{1,\Gamma/\gamma}
+
\delta \calR^{(\scheme)}_{2,\Gamma}
&=&
\sum_\gamma \delta  Z^{(\ms_\scheme)}_{1,\gamma}
\cdot\calA_{1,\Gamma/\gamma}
+
\delta \calR^{(\ms_\scheme)}_{2,\Gamma}
\nonumber\\&&{}+
D^{(\dscheme)}_{1}\,
\left(\calA_{1,\Gamma}+
\delta  \calR^{(\ms_\scheme)}_{1,\Gamma}\right)
\,.
\eea
This scheme-transformation formula can be rewritten more compactly as
\bea
\label{eq:RtransfG}
\delta \calR^{(\scheme)}_{2,\Gamma}
&=&
\delta \calR^{(\ms_\scheme)}_{2,\Gamma}
+
D^{(\dscheme)}_{1}\,
\delta  \calR^{(\ms_\scheme)}_{1,\Gamma}
+
\delta\calK^{(\dscheme)}_{2,\Gamma}
\,,
\eea
with
\bea
\label{eq:straK1c}
\delta\calK^{(\dscheme)}_{2,\Gamma}
&=&
D^{(\dscheme)}_{1} \amp{1}{\Gamma}{}{}
-\sum  \limits_{\gamma} 
\delta Z^{(\dscheme)}_{1,\gamma}\cdot \amp{1}{\Gamma/\gamma}{}{}\,.
\eea
The term 
$D^{(\dscheme)}_{1}\,\delta \calR^{(\ms_\scheme)}_{1,\Gamma}$
in~\refeq{eq:RtransfG} 
corresponds to the 
multiplicative  renormalisation of the
one-loop rational term, which requires, according to~\refeq{eq:frc3a},
the first derivative of
$\delta \calR^{(\ms_\scheme)}_{1,\Gamma}$
with respect to all relevant parameters, including
the gauge-fixing parameter $\lambda$.
The remaining term~\refeq{eq:straK1c}
represents a nontrivial source of scheme dependence
that originates from subtleties related to the 
subtraction of subdivergences in $\numdim=4$ dimensions.
As discussed in detail in the next subsections,
also this latter source of scheme dependence 
can be described in a general and 
process-independent way. 
In particular we will show that~\refeq{eq:straK1c}
can be expressed as a linear combination of one-loop renormalisation
constants,
\bea
\label{eq:straKdec}
\delta\calK^{(\dscheme)}_{2,\Gamma}
&=&
\sum_\chi 
\delta \calZ^{(\dscheme)}_{1,\chi}
\,
\delta\hat\calK^{(\chi)}_{1,\Gamma}\,,
\eea
where $\delta\hat\calK^{(\chi)}_{1,\Gamma}$ are scheme- and process-independent 
one-loop counterterms. Their explicit 
expressions are presented in Section~\ref{se:fullfschemedep}.

\subsection{Nontrivial scheme dependence of two-loop rational terms}
\label{se:nontrschdep4d}

The origin of the scheme-dependent contribution~\refeq{eq:straK1c} 
lies in the fact that the 
subtraction of UV subdivergences through multiplicative
renormalisation
in $\numdim=4$ dimensions
yields different results depending on whether 
the renormalisation is carried out 
before projection to $\numdim=4$ dimensions, as defined
in~\refeq{eq:subdivprescX}, or after.

The first term on the rhs of~\refeq{eq:straK1c}
corresponds to the multiplicative renormalisation 
of a one-loop amplitude {\it after} projection to $\numdim=4$ dimensions.
Schematically, regarding the integrand of the 1PI one-loop diagram
$\Gamma$ as a product of internal propagators $G_a$ 
and vertices $\calV_b$, 
one can write 
\bea
\label{eq:straK2d}
D^{(\dscheme)}_{1} \amp{1}{\Gamma}{}{}
&=&
\sum_{G_a}
\left[
D^{(\dscheme)}_{1}\, G_a\,
\right]
\frac{\delta}{\delta G_a}\, \amp{1}{\Gamma}{}{}+
\sum_{\calV_b}
\left[
D^{(\dscheme)}_{1}\, \calV_b
\right]
\, \frac{\delta}{\delta \calV_b}\, \amp{1}{\Gamma}{}{}
\,.
\qquad
\eea
This schematic notation indicates that 
the loop propagators/vertices 
inside $\calA_{1,\Gamma}$
are removed (one by one)
by the derivative operators and 
replaced by the corresponding 
counterterms 
within square brackets.\footnote{More explicitly, 
let us consider a generic one-loop diagram $\Gamma$ containing 
$n_a$ propagators of type $a$, and let us denote as
$k_a^{(j)}$ the loop momentum flowing through the 
$j^{\mathrm{th}}$ type-$a$ propagator.
In this case, the propagator-renormalisation operators on the 
rhs of~\refeq{eq:straK2d} 
should be understood as 
\bea
\big[D^{(\dscheme)}_{1}\, G_a\big]\frac{\delta}{\delta G_a} 
\calA_{1,\Gamma}
&=&
\sum_{j=1}^{n_a}\calA_{1,\Gamma}\Bigg|_{G_a(k_a^{j},m_a)\,\to\, D^{(\dscheme)}_{1}\,
G_a(k_a^{(j)},m_a)},
\eea
where the replacements are applied at the integrand level.
The same holds also for vertex-renormalisation operators.
} 
The sums on the rhs run over all relevant types of 
propagators (fermions, gauge bosons, ghosts, scalars) 
and vertices. %
Similarly as for derivatives, 
the operator $D^{(\dscheme)}_{1}$ is linear, \ie its effect on 
a set of diagrams amounts to the sum of the contributions of individual
diagrams.
For what concerns the renormalisation of parameters,
\ie the terms proportional to $\partial/\partial \param_i$ in~\refeq{eq:frc3a},
the above identity corresponds to the chain rule.
As for the renormalisation of fields, the overall effect 
amounts to a factor 
$\frac{1}{2}\delta \calZ_{1,\varphi}^{(\dscheme)}$ 
for each external line, while all field-renormalisation factors 
associated with the internal loop lines
cancel between vertices and propagators as usual.

The second term on the rhs of~\refeq{eq:straK1c}
is defined through the prescription~\refeq{eq:subdivprescX}, 
where the renormalisation 
of  $\bar\calA_{1,\Gamma}$ is carried out 
{\it before} projecting to $\numdim=4$ dimensions.
It can  be expressed as follows in the from of
one-loop insertions into the internal propagators $G_a$ 
and vertices $\calV_b$
of $\calA_{1,\Gamma}$,
\bea
\label{eq:straK1d}
\sum  \limits_{\gamma} 
\delta Z^{(\dscheme)}_{1,\gamma}\cdot \amp{1}{\Gamma/\gamma}{}{}
&=&
\sum_{G_a}
\Big[G_a\, 
\delta Z^{(\dscheme)}_{1,\Gamma_a}\,
G_a\Big]\,
\frac{\delta}{\delta G_a}\, \amp{1}{\Gamma}{}{}
+
\sum_{\calV_b}
\delta Z^{(\dscheme)}_{1,\calV_b}
\, \frac{\delta}{\delta \calV_b}\, \amp{1}{\Gamma}{}{}
\,.
\eea
The two types of counterterm on the rhs correspond
to the 
amputated 
1PI two-point functions $\Gamma_a$,
which are related to the propagators $G_a$
via~\refeq{eq:fieldren4} and~\refeq{eq:fieldren5},
and to the 
amputated 1PI vertex functions $\calV_b$.
Such counterterms can be generated 
from the corresponding tree-level objects
via multiplicative renormalisation, \ie
\bea
\label{eq:straK1g}
\sum  \limits_{\gamma} 
\delta Z^{(\dscheme)}_{1,\gamma}\cdot \amp{1}{\Gamma/\gamma}{}{}
&=&
\sum_{G_a}
\left[
G_a\, 
\left(D^{(\dscheme)}_{1}\, \Gamma_a\right) 
G_a\,
\right]
\frac{\delta}{\delta G_a}\, \amp{1}{\Gamma}{}{}
+
\sum_{\calV_b}
\left[
D^{(\dscheme)}_{1}\, \calV_b
\right]
\, \frac{\delta}{\delta \calV_b}\, \amp{1}{\Gamma}{}{}
\,.
\nonumber\\
\eea
Note that in~\refeq{eq:straK2d} and~\refeq{eq:straK1d}--\refeq{eq:straK1g} 
all $G_a$, $\Gamma_a$ and $\calV_a$ 
are in 
$\numdim=4$ dimensions, as indicated by the absence of a bar.
As a consequence, in~\refeq{eq:straK1d}--\refeq{eq:straK1g}
all loop numerators are strictly four-dimensional.
On the contrary,  the term $D^{(\dscheme)}_{1}\, G_a$
in~\refeq{eq:straK2d} can give rise 
to extra contributions proportional to $\tilde q^2$ 
in the loop numerator. This is due to the fact that the 
renormalisation 
of $G_a$
is carried out with 
$D$-dimensional denominator and
four-dimensional numerator, \ie {\it after} 
projection to $\numdim=4$.
As we will see,
the interplay of such  $\tilde q^2$ terms with UV poles 
is at the origin of the auxiliary counterterm~\refeq{eq:straK1c}.

Comparing~\refeq{eq:straK2d} to~\refeq{eq:straK1g}
we observe that~\refeq{eq:straK1c} receives contributions only from the renormalisation
of loop propagators and can be expressed as
\bea
\label{eq:straK2e}
\delta\calK^{(\dscheme)}_{2,\Gamma}
&=&
\sum_{G_a}
\left(
\calP^{(\dscheme)}_{1}\, G_a \right)
\frac{\delta}{\delta G_a}\, \amp{1}{\Gamma}{}{}\,,
\eea
with
\bea
\label{eq:straK2f}
\calP^{(\dscheme)}_{1}\, G_a
&=&
\left[
\left( D^{(\dscheme)}_{1}\, G_a\right) 
-
G_a\, 
\left(D^{(\dscheme)}_{1}\, \Gamma_a\right) 
G_a\,
\right]\,.
\eea
Here the multiplicative renormalisation~\refeq{eq:frc3a}
of the 1PI two-point function $\Gamma_a$
yields
\bea
\label{eq:straK2q}
D^{(\dscheme)}_{1}\,\Gamma_a &=&
\left(
\delta \calZ^{(\dscheme)}_{1,\varphi_a}
+
\delta \calZ^{(\dscheme)}_{1,m_a}\, m_a \partial_{m_a}
+
\delta \calZ^{(\dscheme)}_{1,\lambda}\, \lambda \partial_{\lambda}
\right) \Gamma_a\,,
\eea
where the renormalisation of the gauge
parameter $\lambda$ is relevant only when $\Gamma_a$ is a gauge-boson
two-point function.
The first term on the rhs of~\refeq{eq:straK2f} 
corresponds to the renormalisation of a generic propagator 
and yields\footnote{Note 
that the sign of the field-renormalisation constants in~\refeq{eq:frc3a}
is meant for the renormalisation of amputated Green's functions, 
while the renormalisation of propagators requires the opposite sign.}
\bea
\label{eq:straK2p}
D^{(\dscheme)}_{1}\,G_a &=&
\left(
-\delta \calZ^{(\dscheme)}_{1,\varphi_a}
+
\delta \calZ^{(\dscheme)}_{1,m_a}\, m_a \partial_{m_a}
+
\delta \calZ^{(\dscheme)}_{1,\lambda}\, \lambda \partial_{\lambda}
\right) G_a\,.
\eea
As discussed in the context of~\refeq{eq:straK2d},
the factor $-\delta \calZ^{(\dscheme)}_{1,\varphi_a}$ compensates
the factors $\frac{1}{2}\delta \calZ^{(\dscheme)}_{1,\varphi_a}$ associated
with the vertices connected to the two ends of the propagator
in such a way that, at the amplitude level, the net effect of field renormalisation 
amounts to a factor $\frac{1}{2}\delta \calZ^{(\dscheme)}_{1,\varphi_a}$ per external leg.

\subsubsection*{Gauge-independent propagators}

In the following we
 work out explicit expressions for the propagator corrections~\refeq{eq:straK2f}
starting from propagators that are gauge independent at tree level, 
\ie the propagators of fermions, ghosts and physical scalar fields.
In this case, with~\refeq{eq:straK2q}--\refeq{eq:straK2p} we can express 
\refeq{eq:straK2f} as a linear combination of field and mass
renormalisation constants,
\bea
\label{eq:K2r}
\calP^{(\dscheme)}_{1}\, G_a
&=&
\left[\delta \calZ^{(\dscheme)}_{1,\varphi_a}
\,\hat\calP_{1,\varphi}
+
\delta \calZ^{(\dscheme)}_{1,m_a}
\,\hat\calP_{1,m}\right] G_a\,,
\eea
with scheme-independent operators
\bea
\label{eq:fieldren3}
\hat\calP_{1,\varphi}\, G_a
&=&
-G_a-G_a\, \Gamma_a\,G_a\,,
\eea
and 
\bea
\label{eq:mren3}
\hat\calP_{1,m}\, G_a 
&=&
m_a \partial_{m_a}\,
G_a 
-
G_a 
\, 
\Big[m_a \partial_{m_a}\,\Gamma_a 
\Big]
G_a 
\,.
\eea
In order to simplify the above identities, let us first discuss the general
form of the tree-level propagator,
\bea
\label{eq:straK1h}
G_a
\,\equiv\,
G_a(\bar k,m_a)
&=&
\frac{g_a(k,m_a)}{
\bar k^2-m_a^2}\,, 
\eea
where in $\numdim=4$  the numerator $g_a$ and the denominator 
are, respectively, in four and $D$ dimensions.
This different dimensionality leads to a nontrivial relation between
$g_a$, $G_a$ and $\Gamma_a$. The usual relation in 
$\numdim=\dendim$ dimensions is 
\bea
\label{eq:fieldren4a}
\bar g_a(\bar k,m_a)\, \bar \Gamma_a(\bar k,m_a) 
&=&
-(\bar k^2-m_a^2)\,,
\eea
or equivalently,
\bea
\label{eq:fieldren4}
\bar G_a(\bar k,m_a)\, \bar \Gamma_a(\bar k,m_a) &=& -1\,,
\eea
\ie  $\bar G_a$ and $\bar \Gamma_a$ are the inverse of each other
up to a minus sign.
In contrast, for gauge-independent propagators in $\numdim=4$ dimensions we
have
\bea
\label{eq:fieldren5a}
g_a(k,m_a)\, \Gamma_a(k,m_a) 
&=&
-(k^2-m_a^2)\,,
\eea
and
\bea
\label{eq:fieldren5}
G_a(\bar k,m_a)\, \Gamma_a(k,m_a) &=& 
{}-\frac{k^2-m_a^2}{\bar k^2-m_a^2}
\,=\,
\frac{\tilde q^2}{\bar k^2-m_a^2}-1\,,
\eea
where $\bar q = q+\tilde q$ is the loop momentum, and 
$\bar k=\bar q + p$ is the momentum that flows through the
loop propagator $G_a$.
With this latter identity at hand we find that the
field-renormalisation operator~\refeq{eq:fieldren3} corresponds to 
\bea
\label{eq:fieldren6}
\hat \calP_{1,\varphi}\, G_a(\bar k, m_a) &=& {}-
G_a(\bar k, m_a)\,
\frac{\tilde k^2}{\left(\bar
k^2-m_a^2\right)}\,
\,.
\eea
As for the mass-renormalisation operator,
the two terms on the rhs of~\refeq{eq:mren3} can be simplified using
\bea
\label{eq:mren7}
(\bar k^2-m_a^2)\,
m_a \partial_{m_a}\,
G_a
&=&
m_a \partial_{m_a}\,
\Big[(\bar k^2-m_a^2)
G_a
\Big]
-
G_a
m_a \partial_{m_a}\,
(\bar k^2-m_a^2)
\nonumber\\
&=&
m_a \partial_{m_a}\,g_a
+2m_a^2\, G_a\,,
\eea
and
\bea
\label{eq:mren8}
(\bar k^2-m_a^2)
\,G_a 
\, 
\Big[m_a \partial_{m_a}\,\Gamma_a 
\Big]
G_a
&=&
G_a 
\, 
\Big[m_a \partial_{m_a}\,\Gamma_a 
\Big]
g_a
\,=\,
G_a 
\, 
m_a \partial_{m_a}
\big(
\Gamma_a\,
g_a 
\big)
-
G_a \,\Gamma_a 
m_a \partial_{m_a}\,
g_a
\nonumber\\
&=&
2m_a^2\,
G_a 
-
\bigg( \frac{\tilde q^2}{\bar k^2-m_a^2}-1\bigg)
m_a \partial_{m_a}\,g_a\,,
\eea
where we have exploited 
\refeq{eq:straK1h} and~\refeq{eq:fieldren5a}--\refeq{eq:fieldren5}.
Combining~\refeq{eq:mren3} with~\refeq{eq:mren7}-\refeq{eq:mren8}
one finds
\bea
\label{eq:mren9}
\hat \calP_{1,m}\, G_a(\bar k, m_a) &=&
\frac{\tilde q^2}{\left(\bar
k^2-m_a^2\right)^2}\,m_a\partial_{m_a} g_a \,.
\eea
In renormalisable gauge theories without symmetry breaking, 
the term $m_a\partial_{m_a} g_a$ is non-zero 
only for the propagators of massive 
fermions, for which 
$g_a(k, m_a) = \ri({\slashed k} +m_a)$.
Thus
\bea
\label{eq:mren10}
\hat\calP_{1,m}\, G_a(\bar k, m_a)
\,=\,
\frac{\tilde q^2 }{\left(\bar
k^2-m_a^2\right)^2}\times
\begin{cases}
\;{\displaystyle \ri m_a  }
& \mbox{for fermion propagators,}\\[3mm]
\;0 & \mbox{for ghost, scalar, and massless}\\[-2mm]
& \mbox{gauge-boson propagators.}
\end{cases}
\eea

\subsubsection*{Gauge-dependent propagators}

Let us now consider the two-point function of massless gauge bosons 
in the so-called \mbox{$\xi$-gauge}.
The relevant part of the renormalised Lagrangian reads
\bea
\label{eq:gbpro1}
\mathcal L_{\text{gauge}} &=& 
-\frac{1}{2}\left[\calZ_A\,\partial_\mu A_\nu
\left(\partial^\mu A^\nu-\partial^\nu A^\mu\right) +
\frac{\calZ_A}{\lambda\calZ_\lambda}
(\partial^\mu A_\mu)^2\right]\,.
\eea
Due to Ward identities, the renormalisation constants associated with 
the gauge field and the gauge-fixing parameter 
have identical UV poles.
Thus it is convenient to define the latter as
\bea
\label{eq:gbpro1b}
\calZ_\lambda &=&
\frac{\calZ_A}{\calZ_{\gpar}}\,,
\eea
where $\calZ_{\gpar}$ is a finite renormalisation constant that is 
typically set equal to one.

At tree level in $\numdim=4$, the gauge-boson two-point function
and the associated propagator read
\bea
\label{eq:gbpro2}
\Gamma^{\mu\nu}_A(k,\lambda)
&=&
{}-\ri\left[
k^2 g^{\mu\nu}+\left(\frac{1}{\lambda}-1\right)k^\nu k^\nu
\right]\,,
\eea
and
\bea
\label{eq:gbpro3}
G^{\mu\nu}_A(\bar k,\lambda)
&=&
{}-\frac{\ri}{\bar k^2}\left[
g^{\mu\nu}+(\lambda-1)\frac{k^\nu k^\nu}{\bar k^2}
\right]\,.
\eea
Since the two-point function is free from denominators, 
all objects on the rhs of~\refeq{eq:gbpro2} are projected to four dimensions,
while all denominators in~\refeq{eq:gbpro3} are in $\dendim$ dimensions. 
Due to the different dimensionality of numerators and denominators, the
transverse and longitudinal tensors in~\refeq{eq:gbpro3}, 
\bea
P^{\mu\nu}_{\rT}(\bar k)&=&
g^{\mu\nu}-\frac{k^\nu k^\nu}{\bar k^2}\,,
\qquad
P^{\mu\nu}_{\rL}(\bar k)\,=\,
\frac{k^\nu k^\nu}{\bar k^2}\,,
\eea
do not fulfil the usual projector properties. In particular,
\bea
\label{eq:gbpro4}
P_{\rT}(\bar k)\,
P_{\rL}(\bar k)
\,=\,
\frac{\tilde k^2}{\bar k^2}\,
P_{\rL}(\bar k) \,\neq\,0\,.
\eea
As a consequence, instead of~\refeq{eq:fieldren5}, for gauge-boson propagators we have
\bea
\label{eq:gbpro5}
G^{\mu}_{A\rho}
(\bar k,\lambda)\,
\Gamma^{\rho\nu}_{A}
(\bar k,\lambda)
&=&
{}-g^{\mu\nu}
+\frac{\tilde k^2}{\bar k^2}
\left[ g^{\mu\nu}+\left(1-\frac{1}{\lambda}\right)\frac{k^\nu k^\nu}{\bar
k^2}\right]\,.
\eea

Let us now work out the auxiliary counterterms~\refeq{eq:straK2f}
for gauge-bosons propagators.
Along the same lines as in~\refeq{eq:K2r}--\refeq{eq:mren3},
we can express the operators~\refeq{eq:straK2f}--\refeq{eq:straK2p}
as linear combinations of the one-loop renormalisation constants $\delta Z_{1,A}$
and $\delta Z_{1,\gpar}=\delta Z_{1,A}-\delta Z_{1,\lambda}$. This yields
\bea
\label{eq:gbpro7}
D^{(\dscheme)}_{1}\,\Gamma_A &=&
\left[
\delta \calZ^{(\dscheme)}_{1,A}\left(1+\lambda \partial_{\lambda}\right)
-
\delta \calZ^{(\dscheme)}_{1,\gpar}\, \lambda \partial_{\lambda}
\right] \Gamma_A\,,
\eea
\bea
\label{eq:gbpro8}
D^{(\dscheme)}_{1}\,G_A &=&
\left[
\delta \calZ^{(\dscheme)}_{1,A}\left(-1+\lambda \partial_{\lambda}\right)
-
\delta \calZ^{(\dscheme)}_{1,\gpar}\, \lambda \partial_{\lambda}
\right] G_A\,,
\eea
and
\bea
\label{eq:gbpro6}
\calP^{(\dscheme)}_{1}\, G_A
&=&
\left[\delta \calZ^{(\dscheme)}_{1,A}
\,\hat\calP_{1,A}
+
\delta \calZ^{(\dscheme)}_{1,\gpar}
\,\hat\calP_{1,\gpar}\right] G_a\,,
\eea
with scheme-independent operators
\bea
\label{eq:gbpro9}
\hat\calP_{1,A}\, G_A
&=&
\left(-1+\lambda \partial_{\lambda}\right)G_A\, 
\,-\,G_A
\Big[\left(1+\lambda \partial_{\lambda}\right)
\Gamma_A\Big]
\,G_A\,,
\eea
and 
\bea
\label{eq:gbpro10}
\hat\calP_{1,\gpar}\, G_A 
&=&
-\lambda \partial_{\lambda}\,
G_A 
+
G_A 
\, 
\Big[\lambda \partial_{\lambda}\,\Gamma_A 
\Big]
G_A 
\,.
\eea
Inserting the explicit expressions~\refeq{eq:gbpro2}--\refeq{eq:gbpro3}
and using~\refeq{eq:gbpro5}
we find
\bea
\label{eq:gbpro11}
\hat\calP_{1,A}\, G^{\mu\nu}_A (\bar k,\lambda)
&=&
\frac{\,\tilde k^2\,}{\bar k^2}
\, \left(\frac{\ri\, g^{\mu\nu}}{\bar k^2}\right)\,,
\eea
and
\bea
\label{eq:gbpro12}
\hat\calP_{1,\gpar}\, G^{\nu\nu}_A (\bar k,\lambda)
&=&
\frac{\,\tilde k^2\,}{\bar k^2}
(\lambda-1)\left[2+\left(\frac{1}{\lambda}-1\right)
\frac{\,\tilde k^2\,}{\bar k^2}
\right]
\left(\frac{\ri\,k^\nu k^\nu}{\bar k^4}\right)\,.
\eea
Similarly as for the case of gauge-independent propagators, the above auxiliary
counterterms are
proportional to $\tilde k^2/\bar k^2$. Note also that the term~\refeq{eq:gbpro12} 
associated with the finite renormalisation  of the gauge parameter
vanishes in the Feynman gauge.

\subsection{Full scheme dependence of two-loop rational terms}
\label{se:fullfschemedep}

Combining the various results derived in~\refse{se:nontrschdep4d}
one can express the nontrivial part
of the two-loop scheme dependence~\refeq{eq:straK1c} 
as a linear 
combination of field, mass 
and gauge-parameter
renormalisation constants,
\bea
\label{eq:strarec1}
\delta\calK^{(\dscheme)}_{2,\Gamma}
&=&
\sum_{a}
\left[
\delta \calZ^{(\dscheme)}_{1,\varphi_a}
\,
\delta\hat\calK^{(\varphi_a)}_{1,\Gamma}
+
\delta \calZ^{(\dscheme)}_{1,m_a}
\,
\delta\hat\calK^{(m_a)}_{1,\Gamma}
\right]
+
\delta \calZ^{(\dscheme)}_{1,\gpar}
\,
\delta\hat\calK^{(\gpar)}_{1,\Gamma}\,,
\eea
where the sum extends over all kinds of
fields $a$ 
(gauge bosons, fermions, ghosts or scalars) that 
propagate inside the loop diagrams contributing to the
one-loop amplitude $\calA_{1,\Gamma}$. 
The various coefficients $\delta\hat\calK_{1,\Gamma}$
can be regarded as scheme-independent one-loop
counterterms. The ones associated with the renormalisation of fields
are given by
\bea
\label{eq:strarec2a}
\delta\hat\calK^{(\varphi_a)}_{1,\Gamma}
&=&
\left(\hat\calP_{1,\varphi}\, G_a \right)
\frac{\delta}{\delta G_a}\, \amp{1}{\Gamma}{}{}\,,
\eea
with
\bea
\label{eq:strarec2b}
\hat\calP_{1,\varphi}\, G_a
\,=\,
\frac{\tilde q^2
}{\left(\bar
k^2-m_a^2\right)}
\times
\begin{cases}
\;{\displaystyle {}-G_a(\bar k, m_a)}
& \mbox{for fermion, ghost, and scalar propagators,}\\[3mm]
\;\;{\displaystyle \frac{\ri\, g^{\mu\nu}}{\bar k^2}}
& \mbox{for massless gauge-boson propagators.}
\end{cases}
\nonumber\\
\eea
The counterterms associated with the mass renormalisation 
are given by 
\bea
\label{eq:strarec2e}
\delta\hat\calK^{(m_a)}_{1,\Gamma}
&=&
\left(\hat\calP_{1,m}\, G_a \right)
\frac{\delta}{\delta G_a}\, \amp{1}{\Gamma}{}{}\,,
\eea
where $\hat\calP_{1,m}$ is defined in~\refeq{eq:mren10}.
Finally, the counterterm associated with 
the finite renormalisation of the gauge parameter 
originates only from
 gauge-boson propagators 
$G_A$ and is given by 
\bea
\label{eq:strarec2c}
\delta\hat\calK^{(\gpar)}_{1,\Gamma}
&=&
\left(\hat\calP_{1,\gpar}\, G_A \right)
\frac{\delta}{\delta G_A}\, \amp{1}{\Gamma}{}{}\,,
\eea
where $\hat\calP_{1,\gpar}$ is defined in~\refeq{eq:gbpro12}.
For the standard choice $\calZ_\gpar=1$
this counterterm is irrelevant.

Note that all quantities in~\refeq{eq:strarec1}--\refeq{eq:strarec2c}
are free from UV poles. In particular, 
the renormalisation constants $\delta \calZ^{(\dscheme)}_{1,\rcarg}$ 
in~\refeq{eq:strarec1} are UV finite.
For this reason, the $\eps$-expansion 
of the auxiliary counterterms~\refeq{eq:strarec2a}--\refeq{eq:strarec2c}
will be truncated {\it by 
definition} at order $\eps^0$.
With other words, terms of $\ord(\eps)$ in 
the above auxiliary counterterms
will be discarded also when~\refeq{eq:strarec1} is 
split into UV-divergent parts using 
\bea
\label{eq:strarec4}
\delta \calZ^{(\dscheme)}_{1,\rcarg}= \delta \calZ^{(\scheme)}_{1,\rcarg}
-\delta \calZ^{(\ms_\scheme)}_{1,\rcarg}\,.
\eea
The counterterms~\refeq{eq:strarec2a}--\refeq{eq:strarec2c}
involve only finite parts that originate 
from the interplay of $\tilde q^2$ numerator terms with UV poles.
Thus they are universal in the same sense as the usual 
renormalisation constants and rational terms.

From the viewpoint of UV power counting,
the counterterms~\refeq{eq:strarec2a} and~\refeq{eq:strarec2c}
correspond to the insertion of a term of order 
$\ord(\tilde q^2/\bar q^2)$ into the original one-loop 
amplitude. Thus when $\calA_{1,\Gamma}$ is 
UV divergent
$\delta\hat\calK^{(\varphi_a)}_{1,\Gamma}$ and
$\delta\hat\calK^{(\gpar)}_{1,\Gamma}$
are expected to 
be non-zero. 
In contrast, the counterterm~\refeq{eq:strarec2e} 
replaces a fermion propagators of $\ord(1/\bar q)$ 
by objects of $\ord(\tilde q^2/\bar q^4)$,
thereby reducing the degree of UV divergence by one.
Thus non-vanishing $\delta\hat\calK^{(m_a)}_{1,\Gamma}$
contributions are expected only when $\calA_{1,\Gamma}$
involves non-logarithmic UV divergences.

In summary, the identities~\refeq{eq:RtransfG} and
\refeq{eq:strarec1}--\refeq{eq:strarec2c} 
make it possible to transform two-loop
rational terms from the minimal subtraction scheme
(or any other reference scheme) 
to a generic renormalisation scheme $\scheme$
using only universal one-loop quantities.
Since all scheme-dependent parts in~\refeq{eq:RtransfG} 
are linear combinations of the finite renormalisation constants~\refeq{eq:strarec4},
the two-loop rational terms in the scheme $\scheme$ 
can be expressed as 
\bea
\label{eq:strategy1}
\delta \calR^{(\scheme)}_{2,\Gamma}
&=& \delta \calR^{(\ms_\scheme)}_{2,\Gamma} +
\sum_{\rcarg}
\delta \calZ^{(\dscheme)}_{1,\rcarg}
C^{(\rcarg)}_{1,\Gamma}\,,
\eea
where the sum over $\rcarg$ includes all relevant coupling-, gauge-, mass-,
and field-renormalisation
constants. Here the scheme dependence is isolated in the renormalisation
constants $\delta \calZ^{(\dscheme)}_{1,\rcarg}$,
while their coefficients $C^{(\rcarg)}_{1,\Gamma}$, 
which are dictated
by~\refeq{eq:RtransfG}, are scheme independent.
More precisely, their scheme dependence consist only of a 
trivial scale factor $t_\scheme^{\eps}$.

Contrary to what is suggested by the representation~\refeq{eq:strategy1},
the $\delta \calR^{(\scheme)}_{2,\Gamma}$ terms do not depend on the
corresponding rational term in the $\ms_\scheme$ scheme.
This becomes evident by recasting~\refeq{eq:strategy1},
through~\refeq{eq:strarec4}, 
as a 
linear combination of the full renormalisation constants
in the $\scheme$ scheme,
\bea
\label{eq:strategy2}
\delta \calR^{(\scheme)}_{2,\Gamma}
&=& \delta \calR^{(\inv)}_{2,\Gamma} +
\sum_{\rcarg} \delta \calZ^{(\scheme)}_{1,\rcarg}\,
C^{(\rcarg)}_{1,\Gamma}\,.
\eea
In this representation $\delta \calR^{(\inv)}_{2,\Gamma}$ 
consists, apart from an overall scale factor $t_\scheme^{2\eps}$,
of terms of order $\eps^{-1}$ and $\eps^0$ that 
are independent of the schemes $\scheme$ and $\ms_\scheme$.
In fact~\refeq{eq:strategy2} and~\refeq{eq:strategy1} imply that
\bea
\label{eq:strategy2b}
\delta \calR^{(\inv)}_{2,\Gamma}
&=& 
\delta \calR^{(\scheme)}_{2,\Gamma}
-
\sum_{\rcarg} \delta \calZ^{(\scheme)}_{1,\rcarg}\,
C^{(\rcarg)}_{1,\Gamma}
\,=\,
\delta \calR^{(\ms_\scheme)}_{2,\Gamma}
-
\sum_{\rcarg} \delta \calZ^{(\ms_\scheme)}_{1,\rcarg}\,
C^{(\rcarg)}_{1,\Gamma}\,,
\eea
and, in practice, $\delta \calR^{(\inv)}_{2,\Gamma}$ can be derived from 
existing results in a minimal-subtraction scheme, or in any other scheme.

The general scheme-transformation properties derived in this section have been
validated through a direct calculation of all QED and QCD two-loop rational terms
in a generic scheme $\scheme$.

\newcommand{\shiftleft}{\!\!\!\!\!\!\!\!\!\!\!\!\!\!\!\!\!\!\!}

\section{Two-loop rational counterterms for 
SU(N) and U(1) gauge theories}
\label{sec:qcdres}

In this section we present the full set of one- and two-loop rational counterterms 
for the generic Yang--Mills theory defined by the Lagrangian~\refeq{eq:ymlag}, 
which describes both SU(N) and U(1) gauge theories as special cases.

\subsection{Technical details of the calculations}
\label{se:techdetails}

To compute all relevant $\delta\calR_1$ and $\delta\calR_2$
counterterms we have applied the master formula~\refeq{eq:proofsketchG}
to the full set of globally divergent Feynman diagrams 
that contribute to the various 1PI vertices
with two, three and four external lines.

Based on the general scheme-transformation properties
derived in \refse{se:schemetrans}, we have computed the 
$\delta \calR_{2,\Gamma}$ counterterms
in a generic renormalisation scheme. 
To this end we have 
recast the master formula~\refeq{eq:genmform2} 
in the form~\refeq{eq:proofsketchG}. For the 
calculation of the relevant loop integrals we have 
employed the tadpole expansions presented in~\refapp{se:tadexp}.
The one-loop counterterms $\delta Z_{1,\gamma}$,
$\delta \tilde Z_{1,\gamma}$ and
$\delta \calR_{1,\gamma}$
that are required for the derivation of two-loop 
rational counterterms
can be found in~\refse{se:R2results}.
In order to keep the scheme choice fully flexible we have
decomposed all one-loop 
renormalisation constants as
\bea
\label{eq:}
\delta \calZ^{(\scheme)}_{1,\rcarg}\,=\
\delta \calZ^{(\ms_\scheme)}_{1,\rcarg}+
\delta \calZ^{(\dscheme)}_{1,\rcarg}\,,
\eea
where the generalised minimal-subtraction
constant $\delta \calZ^{(\ms_\scheme)}_{1,\rcarg}$, defined 
in~\refeq{eq:genrsF}, contains all explicit UV poles
(see~\refapp{app:msbarRCs}), while the 
finite remainder $\delta \calZ^{(\dscheme)}_{1,\rcarg}$ is treated as a free parameter. 
The final results are presented in the form~\refeq{eq:strategy2},
\ie as linear combinations of the full renormalisation constants 
$\delta \calZ^{(\scheme)}_{1,\rcarg}$.
In the rest of this section the scheme label $X$ will be kept implicit.

All calculations have been performed twice and independently using different
tools. 
On the one hand we have used {\sc Geficom}~\cite{GEFICOM},
which is based on {\sc Qgraf}~\cite{QGRAF}, {\sc Q2E} and
{\sc Exp}~\cite{Seidensticker:1999bb,Harlander:1997zb} 
for the generation and topology
identification of Feynman diagrams.
Within {\sc Geficom} algebraic manipulations, one-loop insertions
and tadpole decompositions are implemented 
in {\sc Form}~\cite{Vermaseren:2000nd,Tentyukov:2007mu}. 
Massive tadpole integrals are computed
with {\sc Matad}~\cite{MATAD}, and gauge-group factors with {\sc Color}~\cite{COLOR}, 
both of which are based on {\sc Form}.
Tadpole expansions are carried out with the methods 
described in Appendices~\ref{se:naivetadpoledec} and~\ref{se:TexpwithCT}.
The algebraic structures of the result are expressed in terms of a minimal set of
independent Lorentz, Dirac and colour tensors, which
are isolated in the beginning of the calculation using projectors 
that saturate all external indices.

To cross check all calculations we have developed a second in-house framework
implemented in {\sc Python} that uses {\sc Qgraf}~\cite{QGRAF} for the amplitude generation and
{\sc Form}~\cite{Vermaseren:2000nd,Tentyukov:2007mu} as well as {\sc
python-Form}\footnote{\url{https://github.com/tueda/python-form}} for the
amplitude manipulations. 
In this framework all algebraic objects are directly reduced in the form of
Dirac, Lorentz and group-theoretical tensors, \ie without applying any projection
to the indices associated with external lines.
The gauge-group algebra is handled as described in~\refse{sec:renlag} or,
alternatively, based on the colour-flow representation \cite{tHooft:1973alw} for
the SU(N) case. 
The tadpole expansions are implemented in the four different 
versions described 
in Appendices~\ref{se:naivetadpoledec}--\ref{se:TexpwithCT}.
The resulting tensorial tadpole integrals 
are expressed as combinations of metric tensors, and
the coefficients are automatically reduced to master integrals
with an in-house algorithm based on IBP
identities~\cite{Tkachov:1981wb,Chetyrkin:1981qh}.  

In both frameworks, two-loop amplitudes are directly
decomposed into loop chains and connecting vertices according to~\refeq{eq:twoloopnumA} 
in such a way that enables the relevant power-counting operations 
and the further processing of the (sub-)diagrams.
For what concerns dimensional regularisation, all calculations are 
carried out by handling $\dendim=4-2\eps$ and 
the loop-numerator dimension $\numdim$ as independent free parameters.
In this way all relevant UV poles and rational terms
can be determined a posteriori by setting $\numdim=\dendim$ and $\numdim=4$.

All results presented in~\refse{se:R2results} have been derived
independently in the two computing frameworks. In addition the following 
consistency checks have been carried out.

\begin{enumerate}

\item We have checked that all $\delta\calR_2$ results
are independent of the 
auxiliary mass $M$.

\item For all 1PI vertices in~\refse{se:R2results}  
we have verified the cancellation of 
UV poles in the two-loop master formula~\refeq{eq:genmform2}.
Note that the $\delta\calR_2$ terms involve $1/\eps$ poles, and
finite results are obtained only when all one- and two-loop 
counterterm contributions of UV and rational type are combined.

\item To validate the consistency of the employed tadpole expansions, all
calculations have been repeated using the four types of expansions presented
in Appendices~\ref{se:naivetadpoledec}--\ref{se:TexpwithCT} finding
consistent results.
Note that changing the tadpole expansion method 
shifts the finite parts of the expanded amplitudes.
Thus the validation at hand corresponds to a test 
of the master formula~\refeq{eq:genmform2} 
at the level of the finite parts of the amplitudes.

\item We have checked that the Taylor-expansion method 
of \refapp{se:taylorexp}
is independent of the choice of parametrisation.
To this end we have carried out all $\delta \calR_2$ calculations using
independent parametrisations for one- and two-loop integrals.

\item We have verified that the renormalisation-scheme dependent parts of
all $\delta\calR_2$ counterterms are consistent
with~\refeq{eq:RtransfG}--\refeq{eq:straK1c}.
To this end we have explicitly derived 
the $\delta\calK_{2}$ parts
using~\refeq{eq:strarec1}--\refeq{eq:strarec2c}.

\item The one-loop counterterms 
$\delta Z_{1,\gamma}$,
$\delta \tilde Z_{1,\gamma}$ and
$\delta \calR_{1,\gamma}$
that enter the calculation of 
$\delta\calR_{2,\Gamma}$ have been treated in two alternative ways.
On the one hand we have used 
available results at the level of full one-loop vertex functions $\gamma$.
Alternatively, we have generated such counterterms
at the level of individual two-loop diagrams 
by applying tadpole expansions to the relevant subdiagrams.

\end{enumerate}

\subsection{Renormalised Lagrangian}
\label{sec:renlag}

We have computed the rational counterterms 
for the generic Yang--Mills theory defined by the 
renormalised Lagrangian
\bea
\label{eq:ymlag}
\mathcal L &=& 
\sum_{f\in \calF} \calZ_{f} \,\bar{\psi}_f 
\Big(\ri
\gamma_\mu
D^\mu - \calZ_{m_f}\,\mf\Big)\psi_f 
-\frac{1}{4}
\,
F^a_{\mu\nu}F^{a,\mu\nu} 
-
\frac{\calZ_{\gpar}}{2 \, \lambda} 
\Big(\partial^\mu A^a_\mu\Big)^2
- \, \calZ_{u} \,
\bar{u}^a \partial_\mu D^\mu_{ab}\, u^b\,,
\nonumber\\
\eea
with the field-strength tensor and the covariant derivatives
\bea
F^a_{\mu\nu}&=&
\calZ_A^{1/2}
\left[
\partial_\mu A^a_\nu -
\partial_\nu A^a_\mu
+ 
\left(\calZ_{\als}
\calZ_A
\right)^{1/2}
g\, f^{abc}A_\mu^b A_\nu^c
\right]
\,,
\nonumber\\
D^\mu &=& \partial^\mu - \ri \left(\calZ_{\als} \calZ_A\right)^{1/2} \gs \,
T^a \, A^{a\mu}\,,\nonumber\\ 
D^\mu_{ab} &=& \partial^\mu \delta_{ab} - \left(\calZ_{\als} \calZ_A\right)^{1/2} \gs \,
f^{abc}\, A^{c\mu}\,,
\eea 
where 
$t^a_\rF=T^a$ 
and 
$(t^a_\rA)_{bc} = -\ri f^{abc}$
are the generators of the gauge group in the fundamental and adjoint
representations,
while $\gs=\sqrt{4\pi \als}$ is the gauge coupling.
For the gauge fixing we adopt the Feynman gauge, which corresponds to 
$\lambda=1$.
The gauge interaction acts on a certain number $\nf$ of 
fermions, $f\in \calF$, 
which belong to the fundamental representation, and the various fermion 
masses can have arbitrary values $\mf\ge 0$.

In the fundamental ($r=\rF$) and adjoint
($r=\rA$)
representations, the generators 
satisfy the identities
\bea
 \left[t_r^a, t_r^b\right] \,=\, \ri \, f^{abc} \, t_r^c\,,\qquad
\Tr \left( t_r^a t_r^b\right)\, =\,
 T_r\,\delta^{ab}\,,
\eea
and the quadratic Casimir operators have eigenvalues
\bea
\CF\,=\, \frac{\TF\dA}{\dF}\,,\qquad
\CA\,=\TA\,,
\eea
where  $d_r$ denotes the dimension of the $r$ representation, and
$\dF=\Nc$.
Our results are expressed in terms of 
the invariants $\CF$, $\CA$, $\Nc$ and $\TF$.
Note that the normalisation of all generators and combinations thereof
is controlled by $\TF$.  
In particular, $T^a$ and $f^{abc}$ scale like $\TF^{1/2}$,
while $\CF$ and $\CA$ scale like $\TF$. 

For all two- and three-point counterterms presented in~\refse{se:R2results} 
we have obtained compact 
expressions using generic identities that are valid for any simple 
or abelian
gauge group,
while for the four-point counterterm~\refeq{eq:R24g} we have employed
identities like
\bea
T_{ij}^a\, T_{kl}^a &=&
\CA\TF\left(\frac{1}{\Nc}\delta_{il}\delta_{kj}-\delta_{ij}\delta_{kl}\right)
+\CF\delta_{ij}\delta_{kl}\,,
\eea
which are valid for SU(N) and U(1) groups.
The explicit expressions of the rational counterterms for SU(N) and U(1) 
gauge theories can be obtained from the results of~\refse{se:R2results} 
by applying the substitutions listed in~\refta{tab:uonesun}.
The SU(N) case with $N=3$ and $\alpha=\alpha_\rS$ corresponds to QCD with $\nf$ active quarks with
masses $m_f\ge 0$, while 
the U(1) case with $\alpha=\alpha_{\mathrm{EM}}$ 
corresponds to QED with $\nf$ fermions with charges
$Q_f$ and masses $m_f\ge 0$.

\begin{table}
\renewcommand{\arraystretch}{1.3}
\centering
\begin{tabular}{c|cccccccc}
        & $\dF$ & $T^a$ & $f^{abc}$ & $\delta^{ab}$ & $\dA$  & $\TF$         & $\CF$                   & $\CA$  \\\hline
SU(N)   & $N$ & $T^a$ & $f^{abc}$ & $\delta^{ab}$ & $\Nc^2-1$  & $\frac{1}{2}$ & $\frac{\Nc^2-1}{2\Nc}$  &  $\Nc$ \\\hline 
U(1)    & 1   & $Q_f$ & 0         & 1             & 1      &
$Q_f^2$             & $Q_f^2$                 &  0  
\end{tabular}
\label{tab:uonesun}
\caption{Values of the various group-theoretical quantities
for SU(N) and U(1) gauge theories. In the U(1) case
the replacements $T^a \to Q_f$, $\TF \to Q_f^2$  and $\CF\to Q_f^2$ involve the
charge $Q_f$, where $f$ is the 
fermion on which $T^a$, $\TF$ or $\CF$ acts.
For diagrams that involve external fermions
one should use the substitution $\CF \to Q_f^2$ at one loop, 
and the two-loop substitutions 
$\CF^2 \to Q_f^4$ and $\nf\TF\CF \to Q_f^2\sum_{f'} Q_{f'}^2$,
where the sum runs over all fermions $f'\in \calF$ that circulate in closed loops.
Instead, for vertices without external fermions one should use the 
one-loop substitutions
$\nf\TF \to \sum_{f'} Q_{f'}^2$ and 
$\TF \sum_{f'} m^2_{f'} \to \sum_{f'} Q^2_{f'} m^2_{f'}$,
and the two-loop substitutions 
$\nf\TF\CF \to \sum_{f'} Q_{f'}^4$ and
$\TF \sum_{f'} \CF  m^2_{f'}\to \sum_{f'} Q^4_{f'} m^2_{f'}$\,.
}
\end{table}

The renormalisation scheme is specified through generic 
renormalisation constants using the formalism of~\refse{se:schemetrans}.
The constants $\calZ_\alpha$ and $\calZ_{m_f}$ renormalise $\alpha$ and the 
fermion masses, while $\calZ_{\gpar}=\calZ_A/\calZ_\lambda$ is a finite 
parameter that renormalises the gauge-fixing term, see~\refeq{eq:gbpro1b}.
Finally, the constants $\calZ_f$, $\calZ_A$ and $\calZ_u$ control the 
renormalisation of the fermion, gauge-boson and ghost fields. 
\def\ext{\mathrm{ext}}
\def\inte{\mathrm{in}}
At the level of renormalised amplitudes, the net effect of field renormalisation 
amounts to a factor $(\calZ_{\varphi_{\ext}})^{1/2}$ for each external leg
associated with the field $\varphi_{\ext}$.
Note, however, that the $\delta \calR_{2,\Gamma}$ counterterms 
depend also on other field-renormalisation constants.
This dependence originates from the contributions 
$\delta\calK^{(\dscheme)}_{2,\Gamma}$, which are defined in~\refeq{eq:RtransfG}--\refeq{eq:straK1c}
and depend, see~\refeq{eq:strarec1}, on the $\delta\calZ_{1,\varphi_{\inte}}$ factors
associated with the renormalisation of one-loop
$\varphi_{\inte}$-selfenergy subdiagrams.
In the renormalised two-loop amplitude~\refeq{eq:genmform2}
the dependence on $\delta\calZ_{1,\varphi_{\inte}}$ 
cancels when the $\delta \calR_{2,\Gamma}$ counterterm is 
combined with the contribution of the
$\delta Z_{1,\gamma}$ counterterms associated with 
$\varphi_{\inte}$ selfenergies.
This nontrivial cancellation mechanism 
can be exploited to validate the implementation 
of the master formula~\refeq{eq:genmform2}.

The perturbative expansion of the various renormalisation constants
is written in the form 
\bea
\label{eq:qed2}
\calZ_{\rcarg} & = &  1+\sum_{k=1}^\infty
\lb\frac{\alpha\, t^{\eps}}{4\pi} \rb^k 
\delta \hat \calZ_{k,\rcarg}
\qquad\mbox{for}\qquad
\rcarg= \alpha,\, m_f,\,f,\, A,\,u\,, \gpar\,,
\eea
where $t= \msfact \mu_0^2/\mu_\rR^2$
embodies the dependence on the
regularisation scale $\mu_0$,
the renormalisation scale $\mu_\rR$, and the 
rescaling factor $\msfact$ (see \refse{se:msmsbarconv}).
At variance with~\cite{Pozzorini:2020hkx}, where
$\mu_0$ was set equal to $\mu_\rR$, 
here $\mu_0, \mu_\rR$ and $\msfact$
are treated as independent parameters.
Note that the renormalisation-scheme label $X$ used in~\refse{se:schemetrans} 
is kept implicit in this section.
Still, the renormalisation constants~\refeq{eq:qed2} 
describe a fully generic renormalisation scheme, 
which may be the minimal subtraction scheme, the
on-shell scheme, or any other scheme.
It is implicitly understood that the renormalised parameters depend on the 
renormalisation scale $\mu_\rR$, but, depending on the scheme, $\mu_\rR$
may be replaced by a physical mass scale, such as $m_e$ or $M_Z$.
Explicit expressions for the various renormalisation constants in the $\msbar$ 
scheme are reported in~\refapp{app:msbarRCs}.

\subsection{Rational counterterms}
\label{se:R2results}

In the following we present the rational and UV
counterterms for the Yang--Mills Lagrangian~\refeq{eq:ymlag}
at order $\alpha$ and $\alpha^2$.
As usual UV singularities are regularised in $\dendim=4-2\eps$ dimensions. 
The rational terms associated with a certain 1PI vertex function 
$\Gamma$ are presented in the form 
\bea
\label{eq:qed4}
\ratamp{k}{\Gamma}{\alpha_1\dots\alpha_N}{} &=& \ri
\lb\frac{\alpha\, t^\eps}{4\pi}\rb^k 
\sum_{a} 
\delta \hat\calR^{(a)}_{k,\Gamma}\,\,
\calT_{a,\Gamma}^{\alpha_1\dots\alpha_N}\,
\,,
\eea
where $k=1,2$ is the loop order, 
and
$\calT_{a,\Gamma}^{\alpha_1\dots\alpha_N}$ are independent tensor structures
carrying the indices $\alpha_1\dots\alpha_N$ of the external lines of the
vertex function at hand.
A similar decomposition is used also for the full $k$-loop counterterm
$\deltaZ{k}{\Gamma}{\alpha_1\dots\alpha_N}{}$ 
associated with $\Gamma$.

We recall that, when one-loop counterterms
$\delta Z_{1,\gamma}^{\alpha_1\dots \alpha_N}(q_1)$ are inserted into one-loop diagrams
in the context of two-loop calculations,
the associated tensor
structures and their dependence on the loop momentum $q_1$ 
have to be adapted to the dimensionality of the
loop numerator, \ie in $\numdim=\dendim$ and $\numdim=4$
numerator dimensions
$\delta Z_{1,\gamma}^{\bar \alpha_1\dots \bar
\alpha_N}(\barq_1)$ and $\delta Z_{1,\gamma}^{\alpha_1\dots
\alpha_N}(q_1)$ have to be used, respectively.

\vskip 5mm
\subsubsection*{Fermion two-point function}
For the two-point function of a fermion $f$ with mass $m_f$ we have 
\begin{align}
&\vcenter{\hbox{\raisebox{10pt}{\includegraphics[width=0.22\textwidth]{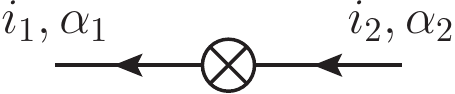}} }}   
 = \; \ri  \,  \delta_{i_1 i_2} \bigg\{ \,\lb \slashed p - m_f \rb_{\alpha_1
\alpha_2} \nonumber \\[2mm]
&\qquad + \,
\sum_{k=1}^2 \lb\frac{\als \, t^{\eps}}{4\pi} \rb^k 
\bigg[
\lb \delta \hat Z^{(\srp)}_{k,\ff} + \delta \hat\calR^{(\srp)}_{k,\ff} \rb \,
\slashed p_{\alpha_1\alpha_2} \,
\,+\,
\lb \delta \hat Z^{(\srm)}_{k,\ff} + \delta \hat\calR^{(\srm)}_{k,\ff} \rb \,
  m_f\,\delta_{\alpha_1\alpha_2}
\bigg] \bigg\} \,,
\label{eq:R2quark}
\end{align}
with UV counterterms
\renewcommand{\arraystretch}{1.5}
\begin{align}
\delta \hat Z_{1,\ff}^{(\srp)} &  \;=\;  
\delta\hat\calZ_{1,\psif}\,,                            &\quad  
\delta \hat Z_{2,\ff}^{(\srp)}   \;=\; & 
\delta\hat \calZ_{2,\psif}\,,    \nonumber \\[2mm]
\delta \hat Z_{1,\ff}^{(\srm)} &  \;=\;  
-\delta\hat\calZ_{1,\psif}-\delta\hat\calZ_{1,m_f}\,,  &\quad  
\delta \hat Z_{2,\ff}^{(\srm)}   \;=\; & 
-\delta\hat\calZ_{2,\psif}-\delta\hat\calZ_{2,m_f}
-\delta\hat\calZ_{1,\psif}\,\delta\hat\calZ_{1,m_f}\,,
\end{align}
and rational counterterms
\renewcommand{\arraystretch}{1.5}
\begin{align}
\delta \hat \calR_{1,\ff}^{(\srp)}   \;=\; & - \CF \,, \nonumber\\[2mm] 
\delta \hat \calR_{2,\ff}^{(\srp)}   \;=\; &
\left(\frac{7}{6}\, \CF^2 -\frac{61}{36} \, \CA \, \CF+\frac{5}{9} \,  \TF \, \nf \, \CF  \right)  \eps^{-1}  +\frac{43}{36}\, \CF^2 - \frac{1087}{216} \, \CA \, \CF+\frac{59}{54} \,  \TF \, \nf \, \CF   \nonumber \\
& - \CF \left( \dcalZ_{1,\als} + \frac{2}{3} \, \dcalZ_{1,\psif}-\frac{2}{3} \, \dcalZ_{1,\gpar}\right) \,,
\end{align}
\begin{align}
\delta \hat \calR_{1,\ff}^{(\srm)}   \;=\; &  2 \, \CF \,, \nonumber\\[2mm] 
\delta \hat \calR_{2,\ff}^{(\srm)}   \;=\; &  
\left(-2 \, \CF^2+\frac{61}{12} \, \CA \, \CF - \frac{5}{3} \, \TF \, \nf \, \CF  \right)  \eps^{-1} + \CF^2+\frac{199}{24} \, \CA \, \CF-\frac{11}{6} \, \TF \, \nf \, \CF \nonumber \\
& +\CF \left( 2 \, \dcalZ_{1,\als}+ 4 \, \dcalZ_{1,m_f} -\frac{3}{2} \, \dcalZ_{1,A}-\frac{1}{2} \, \dcalZ_{1,\gpar}\right) 
 \,.
\label{eq:R2quarklast}
\end{align}
As usual the direction of the momentum $p$ in~\refeq{eq:R2quark} 
coincides with the fermion flow.

\vskip 5mm
\subsubsection*{Gauge-boson two-point function}
For the 
gauge-boson
two-point function we have
\begin{align}
&\vcenter{\hbox{\raisebox{0pt}{\includegraphics[width=0.2\textwidth]{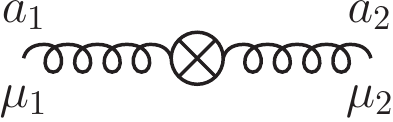}} }} 
\;=\; {}\ri \, \delta^{a_1 a_2} \bigg\{  -p^2 g^{\mu_1\mu_2} \nonumber\\[2mm]
&\qquad
\hspace{-8mm}+ \, \sum_{k=1}^2 \lb \f{\als \, t^\eps}{4\pi}\rb^k  
\bigg[ \lb \delta \hat Z^{(\srp)}_{k,\mathrm{gg}} + \delta \hat \calR^{(\srp)}_{k,\mathrm{gg}}  \rb
\, p^{\mu_1}p^{\mu_2}  
+\lb \delta \hat Z^{(\srG)}_{k,\mathrm{gg}}  p^2
+ \delta \hat \calR^{(\srG)}_{k,\mathrm{gg}}\, p^2
+ \delta \tilde{Z}^{(\srG)}_{k,\mathrm{gg}} \, \tilde p^2
\rb \, g^{\mu_1\mu_2} 
\bigg]\bigg\}
\,, 
\label{eq:R2gluon}
\end{align}
with UV counterterms
\renewcommand{\arraystretch}{1.5}
\begin{align}
\delta \hat  Z_{1,\mathrm{gg}}^{(\srp)} &  \;=\;  
\delta\hat\calZ_{1,A} - \delta\hat\calZ_{1,\gpar} \,,                            &\quad  
\delta \hat  Z_{2, \mathrm{gg}}^{(\srp)}   \;=\; & 
\delta\hat\calZ_{2,A} - \delta\hat\calZ_{2,\gpar} \,,    \nonumber \\[2mm]
\delta \hat  Z_{1,\mathrm{gg} }^{(\srG)} &  \;=\;  
{}-\delta\hat\calZ_{1,A}\,,                            &\quad  
\delta \hat  Z_{2,\mathrm{gg} }^{(\srG)}   \;=\; & 
{}-\delta\hat\calZ_{2,A}\,,    
\end{align}
and rational counterterms
\begin{align}
\delta \hat \calR_{1,\mathrm{gg}}^{(\srp)}  \;=\; &  
- \frac{\CA}{3}\,,           \nonumber \\[2mm]
\delta \hat \calR_{2,\mathrm{gg}}^{(\srp)}   \;=\; &
   \left[\frac{19}{36} \, \CA^2 + \TF \, \nf \lb - \frac{32}{9}  \,  \CA  + 2  \, \CF \rb
\right] \eps^{-1} 
 + \TF \, \nf \lb \frac{217}{108}  \, \CA  - \frac{71}{18} \, \CF \rb
\nonumber \\
&    +\frac{1211}{864} \, \CA^2
   + \CA \lb - \frac{1}{3} \,\dcalZ_{1,\als} -\frac{35 }{12} \, \dcalZ_{1,A} +\frac{3}{4} \, \dcalZ_{1,\gpar} + \frac{1}{6} \, \dcalZ_{1,u}\rb
\nonumber\\
& + \frac{4}{3} \, \TF \, \sum_{f\in\calF} \dcalZ_{1,\psif}
         \,,
\end{align}
and
\begin{align}
\delta \hat \calR_{1,\mathrm{gg}}^{(\srG)} \;=\; &   
\left(\frac{\CA}{2} +\frac{2}{3} \,  \TF \, \nf \right)
             -4 \, \TF \sum_{f\in \calF}   \f{m_f^2}{p^2}
             \,,              \nonumber \\[2mm]
\delta \hat \calR_{2,\mathrm{gg}}^{(\srG)}  \;=\;  &  
     \left[-\frac{4}{9} \, \CA^2+ \TF \, \nf \lb \frac{35}{9} \, \CA  - 2  \, \CF \rb
\right] \eps^{-1} 
    + \TF \, \nf \lb - \frac{193}{108} \, \CA  + \frac{109}{36} \, \CF \rb
 \nonumber \\
 &      -\frac{541}{432} \, \CA^2 
    -\TF \sum_{f\in \calF}\,  \left[ 
    \left(  \CA  + 6 \,  \CF  \right) \eps^{-1} 
    + \frac{13}{6}  \, \CA  - 7 \, \CF
    \right] \, \f{m_f^2}{p^2}
  \nonumber \\
&        +\lb \f{\CA}{2} + \f{2}{3}\, \TF \, \nf \rb  \, \dcalZ_{1,\als}
         + \lb \f{71}{24} \, \CA + \f{2}{3} \, \TF \, \nf \rb  \, \dcalZ_{1,A} 
          - \f{7}{8} \, \CA  \, \dcalZ_{1,\gpar}
\nonumber \\
&
          + \f{\CA}{12}\, \dcalZ_{1,u}  
        	-4 \, \TF  \sum_{f\in\calF} 
          \left[\frac{1}{3}\dcalZ_{1,\psif} 
          +\lb 
           \dcalZ_{1,\als} 
          +  \delta \hat \calZ_{1,A}
          + \delta \hat \calZ_{1,m_f} 
           \rb \, \f{m_f^2}{p^2}\right] \,. \nonumber \\
\end{align}
In addition, due to the presence of a quadratic divergence, the 
usual UV counterterm for the 
gluon two-point function needs to be supplemented by
\begin{align}
\label{eq:qtildeggterm}
\delta \tilde Z^{(\srG)}_{1,\mathrm{gg}}  \;=\;  
\left( \frac{2}{3} \, \CA + \frac{2}{3} \, \TF \, \nf \right)  \eps^{-1}\,.
\end{align}
This extra term is relevant only 
when it is inserted into a 
one-loop diagram 
in the context of two-loop 
calculations,
and its two-loop extension $\delta \tilde Z^{(\srG)}_{2,\mathrm{gg}}$
is required only for calculations beyond two loops.

\vskip 5mm
\subsubsection*{Ghost two-point function}
For the ghost two-point function we have
\begin{align}
&\vcenter{\hbox{\raisebox{0pt}{\includegraphics[width=0.2\textwidth]{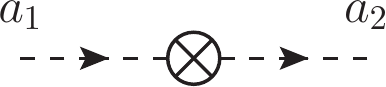}} }}   
\;  = \; {}
\ri\,  \delta^{a_1 a_2} \bigg\{ \, p^2 +
\sum_{k=1}^2 \lb\frac{\als \, t^{\eps} }{4\pi} \rb^k  \,
\lb \delta \hat Z^{(\srp)}_{k,\mathrm{uu}} + \delta
  \hat\calR^{(\srp)}_{k,\mathrm{uu}}  \rb \, p^2   \bigg\} \,,
\label{eq:R2ghost}
\end{align}
with UV counterterms
\begin{align}
\delta \hat  Z^{(\srp)}_{1,\mathrm{uu}} &  \;=\;  
\delta\hat\calZ_{1,u} \,,                            &\quad  
\delta \hat  Z^{(\srp)}_{2, \mathrm{uu}}   \;=\; & 
\delta \hat \calZ_{2,u}  \,,    
\end{align}
and rational counterterms
\begin{align}
\delta \hat\calR^{(\srp)}_{1,\mathrm{uu}}  \;=\; & 0  \,, \nonumber \\[2mm]
\delta \hat\calR^{(\srp)}_{2,\mathrm{uu}}  \;=\; & 
\left(\f{7}{18} \, \CA^2 - \f{5}{18} \, \TF \, \nf \, \CA   \right) \eps^{-1} 
+ \f{10}{27} \, \CA^2 - \f{17}{108} \, \TF \, \nf \, \CA  \nonumber \\
&  - \CA \lb \f{1}{2}   \, \dcalZ_{1,A}  
	- \f{1}{6}   \, \dcalZ_{1,\gpar}
	 + \f{1}{6}   \, \dcalZ_{1,u} \rb
\,. 
\end{align}
The vanishing of the one-loop rational term is due to the fact that, apart
from coupling factors, the 
numerator of the  ghost one-loop selfenergy is simply given by 
$\barq_{\bar \mu}\, p^\mu = q_\mu \,p^\mu$, where $p$ is the external
momentum, and is thus free form 
$(D-4)$-dimensional parts. 
Note also that the quadratic mass dimension of the
ghost two-point function may require a $\tilde p^2/\eps$
counterterm of type~\refeq{eq:4dimsubdiagE8}.
However, this is not the case since,
due to the absence of 
quadratic terms in $q$ in the loop numerator,
the one-loop ghost  selfenergy is free from
quadratic divergences.

\vskip 5mm
\subsubsection*{Gauge-boson--fermion three-point vertex}
For the gauge-boson--fermion--antifermion vertex
we have
\begin{align}
& \vcenter{\hbox{\raisebox{0pt}{\includegraphics[width=0.20\textwidth]{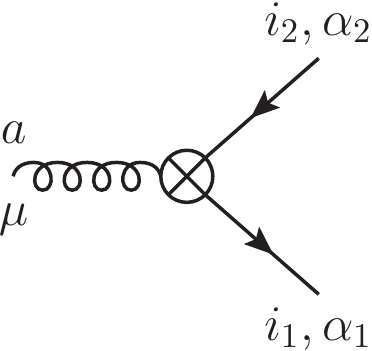}} }} 
\;  = \; {}
\ri\,\gs  \, \gamma_{\alpha_1\alpha_2}^{\mu} \, T^{a}_{i_1 i_2}\,  \bigg\{ \, 1 +
\sum_{k=1}^2 \lb \f{\als \, t^{\eps}}{4\pi}  \rb^k 
\lb \delta \hat Z^{(\srV)}_{k,\mathrm{ffg}} + \delta \hat \calR^{(\srV)}_{k, \mathrm{ffg}} \rb \bigg\}
\,,
\label{eq:R2qqg}
\end{align}
with UV counterterms
\begin{align}
\delta \hat  Z_{1,\mathrm{ffg}}^{(\srV)}  \;=\; &  
\f{1}{2} \lb \dcalZ_{1,\als} + \dcalZ_{1,A} \rb+ \delta\hat \calZ_{1,\psif}
\,,    
\nonumber \\
\delta \hat  Z_{2, \mathrm{ffg}}^{(\srV)}   \;=\; &
\f{1}{2} \lb \dcalZ_{2,\als} + \dcalZ_{2,A} \rb + \delta\hat \calZ_{2,\psif}  
 -\frac{1}{8} \lb \dcalZ_{1,\als}^2 + \delta\hat \calZ_{1,A}^2 \rb
+\f{1}{2}\dcalZ_{1,\psif} \lb \dcalZ_{1,\als} + \dcalZ_{1,A} \rb \nonumber \\
& +\frac{1}{4}\delta\hat \calZ_{1,A} \, \delta\hat \calZ_{1,\als} 
\,,
\end{align}
and rational counterterms
\begin{align}
\delta \hat \calR_{1,\mathrm{ffg}}^{(\srV)}   \;=\; & - 2 \, \CF \,, \nonumber \\[2mm]         
\delta \hat \calR_{2,\mathrm{ffg}}^{(\srV)}   \;=\; &
   \left[
   -\frac{5}{144} \, \CA^2 - \frac{26}{9}  \, \CA \, \CF + \frac{4}{3} \, \CF^2 +\frac{7}{9} \, \TF \, \nf \lb \CA + \CF \rb   
   \right]\eps^{-1} \nonumber  \\
& + 
    \frac{829}{864} \,  \CA^2 -\frac{563}{54} \, \CA \, \CF +\frac{109}{18} \, \CF^2  
-\frac{\TF \, \nf}{27} \lb 7\, \CA - \frac{55}{2} \, \CF \rb
   - 3 \, \CF  \, \dcalZ_{1,\als} 
    \nonumber \\
&          +   \frac{1}{2} \lb \CA - 3 \, \CF \rb  \, \dcalZ_{1,A} \nonumber  +   \frac{1}{6} \lb 2 \, \CA + 5 \, \CF \rb  \, \dcalZ_{1,\gpar}   
         + \frac{1}{6} \lb \CA - 8 \, \CF \rb \, \dcalZ_{1,\psif} \,. \nonumber \\
\end{align}

\subsubsection*{Gauge-boson three-point vertex}
For the triple gauge-boson vertex
we have
\begin{align}
\vcenter{\hbox{\raisebox{0pt}{\includegraphics[width=0.19\textwidth]{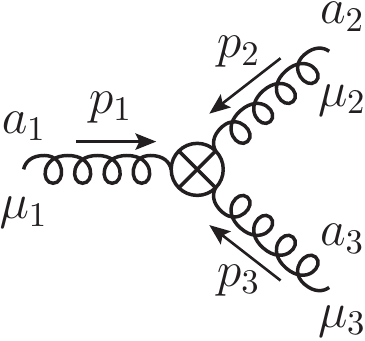}} }} 
\hspace{2mm}
&\;=\;
\gs  \, f^{a_1 a_2 a_3} \, 
\Big[
g^{\mu_1 \mu_2} (p_1-p_2)^{\mu_3} 
+ g^{\mu_2 \mu_3}(p_2-p_3)^{\mu_1} 
\nonumber\\[-9mm]&\qquad
+ g^{\mu_3 \mu_1}(p_3 - p_1)^{\mu_2}
\Big]
\bigg\{ \, 1 +
\sum_{k=1}^2 \lb \f{\als \, t^{\eps} }{4\pi} \rb^k  \lb \delta \hat Z^{(\srV)}_{k,\mathrm{ggg}} +  \delta \hat \calR^{(\srV)}_{k, \mathrm{ggg}} \rb \bigg\}
\,,
\label{eq:R2ggg}
\end{align}
with UV counterterms
\begin{align}
\delta \hat  Z_{1,\mathrm{ggg}}^{(\srV)}  \;=\; &  
\f{1}{2}\dcalZ_{1,\als}+ \frac{3}{2} \dcalZ_{1,A}\,,    
\nonumber \\
\delta \hat  Z_{2, \mathrm{ggg}}^{(\srV)}   \;=\; &
 \f{1}{2}\dcalZ_{2,\als}  + \frac{3}{2}\delta\hat \calZ_{2,A}
 - \f{1}{8}\dcalZ_{1,\als}^2 + \frac{3}{8} \delta\hat \calZ_{1,A}^2
+\frac{3}{4}\delta\hat \calZ_{1,A} \, \delta\hat \calZ_{1,\als} 
\,,
\end{align}
and rational counterterms
\begin{align}
\delta \hat \calR_{1,\mathrm{ggg}}^{(\srV)}   \;=\; &
    -\frac{11}{12} \, \CA - \frac{4}{3}  \, \TF \, \nf  \,, \nonumber \\[2mm]         
\delta \hat \calR_{2,\mathrm{ggg}}^{(\srV)}   \;=\; &
     	-\left[\frac{11}{48} \, \CA^2 + \TF \, \nf \lb \frac{23}{6}  \, \CA
-\frac{8}{3}  \, \CF \rb
\right] \eps^{-1} 
+ \TF \, \nf \lb \frac{25}{9}  \, \CA - \frac{119}{36} \, \CF \rb   \nonumber \\
           &           + \frac{145}{288} \, \CA^2 
- \lb \f{11}{8} \, \CA + 2 \, \TF \,  \nf \rb \, \dcalZ_{1,\als}
          - \lb \f{13}{4} \, \CA + 2 \, \TF \, \nf \rb \, \dcalZ_{1,A}   
\nonumber \\
&          + \f{5}{4} \, \CA \, \dcalZ_{1,\gpar} 
          - \f{\CA}{24} \, \dcalZ_{1,u}  
           + \frac{4}{3} \, \TF \, \sum_{f\in\calF} \dcalZ_{1,\psif}  \,. 
\end{align}

\vskip 5mm
\subsubsection*{Gauge-boson--ghost three-point vertex}
For the gauge-boson--ghost--antighost vertex
we have
\begin{align}
& \vcenter{\hbox{\raisebox{0pt}{\includegraphics[width=0.18\textwidth]{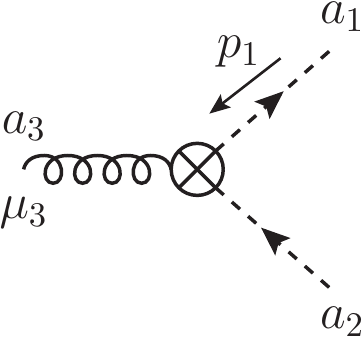}} }} 
\;  = \; {}
{}-\gs  \, f^{a_1 a_2 a_3} \, p_1^{\mu_3}\, \bigg \{ \, 1 +
\sum_{k=1}^2 \lb \f{\als \, t^{\eps}}{4\pi}  \rb^k 
\lb \delta \hat Z^{(\srV)}_{k, \mathrm{uug}} + \delta \hat \calR^{(\srV)}_{k, \mathrm{uug}}  \rb \bigg\}
\,,
\label{eq:R2ccg}
\end{align}
with UV counterterms
\begin{align}
\delta \hat  Z_{1,\mathrm{uug}}^{(\srV)}  \;=\; &  
\f{1}{2}\lb \dcalZ_{1,\als} + \dcalZ_{1,A} \rb+ \delta\hat \calZ_{1,u}
\,,    
\nonumber \\
\delta \hat  Z_{2, \mathrm{uug}}^{(\srV)}   \;=\; &
\f{1}{2} \lb \dcalZ_{2,\als} + \dcalZ_{2,A} \rb  + \delta\hat \calZ_{2,u} 
- \f{1}{8} \lb \dcalZ_{1,\als}^2+ \dcalZ_{1,A}^2\rb  
+\f{1}{2}\dcalZ_{1,u} \lb \dcalZ_{1,\als} + \dcalZ_{1,A} \rb \nonumber \\
& +\frac{1}{4}\dcalZ_{1,A} \, \dcalZ_{1,\als} 
\,,
\end{align}
and rational counterterms
\begin{align}
\delta \hat \calR_{1,\mathrm{uug}}^{(\srV)}   \;=\; &
     -\frac{\CA}{4}  \,, \nonumber \\[2mm]         
\delta \hat \calR_{2,\mathrm{uug}}^{(\srV)}   \;=\; &
     	-\lb \f{7}{36} \, \CA^2 - \f{5}{36} \, \TF \, \nf \, \CA    \rb \eps^{-1} 
	- \f{107}{432} \, \CA^2 - \f{19}{216} \, \TF \, \nf \, \CA    \nonumber \\
	& - \CA \lb  \f{3}{8}  \, \dcalZ_{1,\als} 
          + \f{1}{4}  \, \dcalZ_{1,A} 
          - \f{5}{12}  \, \dcalZ_{1,\gpar} 
          + \f{1}{24}  \, \dcalZ_{1,u}  \rb \,. \nonumber \\
\end{align}

\subsubsection*{Gauge-boson four-point vertex}
For the quartic-gluon vertex we find
\begin{align}
&\vcenter{\hbox{\raisebox{0pt}{\includegraphics[width=0.18\textwidth]{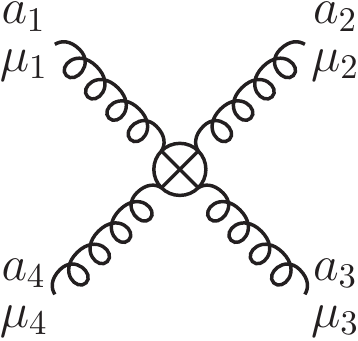} }}} 
\;  = \; {}
\ri \,
\gs^2 \sum_{\pi(234)}\Bigg \{
f^{a_1 a_3 e} f^{a_2 a_4 e}
\,\calV_{\rI}^{\mu_1\mu_2\mu_3\mu_4}
\left[1+\sum_{k=1}^2 \lb \f{\als \, t^{\eps}}{4\pi} \rb^k 
\delta \hat Z_{k, 4\mathrm{g}}^{(\rA\rI)}
\right]
\nonumber\\[4mm]
&
+ \sum_{k=1}^2 \lb \f{\als \, t^{\eps}}{4\pi} \rb^k 
\Bigg[ \,
\sum_{\beta=\rI,\rII}
\Big(
\TF\,
\delta^{a_1 a_2} \delta^{a_3 a_4}
\,\delta \hat \calR^{(\rB\beta)}_{k, 4\mathrm{g}}
+
  \frac{\Tr(T^{a_1}T^{a_3}T^{a_2}T^{a_4})}{\TF}
\,\delta \hat \calR^{(\rC\beta)}_{k, 4\mathrm{g}}
\Big)
\calV_{\beta}^{\mu_1\mu_2\mu_3\mu_4}
\Bigg]
\Bigg\}\,,
\label{eq:R24g}
\end{align}
where
\bea
\label{eq:4glorstr}
\calV_{\rI}^{\mu_1\mu_2\mu_3\mu_4} &=& g^{\mu_1\mu_2}g^{\mu_3\mu_4}\,,
\qquad
\calV_{\rII}^{\mu_1\mu_2\mu_3\mu_4} \,=\, 
g^{\mu_1 \mu_3}g^{\mu_2\mu_4} + g^{\mu_1 \mu_4}g^{\mu_2\mu_3}\,,
\eea
and $\pi(234)$ denotes the six permutations of the particle labels $234$. 
Note that the Lorentz tensors~\refeq{eq:4glorstr} are separately invariant 
wrt $1\leftrightarrow 2$ and  $3\leftrightarrow 4$  . Thus the sum over
$\pi(234)$ generates only three independent Lorentz tensors for each 
$\calV_\beta^{\mu_1\mu_2\mu_3\mu_4}$.
When these Lorentz tensors are combined 
with the various group-theoretical structures in~\refeq{eq:R24g}, \ie
\bea
\label{eq:4gcolstr}
\calC_{\rA}^{a_1 a_2 a_3 a_4} &=& f^{a_1 a_3 e} f^{a_2 a_4 e}\,,
\quad
\calC_{\rB}^{a_1 a_2 a_3 a_4} \,=\,\TF\,\delta^{a_1 a_2} \delta^{a_3 a_4}\,,
\quad
\calC_{\rC}^{a_1 a_2 a_3 a_4} \,=\,
\frac{\Tr(T^{a_1}T^{a_3}T^{a_2}T^{a_4})}{\TF}\,,
\nonumber\\
\eea
the summation of each combination 
$\calC_{\alpha}^{a_1 a_2 a_3 a_4}
\calV_{\beta}^{\mu_1\mu_2\mu_3\mu_4}$ 
over $\pi(234)$ yields,
\bea
\label{eq:colstrsymm}
\sum_{\pi(234)}
\calC_{\alpha}^{a_1 a_2 a_3 a_4}\,
\calV_\beta^{\mu_1\mu_2\mu_3\mu_4}
&=&
\Big[\calC_{\alpha}^{a_1 a_2 a_3 a_4}
+\calC_{\alpha}^{a_1 a_2 a_4 a_3}\Big]
\calV_\beta^{\mu_1\mu_2\mu_3\mu_4}
+(2\leftrightarrow 3)+(2\leftrightarrow 4)\,,\qquad
\eea
where each of the three terms on the rhs is separately invariant
wrt $1\leftrightarrow 2$ and  $3\leftrightarrow 4$. As a consequence~\refeq{eq:colstrsymm}
is totally symmetric in the four particle indices $1234$.
Note that the ordering of the generators in the trace of the gauge-group structure 
$\calC_{\rC}^{a_1 a_2 a_3 a_4}$ in~\refeq{eq:R24g} and~\refeq{eq:4gcolstr}
is $T^{a_1}T^{a_3}T^{a_2}T^{a_4}$. Note also that
in the definition of $\calC_{\rB}$ and $\calC_{\rC}$ we include explicit $\TF$ factors
in such a way that all gauge-group structures in~\refeq{eq:4gcolstr}
scale like $\TF$.

The UV counterterms for the quartic vertex~\refeq{eq:R24g}
read
\begin{align}
\delta \hat  Z_{1,\mathrm{4g}}^{(\rA\rI)}  \;=\; &  
 \dcalZ_{1,\als} + 2\, \delta\hat \calZ_{1,A} \,,    
\nonumber \\
\delta \hat  Z_{2, \mathrm{4g}}^{(\rA\rI)}   \;=\; &
  \delta\hat \calZ_{2,\als} + 2\, \delta\hat \calZ_{2,A} + \delta\hat \calZ_{1,A}^2
+ 2 \, \delta\hat \calZ_{1,\als} \, \delta\hat \calZ_{1,A} \,,
\end{align}
and for the rational counterterms we find
\begin{align}
\label{eq:BI4gres}
  \delta \hat \calR_{1,4\mathrm{g}}^{(\rB\rI)}   \;=\; &
  -\frac{1}{3} \, \frac{\CA}{\Nc}
  \nonumber \\
  \delta \hat \calR_{2,4\mathrm{g}}^{(\rB\rI)}   \;=\;  &
  \frac{\CA}{2}
  \Biggl[
  \left(
  \frac{13}{12} + \frac{1}{6}\,\frac{\nf}{\Nc}
  \right)
  \TF\,
  \varepsilon^{-1}
  - \lb 
   \frac{571}{288}
  - \frac{53}{36}\, \frac{\nf}{\Nc}
  \rb
  \TF
  -\frac{1}{\Nc}
  \bigg(
  \frac{4}{3} \, \dcalZ_{1,\als}
  + \frac{29}{24} \, \dcalZ_{1,A}
  \notag\\  &
  + \frac{1}{4} \, \dcalZ_{1,u}
  - \frac{1}{8} \, \dcalZ_{1,\gpar}
  \bigg)
  \Biggr]\,,
\end{align}

\begin{align}
\label{eq:BII4gres}
 \delta \hat \calR_{1,4\mathrm{g}}^{(\rB\rII)}   \;=\; &
  -\frac{1}{3} \, \frac{\CA}{\Nc} 
  \nonumber \\
  \delta \hat \calR_{2,4\mathrm{g}}^{(\rB\rII)}   \;=\; &
  \frac{\CA}{2}
  \Biggl[
  \left(
  -\frac{23}{12} + \frac{7}{6}\,\frac{\nf}{\Nc}
  \right)
  \TF
  \,\varepsilon^{-1}
  + 
  \lb
  \frac{233}{72}
  +\frac{23}{36}\,
  \frac{\nf}{\Nc}
  \rb
  \TF
-\frac{1}{\Nc}
  \bigg(
  \frac{4}{3} \, \dcalZ_{1,\als}
  + \frac{29}{24} \, \dcalZ_{1,A}
  \notag\\&
  + \frac{1}{4} \, \dcalZ_{1,u}
  - \frac{1}{8} \, \dcalZ_{1,\gpar}
  \bigg)
  \Biggr]\,,
\end{align}

\begin{align}
  \delta \hat \calR_{1,4\mathrm{g}}^{(\rC\rI)}   \;=\; &
  - \frac{8}{3} \, \CA  - \frac{10}{3} \, \TF \, \nf
  \nonumber \\
\delta \hat \calR_{2,4\mathrm{g}}^{(\rC\rI)}   \;=\; 
  &
  -\left[
  \frac{271}{144} \, \CA^2
  + 2 \, \frac{\CA \, \TF}{\Nc}
  +  \TF \, \nf \lb \frac{247}{36} \,  \CA
  -\frac{20}{3} \,  \CF 
  \rb
  \right]
  \varepsilon^{-1}
  -\frac{421}{3456} \, \CA^2
\notag\\ &
   +\frac{167}{48} \, \frac{\CA \, \TF}{\Nc}
   + \TF \, \nf \lb \frac{199}{27} \, \CA
  -\frac{43}{6} \,\CF \rb 
  -\left(\frac{16}{3} \, \CA + \frac{20}{3} \, \TF \, \nf \right) \dcalZ_{1,\als}
\notag\\  &
  -\left( \frac{379}{48} \, \CA + \frac{20}{3} \, \TF \, \nf \right) \dcalZ_{1,A}
-\frac{\CA}{24}
\, \dcalZ_{1,u}
  +\frac{157}{48} \, \CA \, \dcalZ_{1,\gpar}
  +\frac{8}{3} \, \TF \, \sum_{f\in \calF} \dcalZ_{1,\psif}\,,
\end{align}

\begin{align}
  \delta \hat \calR_{1,4\mathrm{g}}^{(\rC\rII)}   \;=\; &
  \frac{7}{6} \, \CA + 2 \, \TF \, \nf
  \nonumber \\
\delta \hat \calR_{2,4\mathrm{g}}^{(\rC\rII)}   \;=\;
  &
  \left[
  \frac{119}{144} \, \CA^2
   +\frac{\CA\, \TF}{\Nc}
  +\TF \, \nf  \lb \frac{143}{36} \, \CA 
  -\frac{10}{3} \, \CF \rb
  \right] \varepsilon^{-1}
  + \frac{857}{3456} \, \CA^2
\notag\\ &
  -\frac{167}{96} \, \frac{\CA \, \TF}{\Nc}
  - \TF \, \nf \lb \frac{1649}{432} \, \CA 
  - \frac{17}{6} \, \CF \rb
  +\left(\frac{7}{3} \, \CA + 4 \, \TF \, \nf \right) \dcalZ_{1,\als}
  \notag\\ &
  +\left(\frac{175}{48}\, \CA + 4 \, \TF \, \nf \right) \dcalZ_{1,A} 
-\frac{\CA}{24} \, \dcalZ_{1,u}
  -\frac{77}{48} \, \CA \, \dcalZ_{1,\gpar}
  -\frac{4}{3} \, \TF \, \sum_{f\in\calF}\, \dcalZ_{1,\psif}\,.
\end{align}
Note that for U(1) gauge theories the coefficients~\refeq{eq:BI4gres}--\refeq{eq:BII4gres}
vanish.

When using the $\msbar$ renormalisation constants
listed in~\refapp{app:msbarRCs} and applying the U(1)
substitutions of~\refta{tab:uonesun},
all results presented in this section
agree with previous results for QED rational counterterms 
in the $\lambda=1$ gauge~\cite{Pozzorini:2020hkx}.

\section{Conclusions}

The most widely used tools for the automated calculation of one-loop
amplitudes are based on numerical algorithms that build the numerators of loop
integrands in 
$\numdim=4$
dimensions, while the remnant parts are reconstructed by
means of rational counterterms.
This approach has proven to be very flexible and efficient
for the automation of NLO calculations, and 
its extension to two loops can become an important ingredient
in the development of automated tools at NNLO.
As a first step in this direction, in~\cite{Pozzorini:2020hkx}
it was shown that
renormalised two-loop amplitudes in $D=4-2\eps$ dimensions
can be related to 
corresponding amplitudes 
in $\numdim=4$ dimensions, i.e.~with
four-dimensional loop numerators,
making use of process-independent rational counterterms.
More precisely,
all two-loop contributions stemming from the interplay
of UV poles with the $(D-4)$-dimensional parts of loop numerators 
can be reconstructed through insertions of the well known one-loop rational counterterms
$\delta \calR_1$ into one-loop amplitudes
and insertions of two-loop rational counterterms
$\delta \calR_2$ into tree amplitudes.
In addition, for the 
subtraction of 
one-loop subdivergences
in $\numdim=4$ dimensions
the usual UV counterterms $\delta Z_1$ 
need to be supplemented by extra counterterms
$\delta \tilde Z_1$ proportional to 
$\tilde q^2/\eps$.

In this paper we have presented a general analysis of
the dependence of two-loop rational terms on the choice of renormalisation scheme.
Specifically we have demonstrated that the form of the 
master formula for renormalised two-loop amplitudes---initially
established within the minimal subtraction scheme---is 
independent of the renormalisation scheme.
Moreover we have derived 
general formulas~\refeq{eq:RtransfG}--\refeq{eq:straK1c}
that describe the scheme dependence of 
$\delta \calR_2$ counterterms as the combination of two contributions:
the naive renormalisation of $\delta \calR_1$ counterterms and
an extra 
nontrivial contribution, which is due to the fact that
the multiplicative renormalisation of 
subdivergences does not commute with the projection of loop numerators 
to 
$\numdim=4$ 
dimensions.
In renormalisable theories, 
such nontrivial contributions
originate only from 
one-loop selfenergy subdiagrams
and can be controlled through auxiliary one-loop counterterms
as specified in~\refeq{eq:strarec1}--\refeq{eq:strarec2c}.

As a consequence, the scheme-dependent part of $\delta \calR_2$ 
counterterms can be expressed 
as a linear combination of one-loop 
renormalisation constants with process- and scheme-independent coefficients.
This makes it possible to derive the $\delta \calR_2$ counterterms 
for a given renormalisable theory in terms of generic 
one-loop renormalisation constants, which can be adapted a posteriori 
to any desired scheme.

Using the above approach we have generalised the known $\delta \calR_2$
counterterms for QED from the minimal subtraction scheme to any
renormalisation scheme.  Moreover, we have presented the first calculation
of the full set of $\delta \calR_2$ counterterms for Yang--Mills theories.  All
calculations have been carried out in the Feynman gauge, and the results are
presented in compact formulas that are applicable to SU(N) or U(1) theories
with $\nf$ 
massless or massive
 fermions.

Technically, for the calculation of $\delta \calR_2$ counterterms 
we have used various expansions that
capture the UV divergences of all relevant one- and two-loop 
diagrams in the form of massive tadpole integrals.
Such tadpole expansions are described in detail in~\refapp{se:tadexp},
including the expansions employed in~\cite{Pozzorini:2020hkx} as well as 
a new variant that reduces the number of 
expansion terms and allows also for a fully flexible parametrisation of 
loop momenta.

In the future we plan to investigate two-loop rational terms 
within
spontaneously broken gauge theories and to 
study the interplay of 
rational terms with infrared divergences.

\subsection*{Acknowledgments}
This research was supported by the Swiss National Science Foundation (SNSF) 
under contract BSCGI0-157722. The work of M.Z. was supported through the
SNSF Ambizione grant PZ00P2-179877.

\appendix

\section{Tadpole expansions}
\label{se:tadexp}

In this Appendix we review the
techniques that have been
used in~\cite{Pozzorini:2020hkx} to express rational terms in the form of
massive tadpole integrals and we present various optimisations.

\subsection{Iterative tadpole decomposition}
\label{se:naivetadpoledec}

The UV poles of multi-loop integrals and the 
associated rational parts can be isolated 
via recursive decomposition of the loop propagators
by means of the partial-fractioning formula~\cite{Misiak:1994zw,beta_den_comp,Zoller:2014xoa}
\bea
\label{eq:dentadexpA}
\frac{1}{D^{(i)}_{a}(\barq_i)}&=& 
\frac{1}{\barqidx{i}{2}-M^2}
+
\frac
{\Delta_{ia}(\barq_i)}
{\barqidx{i}{2}-M^2}
\frac{1}{D^{(i)}_{a}(\barq_i)}\,,
\eea
where the denominator $D^{(i)}_{a}(\barq_i)$ is 
defined in~\refeq{eq:twoloopnotB2}, and
\bea
\label{eq:dentadexpA2}
\Delta_{ia}(\barq_i)&=&
\left(\barqidx{i}{2}-M^2\right) 
-D^{(i)}_a(\barq_i)
\,=\,
-p_{ia}^2-2 \barq_i\cdot p_{ia}+m_{ia}^2-M^2\,,
\eea
while
$M$  is an auxiliary mass scale.
The above formula splits a generic scalar propagator 
into a tadpole propagator $1/(\barqidx{i}{2}-M^2)$
and a remnant part that consists of the original propagator 
times a factor  $\Delta_{ia}(\barq_i)/(\barqidx{i}{2}-M^2)$,
which is suppressed by $\ord(1/\barq_i)$ in the UV 
limit $\barq_i\to \infty$.

Iterating the above identity $X_i+1$ times 
makes it possible to capture all UV contributions of
the propagator $1/D^{(i)}_{a}(\barq_i)$ 
up to relative order $1/\barq^{X_i}_i$ 
in the form of tadpole integrands. 
This procedure can be easily extended to chains of 
$\barq_i$-dependent scalar
propagators~\refeq{eq:twoloopnotB}. To this end
the identity~\refeq{eq:dentadexpA} should be 
iterated on all $N_i$ propagators in the chain, and 
terms with denominators of the form
$\left(\barqidx{i}{2}-M^2\right)^{p}
D^{(i)}_{a_1}(\barq_i)\cdot\cdot\cdot D^{(i)}_{a_r}(\barq_i)$ 
with $p+r>N_i+X_i$  should be discarded.
This algorithm can be encoded in a tadpole expansion operator
$\bfS^{(i)}_{X_i} $, which yields combinations of tadpole integrands
\bea
\label{eq:chaintadexpA}
\bfS^{(i)}_{X_i} \,
\frac{1}{D^{(i)}_0(\barq_i)\cdots D^{(i)}_{N_i-1}(\barq_i)}
&=& 
\sum_{\sigma=0}^{X_i}
\frac
{\Delta_{i}^{(\sigma)}(\barq_i)}
{\left(\barqidx{i}{2}-M^2\right)^{N_i+\sigma}}\,,
\eea
where the numerators on the rhs read 
\bea
\label{eq:tadexptwoB}
{\Delta_i^{(\sigma)}(\barq_i)}
&=& 
\sum_{\sigma_0=0}^\sigma
\ldots
\sum_{\sigma_{N_i-1}=0}^\sigma
\prod_{a=0}^{N_i-1}
\Big[\Delta_{ia}(\barq_i)\Big]^{\sigma_a}
\Bigg|_{\sigma_0+\dots+ \sigma_{N_i-1} = \sigma}\,,
\eea
and correspond to polynomials of homogeneous degree  $\sigma$ in $\barq_i\cdot p_{ia}$ and in the 
squared mass scales $\{p_{ia}^2, m_{ia}^2\}$ and $M^2$.

By construction, the tadpole integrands on the rhs of~\refeq{eq:chaintadexpA}
capture the leading and subleading UV 
contributions of the original propagator chain 
up to relative order $1/\barq_i^{X_i}$.
Thus, formally 
\bea
\label{eq:Soppowcount}
1-\bfS^{(i)}_{X_i} &= & \ord\left({1}/{\barq_i^{\,X_i+1}}\right)\,.
\eea
Note that suppressed contributions of order $1/{\barq_i^{\,X_i+1}}$ 
and beyond are present also in~\refeq{eq:chaintadexpA}. This is 
due to the fact that terms of 
$\ord(\barq_i^{\,1})$ and $\ord(\barq_i^{\,0})$
in~\refeq{eq:dentadexpA2} are treated on the same footing.
Possible
optimisations that minimise the number of 
irrelevant higher-order terms in $1/{\barq_i}$ 
are discussed in the subsequent sections.

For loop chains,
\bea
\label{eq:loopchaindef}
\bar\calF^{(i)}_{\bar\alpha_i}(\barq_i) 
&=&
\frac{\bar\calN^{(i)}_{\bar\alpha_i}(\barq_i)}
{\calD{i}}
\,,
\eea
where $\bar\calN^{(i)}_{\bar\alpha_i}$
and $\calD{i}$ are defined in 
\refeq{eq:twoloopnotB} and~\refeq{eq:twoloopnumA},
the $\bfS^{(i)}_{X_i}$ expansion can be defined as
\bea
\label{eq:chainnaivtadexp}
\bfS^{(i)}_{X_i} \,
\bar\calF^{(i)}_{\bar\alpha_i}(\barq_i) 
&=&
\bar\calN^{(i)}_{\bar\alpha_i}(\barq_i)\,
\bfSX{i} \left(\frac{1}{\calD{i}}\right)
\,,
\eea
where the loop numerator is kept unexpanded.
For one-loop diagrams,
\bea
\label{eq:defgenoneloop}
\ampbar{1}{\Gamma}{}{} 
&=&
\int\rd\barq_1\,
\bar\calF^{(1)}(\barq_1)\,, 
\eea
we define
\bea
\label{eq:oneloopnaivtadexp}
\ampbar{1}{\Gamma_\tad}{}{} 
&=&
\bfS^{(1)}_{X_1}\,
\ampbar{1}{\Gamma}{}{}
\,=\,
\int\rd\barq_1\,
\bfS^{(1)}_{X_1}\, 
\bar\calF^{(1)}(\barq_1)
\,,
\eea
where the order $X_1$ of the expansion should be set equal to the degree of divergence of $\Gamma$.
Due to~\refeq{eq:Soppowcount}, this choice
guarantees that $\bar\calA_{1,\Gamma_\rem}=\bar\calA_{1,\Gamma_\tad}-\bar\calA_{1,\Gamma}$
has negative degree of divergence, which implies that 
all UV divergences of $\Gamma$ are embodied in the
tadpole expansion~\refeq{eq:oneloopnaivtadexp}.

At two loops, UV divergences can be isolated in tadpole integrals
by means of three separate tadpole expansions $\bfS_{X_i}^{(i)}$
with $i=1,2,3$, each of which 
acts exclusively on the $\barq_i$-dependent chain $\calC_i$. 
More explicitly, for the generic two-loop diagram~\refeq{eq:twoloopnotA} one
can define 
\bea
\label{eq:SSSexp}
\ampbar{2}{\Gamma_\tad}{}{} 
&=&
\prod_{i=1}^3\bfS^{(i)}_{X_i}\,
\ampbar{2}{\Gamma}{}{}
\,=\,
\int\rd\barq_1
\int \rd\barq_2\,
\left[
\bar\Gamma^{\bar\alpha_1\bar\alpha_2\bar\alpha_3}(\barq_1,\barq_2,\barq_3)\,
\prod_{i=1}^3
\left({\bf S}^{(i)}_{X_i}\,
\bar\calF^{(i)}_{\bar\alpha_i}(\barq_i) 
\right)\right]_{\barq_3\,=\,-\barq_1-\barq_2}
\hspace{-16mm}\,.\hspace{12mm}
\eea
Here the order of the various
$\bfSX{i}$ expansions should be chosen as
\bea
\label{eq:chainexpord}
X_i &=&
X_i(\Gamma) \,=\,
\text{max}\left\{ X(\Gamma), X_{ij}(\Gamma),
X_{ik}(\Gamma)\right\}\,,
\label{eq:powercountingDirectImp}
\eea
where $X(\Gamma)$ is the global degree of divergence,
$i|jk$ is a partition of $123$, and
$X_{im}(\Gamma)$ with $m=j,k$ are the degrees of divergence of the 
subdiagrams that contain the chain $\calC_i$, \ie the subdiagrams that
are subject to the $\bfSX{i}$ expansion.
With this choice, for each individual $\bfS_{X_i}^{(i)}$
expansion  the discarded $(1-\bfS_{X_i}^{(i)})$
contribution of order $1/\barq^{X_i+1}$ 
has a global degree of divergence
$X(\Gamma)-X_i(\Gamma)<0$, 
and a degree of 
subdivergence 
$X_{im}(\Gamma)-X_i(\Gamma)<0$ 
for the subdiagrams that are subject to the
$\bfS_{X_i}^{(i)}$ expansion. This means that each 
$\bfS_{X_i}^{(i)}$ expansion retains the full local 
divergence as well as the full divergences of the two 
subdiagrams that contain the chain $\calC_i$. 
For what concerns the subdiagram $\gamma_i$, 
which does not contain the chain $\calC_i$, its
subdivergence factorises wrt the $\bfS_{X_i}^{(i)}$ expansion.
This implies that also the $(1-\bfS_{X_i}^{(i)})$  finite remnant
of the expansion of the chain $\calC_i$
factorises wrt the $\gamma_i$ subdivergence.
As a result, the remnant of the complete expansion~\refeq{eq:SSSexp}
still contains the divergent parts
\bea
\label{eq:remdiv}
\bar \calA_{2, \Gamma_{\rem,\div}}
&=&
\sum_{i=1}^3
\left(1-\bfS^{(i)}_{X_i}\right)
\bfS^{(j)}_{X_j}
\bfS^{(k)}_{X_k}\,
\ampbar{2}{\Gamma}{}{}\,
\Bigg|_{\{j,k\}=\{1,2,3\}\backslash{} \{i\}}
\,,
\eea
which involve 
the full subdivergence of the various $\gamma_i$ subdiagrams
combined with the remnants of the expansions of their 
complementary chains $\calC_i$.
These missing UV divergent parts are not globally divergent. Thus,
according to~\refeq{eq:proofsketchC}, they do not contribute 
to $\delta \calR_{2,\Gamma}$. 

As for the tadpole expansion~\refeq{eq:SSSexp},
as discussed above it matches the full local divergence of $\Gamma$ as well as the divergences of
its individual subdiagrams, thereby fulfilling
the requirements~\refeq{eq:proofsketchF} and~\refeq{eq:proofsketchI}.
Thus two-loop rational terms 
can be computed using the formulas~\refeq{eq:proofsketchG}
and~\refeq{eq:proofsketchJ}. More explicitly,
\bea
\label{eq:r2masterformula}
\ratamp{2}{\Gamma}{}{}
&=&
\int\rd\barq_1
\int \rd\barq_2
\left[
\bar\Gamma^{\bar\alpha_1\bar\alpha_2\bar\alpha_3}(\barq_1,\barq_2,\barq_3)\,
\prod_{i=1}^3 
\left({\bf S}^{(i)}_{X_i}\,
\bar\calF^{(i)}_{\bar\alpha_i}(\barq_i) \right)
-\Gamma^{\alpha_1\alpha_2\alpha_3}(q_1,q_2,q_3)\,
\right.
\nonumber\\[3mm]
&&\left.{}\times
\prod_{i=1}^3 
\left({\bf S}^{(i)}_{X_i}\,
\calF^{(i)}_{\alpha_i}(q_i)\right)
\right]_{q_3=-q_1-q_2}
+\;\sum_{i=1}^3
\int\rd\barq_i\,
\Bigg[
\deltaZ{1}{\gamma_i}{\bar\alpha_i}{\barq_i}\,
{\bf S}^{(i)}_{X_i}\,
\bar\calF^{(i)}_{\bar\alpha_i}(\barq_i)
\nonumber\\[3mm]
&&
{}-
\left(\deltaZ{1}{\gamma_i}{\alpha_i}{q_i}
+\deltaZtilde{1}{\gamma_i}{\alpha_i}{\tilde q_i}+\ratamp{1}{\gamma_i}{\alpha_i}{q_i}\right)
{\bf S}^{(i)}_{X_i}\,
\calF^{(i)}_{\alpha_i}(q_i) 
\Bigg]
\,,
\eea 
where
\bea
\label{eq:loopchaindefB}
\calF^{(i)}_{\alpha_i}(q_i) 
&=&
\frac{\calN^{(i)}_{\alpha_i}(q_i)}
{\calD{i}}
\eea
is the 
projection of the
chain~\refeq{eq:loopchaindef} 
to $\numdim=4$ dimensions.
The orders $X_i$ 
for the expansions of the various chains $\calC_i$ 
in~\refeq{eq:r2masterformula}
have to be chosen according to~\refeq{eq:chainexpord}.
In the presence of UV divergent subdiagrams,
their determination can be facilitated by 
observing that\footnote{The relation~\refeq{eq:XboundB} 
is a direct consequence of the inequalities
\bea
\label{eq:XboundA}
X(\Gamma)\ge X_{jk}(\Gamma)+X_{im}(\Gamma)
\qquad\mbox{for}\qquad m=j,k\,,
\eea
where $ijk$ is a permutation of $123$.
See Section~5.2 of~\cite{Pozzorini:2020hkx}.}
\bea
\label{eq:XboundB}
\quad X_i(\Gamma) \,= \,
\text{max}\left\{ X(\Gamma), X_{ij}(\Gamma),
X_{ik}(\Gamma)\right\}\,=\,
X(\Gamma)
\quad
\mbox{if}
\quad X(\gamma_i)\,=\, X_{jk}(\Gamma)
\,\ge\,0\,.\quad
\eea
This relation is especially useful for the  
one-loop integrals with 
$\delta Z_{1,\gamma_i}$, $\delta \tilde Z_{1,\gamma_i}$ and
$\delta \calR_{1,\gamma_i}$ insertions, since such counterterms
are non-vanishing only for divergent subdiagrams $\gamma_i$,
\ie when $X(\gamma_i)\ge 0$. Thus, according to~\refeq{eq:XboundB}, the order of the expansion of the 
complementary chain $\calC_i$ is simply given by 
$X_i=X(\Gamma)$ and does not depend on the 
details of the two chains 
$\calC_j, \calC_k$ inside $\gamma_i$.

In~\refeq{eq:r2masterformula}
rational terms arise from the interplay of the
$(D-4)$-dimensional parts
of the numerators with UV singularities, and by construction
only local divergences contribute, while 
subdivergences cancel out.
This property has various important implications.  
First of all it makes it possible to discard
the divergent parts~\refeq{eq:remdiv}, and thus to reduce $\delta\calR_{2,\Gamma}$
terms to tadpole integrals. 
Moreover, it guarantees that all terms depending on the
auxiliary tadpole mass $M$ cancel in~\refeq{eq:r2masterformula}.
This cancellation mechanism can be understood by observing that, before
applying the $\bfS^{(i)}_{X_i}$ expansion, the original integrals are
independent of $M$. This implies that 
the truncated expansions $\bfS^{(i)}_{X_i}$ 
and their remnants $(1-\bfS^{(i)}_{X_i})$
must have identical $M$-dependent parts with opposite signs.
Moreover, we know that the remnants do not contribute to 
$\delta\calR_{2,\Gamma}$
since their divergent parts~\refeq{eq:remdiv} 
are free from local divergences, and non-divergent parts 
cannot generate rational terms.
For this reason, also the $M$-dependent parts of the 
truncated tadpole expansion~\refeq{eq:SSSexp} 
must cancel
in~\refeq{eq:r2masterformula}.
In practice, the cancellation of the $M$-dependence in~\refeq{eq:r2masterformula}
is guaranteed by the fact that all UV poles and
rational parts stemming from the subdivergences of two-loop amplitudes
are compensated by the corresponding 
counterterm insertions in the last two lines of~\refeq{eq:r2masterformula}.

\subsection{Power counting in $1/\barq_i$ and parametrisation dependence}
\label{se:paramdep}
 
The tadpole expansion defined in the previous section 
makes it possible to capture all terms up to relative order $1/\barq_i^{\,X_i}$ 
in a very simple way, namely by iterating~\refeq{eq:dentadexpA} 
on the denominators of a loop chain. 
However, as mentioned above, 
this naive expansion does not retain only 
the required terms of $\ord(1/\barq_i^{\,X_i})$ and lower, 
but also many unnecessary terms of $\ord(1/\barq_i^{(X_i+1)})$ and higher.

The number of terms, and related tadpole integrals to 
be computed, can be reduced in a drastic way by applying a 
strict power counting in $1/\barq_i$.
In practice the expansion~\refeq{eq:chainnaivtadexp} can be redefined as
\bea
\label{eq:chaintrunctadexp}
\bfS^{(i)}_{X_i} \,
\bar\calF^{(i)}_{\bar\alpha_i}(\barq_i)
&=&
{\bf P}_{X_i}^{(i)}
\left[
\bar\calN(\barq_i)\,
\bfS^{(i)}_{X_i} 
\left(\frac{1}{\calD{i}}\right)
\right]\,,
\eea
where the operator ${\bf P}_{X_i}^{(i)}$ truncates all terms beyond relative
order $1/\barq_i^{X_i}$. With this improvement a large number of irrelevant
terms are discarded, while the entire analysis
of~\refse{se:naivetadpoledec},
including the formula~\refeq{eq:r2masterformula}, remains valid.

When applying the tadpole expansions~\refeq{eq:chaintrunctadexp} 
or~\refeq{eq:chainnaivtadexp}
on the rhs of the 
$\delta \calR_{2}$ formula~\refeq{eq:r2masterformula},
care must be taken that the cancellation of all UV poles and rational terms
stemming from subdivergences is not disturbed.
This requires a one-to-one correspondence between the expansions 
that are applied to the two-loop integrals 
and to the related one-loop integrals with counterterm insertions
in~\refeq{eq:r2masterformula}.
Let us consider, for example, 
the counterterm contribution
\bea
\label{eq:optB}
\sum  \limits_{\gamma_i} \deltaZ{1}{\gamma_i}{}{}
\cdot \ampbar{1}{\Gamma_\tad/\gamma_{i,\tad}}{}{} 
&=&
\sum  \limits_{\gamma_i} 
\int\rd\barq_i\,
\deltaZ{1}{\gamma_i}{\bar\alpha_i}{\barq_i}\,
\bfS^{(i)}_{X_i}\,
\bar\calF^{(i)}_{\bar\alpha_i}(\barq_i)
\,,
\eea 
which embodies the UV singularities of the various
subdiagrams $\gamma_i$ in $\numdim=D$ dimensions. 
Here the expansion $\bfS^{(i)}_{X_i}$ 
of the complementary one-loop chain $\calC_i$
must be identical to the $\bfS^{(i)}_{X_i}$ 
expansion that is applied to the corresponding 
two-loop amplitude with $\numdim=D$ in~\refeq{eq:r2masterformula}.
To this end, the propagators of the two-loop diagrams and 
the corresponding propagators in the one-loop 
insertions need to be parametrised in the same way.
This is mandatory since, in general,
after tadpole expansion the
two-loop diagrams and 
the related
one-loop insertions 
depend on the parametrisation of the loop momenta, and only their combination is 
parametrisation-independent.
This is due to the fact that a loop-momentum shift
$\bar q_i\to \bar q_i+\Delta p_i$, where $\Delta p_i$ is a 
linear combination of the external momenta,
turns each term of fixed order $(1/\bar q_i)^K$ 
into combinations of terms of order 
$(1/\bar q_i)^{K'}$ with $K'\ge K$.
In the case of the counterterm $\deltaZ{1}{\gamma_i}{\bar\alpha_i}{\barq_i}$
in~\refeq{eq:optB}  
all extra higher-order terms resulting from the shift are
retained, 
while in the case of the chain 
$\calF^{(i)}_{\bar\alpha_i}(\barq_i)$ they are 
in part truncated by the $\bfS^{(i)}_{X_i}$ expansion. In general, this 
results into a 
dependence on the shift $\Delta p_i$.
Therefore, changing the parametrisation of two-loop diagrams and 
corresponding one-loop diagrams with 
one-loop counterterms 
independently from one another
can jeopardise
the cancellation of subdivergences 
in~\refeq{eq:r2masterformula}
and give rise to fake $\delta \calR_{2,\Gamma}$ contributions.

In principle,~\refeq{eq:optB} may be rendered parametrisation 
invariant by extending the $1/\barq_i$ expansion
to the full integrand, including also the counterterm
(see~\refapp{app:paraminvariance}). 
However, this is not consistent with the method of
\refse{se:naivetadpoledec}, which requires the expansion of each chain
$\calC_i$ to be independent of the complementary subdiagram $\gamma_i$.  Thus
the counterterm should be excluded from the
$1/\barq_i$ expansion, and when $\delta Z_{1,\gamma_i}(q_i)$ 
depends on $q_i$, \ie when the subdiagram $\gamma_i$ 
involves a non-logarithmic divergence, 
then~\refeq{eq:optB} 
is not invariant wrt shifts of $\barq_i$.

\subsection{Taylor expansion in the external momenta and masses}
\label{se:taylorexp}

In the following we introduce an improved tadpole-expansion approach
that renders the calculations more efficient and makes it possible to 
parametrise two-loop integrals and one-loop counterterm insertions 
independently from each other.
This approach is based on Taylor expansions in the external momenta and
internal masses, which correspond to expansions in the dimensionless
parameters $\{p_{ia}/\barq_i\}$, $\{m_{ia}/\barq_i\}$ and thus to
$1/\barq_i$ expansions at level of loop integrands.

To carry out Taylor expansions in the parameters 
$\{p_{ia}, m_{ia}\}$ associated with a certain chain
$\calC_i$ we introduce the rescaled parameters
\bea
\hat p_{ia} \,=\, \lambda_i\, p_{ia}\,,
\qquad
\hat m_{ia} \,=\, \lambda_i\, m_{ia}\,,
\qquad\mbox{for}\qquad
a=0,\dots,N_i-1\,,
\eea
and the associated expansion operators
\bea
\label{eq:momexpopdef}
\bfT_{K}^{(i)}\, 
\,=\,
\frac{1}{K!}\left(\frac{\rd}{\rd
\lambda_i}\right)^{\hspace{-1mm}K} 
\Bigg|_{\lambda_i=0}
\qquad\mbox{and}\qquad
\bfT_{[0,X_i]}^{(i)} \,=\,  \sum_{K=0}^{X_i}
\bfT_{K}^{(i)}\,.
\eea
For a function $f(\{p_{ia},  m_{ia}\})$ the terms
of fixed order $K$ in $\{p_{ia},m_{ia}\}$ are obtained by 
applying $\bfT_{K}^{(i)}$  to $f(\{\hat p_{ia}, \hat m_{ia}\})$, while
$\bfT_{[0,X_i]}^{(i)}$ corresponds to the truncated Taylor expansion up to order
$X_i$.
In the case of loop integrands, the dependence on the external momenta 
$p_{ia}$ arises only from the internal momenta
\bea
\hat \ell_{ia} &=& \bar q_i +\hat p_{ia}\,.
\eea
Thus the $\bfT_{K}^{(i)}$ operator corresponds to
\bea
\label{eq:lambdaexp}
\frac{1}{K!}
\left(\frac{\rd}{\rd
\lambda_i}\right)^{\hspace{-1mm}K} \Bigg|_{\lambda_i=0}
&=&
\frac{1}{K!}
\left(\sum_{a}
p_{ia}^\mu \frac{\partial}{\partial \hat \ell_{ia}^\mu}
+
\sum_{a}m_{ia}^\mu \frac{\partial}{\partial \hat m_{ia}^\mu}
\right)^{\hspace{-1mm}K} \Bigg|_{\hat p_{ia}\;=\; \hat m_{ia}\;=\;0}\,.
\eea

Contrary to the methods of~\refses{se:naivetadpoledec}{se:paramdep}, 
the above Taylor expansion generates only scaleless tadpole integrals,
since all momenta and masses are set to zero in the denominators.
This can be avoided by supplementing each propagator denominator 
by auxiliary squared mass terms
\bea
\label{eq:auxhatmass}
\hat M_i^2\,=\, (1-\omega^2_i) M^2\,.
\eea
Physical amplitudes correspond to 
$\omega_i=1$, \ie $\hat M_i=0$, but can be 
described through an expansion 
in $\omega_i$ around $\omega_i=0$.
To this end we introduce the operators
\bea
\label{eq:Mexpopdef}
\bfM_{J}^{(i)}\, 
\,=\,
\frac{1}{J!}\left(\frac{\rd}{\rd
\omega_i}\right)^{\hspace{-1mm}J} 
\Bigg|_{\omega_i=0}
\qquad\mbox{and}\qquad
\bfM_{[0,X_i]}^{(i)} \,=\,  \sum_{J=0}^{X_i}
\bfM_{J}^{(i)}\,.
\eea
Note that loop amplitudes with $\omega_i\neq 1$ 
depend only on the squared mass $\hat M_i^2$. Thus
the operator $\bfM_{J}^{(i)}$ yields zero for odd values of $J$, 
while for even values of $J$ it generates terms of relative order 
$(M/\barq_i)^{J}$. 
In practice, renormalisable theories require expansions
only up to order $X_i\le 2$, and, once all $\{p_{ia},m_{ia}\}$
have been set equal to zero in the denominators
as a result of~\refeq{eq:lambdaexp},
only the two following trivial  $\hat M_i$ expansions 
are required,
\bea
\label{eq:pmexp}
\bfM^{(i)}_{[0,X_i]}\,\,
\frac{1}{(\barq_i^2-\hat M_i^2)^{P_i}}
&=&
\frac{1}{(\barq_i^2- M^2)^{P_i}}
\qquad\mbox{for}\quad X_i\le 1\,,
\nonumber\\[2mm]
\quad
\bfM^{(i)}_{[0,2]}\,\,
\frac{1}{(\barq_i^2-\hat M_i^2)
^{P_i}
}
&=&
\frac{1}{(\barq_i^2- M^2)^{P_i}}
\left(1- P_i\,
\frac{M^2}{\bar q_i^2-M^2}\right)
\qquad\mbox{for}\quad X_i = 2\,.\quad
\eea
Here $1/(\bar q_i^2-\hat M_i^2)^{P_i}$
is a tadpole denominator that 
results from the $\{p_{ia},m_{ia}\}$ expansion of a certain chain 
$\calC_i$ with auxiliary mass~\refeq{eq:auxhatmass}.

In order to express the expansion of a generic loop chain 
in terms of the operators~\refeq{eq:momexpopdef} and
\refeq{eq:Mexpopdef}, let us define the modified loop chain
\bea
\label{eq:rescchain}
\bar \calF^{(i)}(\{\hat \ell_{ia},\hat m_{ia}\}, \hat M_i)
&=&
\frac{\bar\calN^{(i)}\big(\{\hat \ell_{ia},\hat m_{ia}\}\big)}
{{\cal D}^{(i)}\big(\{\hat \ell_{ia},\hat m_{ia}\}, \hat M_i\big)}\,,
\eea
where we explicitly indicate the dependence on 
$\hat \ell_{ia}$,  $\hat m_{ia}$ and $\hat M_i$,
and the modified chain denominator
is defined as 
\bea
\label{eq:pioB}
{\cal D}^{(i)}\big(\{\hat \ell_{ia},\hat m_{ia}\}, \hat M_i\big)
&=&
\prod_{a=0}^{N_i-1}\left(\hat\ell_{ia}^2-\hat m_{ia}^2-\hat M_i^2\right)\,,
\eea
while the associated numerator
in~\refeq{eq:rescchain}
corresponds to the usual chain numerator 
$\bar \calN^{(i)}_{\bar\alpha_i}(\bar q_i)$ with $p_{ia}\to \hat p_{ia}$,
$m_{ia}\to \hat m_{ia}$ and with the  multi-index $\bar\alpha_i$ kept
implicit. 
For $\lambda_i=\omega_i=1$ the modified chain~\refeq{eq:rescchain}
is equivalent to~\refeq{eq:loopchaindef}, while 
applying $\bfT_{K}^{(i)}\bfM_{J}^{(i)}$ to~\refeq{eq:rescchain}  
generates a massive tadpole chain 
of order $K$ in $\{p_{ia}/\barq_i,m_{ia}/\barq_i\}$ and order $J$  in
$M/\barq_i$.
Thus the 
truncated tadpole expansion~\refeq{eq:chaintrunctadexp} 
can be generated by applying~\refeq{eq:rescchain}  to all $\bfT_{K}^{(i)}\, \bfM_{J}^{(i)}$ 
combinations with $K+J\le X_i$, \ie
\bea
\label{eq:fochainexp}
\bfS^{(i)}_{X_i}\,\bar\calF^{(i)}(\bar q_i)
&=& 
\sum_{J=0}^{X_i}
\sum_{K=0}^{X_i-J}
\bfT_{K}^{(i)}\, 
\bfM_{J}^{(i)}\, 
\bar\calF^{(i)}\big(\{\hat \ell_{ia},\hat m_{ia}\}, \hat
M_i^2\big)\,.
\eea

In order to arrive at a more efficient expansion, we first observe that 
the expansions in $\lambda_i$ and $\omega_i$ can be decoupled from each
other by replacing 
\bea
\sum_{J=0}^{X_i}
\sum_{K=0}^{X_i-J}
\bfT_{K}^{(i)}\, 
\bfM_{J}^{(i)}\, 
\;\;&\to&\;\;
\bfT_{[0,X_i]}^{(i)}\, 
\bfM_{[0,X_i]}^{(i)}\,.
\eea
Such a modified expansion generates unnecessary extra terms 
up to order $(1/\bar q_i)^{2 X_i}$ for each chain. However it makes it
possible to combine 
the $\bfT_{[0,X_i]}^{(i)}$ expansions of all chains into
a global expansion  in $\{p_{ia}, m_{ia}\}$ that
extends to the full
integral, including also the vertices that connect the 
various chains.
In this way, 
exploiting the fact that $\delta \calR_2$
terms are homogeneous polynomials of order $X$ in $\{p_{ia}, m_{ia}\}$,
all contributions of lower and higher order 
can be discarded.
Thus one can replace
\bea
\prod_{i}\,
\bfT_{[0,X_i]}^{(i)}\, 
\bfM_{[0,X_i]}^{(i)}
\;\;&\rightarrow &\;\;
\bfT_{X}\,
\prod_{i}\,\bfM_{[0,X_i]}^{(i)}
\,, 
\eea
where $\bfT_{X}$ extracts terms of fixed order $X$
in all $\{p_{ia}, m_{ia}\}$.
More precisely, one can apply a global rescaling parameter 
$\lambda$ to all momenta and masses inside loop chains, connecting vertices
and counterterms, 
\bea
\label{eq:globrescaling}
\hat p_{ia} \,=\, \lambda\, p_{ia}\,,
\qquad
\hat m_{ia} \,=\, \lambda\, m_{ia}
\qquad\forall\;\;
i,a\,,
\eea
and define the expansion operator
\bea
\label{eq:pioL}
\bfT_{X} &=&
\frac{1}{K!}
\left(\frac{\rd}{\rd
\lambda}\right)^{\hspace{-1mm}K} \Bigg|_{\lambda=0}\,.
\eea
With these conventions one can define the optimised tadpole expansion\footnote{Here and in the following is is 
implicitly understood that
the parameters of the amplitude 
$\bar\calA_{k,\Gamma}$ on the rhs should be rescaled according 
to~\refeq{eq:auxhatmass} and~\refeq{eq:pioL}
before applying the $\bfT_{X}$  and $\bfM_{[0,X_i]}^{(i)}$ operators.}
\bea
\label{eq:genoptexp}
\bar\calA_{k,\Gamma_\tad} &=&
\bfT_{X}\, 
\bigg(\prod_{i}
\bfM_{[0,X_i]}^{(i)}\bigg)\, 
\bar\calA_{k,\Gamma}\,,
\eea
where $\bar\calA_{k,\Gamma}$ is the amplitude of a generic $k$-loop diagram,
and the product includes all relevant loop chains, \ie one chain at one loop
and three chains at two loops.
For one-loop diagrams with a counterterm insertion 
the expansion should be carried out as for bare one-loop 
diagrams, applying $\bfT_{X}$ also to 
the momentum and mass dependence of the counterterm.

When using the optimised tadpole expansion~\refeq{eq:genoptexp}
for the calculation of $\delta \calR_2$ terms,
on the rhs of~\refeq{eq:r2masterformula} one should 
apply a global Taylor expansion $\bfT_{X}$,
where $X=X(\Gamma)$ is the global degree
of divergence of the two-loop diagram at hand,
and replace $\bfS_{X_i}^{(i)}$ by $\bfM_{X_i}^{(i)}$, where
$X_i=X_i(\Gamma)$ is defined in~\refeq{eq:chainexpord}.
More explicitly, for an individual two-loop
diagram the above expansion amounts to the following operations.

\begin{enumerate}
\item Rescale all external masses and internal momenta 
according to~\refeq{eq:globrescaling}
and insert the auxiliary mass term $\hat M^2_i$ 
in every propagator denominator that depends on the loop momentum
$\bar q_i$.

\item Apply the operator $\bfT_{X}$, which selects terms of fixed total order
$X=X(\Gamma)$
in $\{p_{ia}, m_{ia}\}$,
at the level of the full two-loop diagram. This yields tadpole integrals with denominators of the form
$\prod_i (\bar q_i^2-\hat M_i^2)^{P_i}$, where $P_i\in [N_i, N_i+X]$.

\item  Apply the auxiliary-mass expansions $\prod_i \bfM_{[0,X_i]}$
using~\refeq{eq:pmexp}.

\end{enumerate}
The same procedure should be used for one-loop diagrams with counterterm 
insertions.
In that case, according to~\refeq{eq:XboundB}, the
order of the $\bfM^{(i)}_{[0,X_i]}$ expansions is simply 
$X_i=X(\Gamma)$.
Note that steps~2 and~3 of the above algorithm may be inverted. Alternatively,
they may be
implemented by means of the recursive tadpole 
decomposition~\refeq{eq:dentadexpA} 
with subsequent selection of terms of total order 
$X(\Gamma)$
in $\{p_{ia}, m_{ia}\}$ and from order zero to 
$X_i$ in $\hat M_i$.

Contrary to the naive tadpole expansion 
discussed in~\refses{se:naivetadpoledec}{se:paramdep},
the optimised expansion~\refeq{eq:genoptexp} 
and its further simplification described 
in~\refse{se:TexpwithCT}
are invariant wrt shifts of the loop momenta.
This is demonstrated in~\refse{app:paraminvariance}. 

\subsection{Taylor expansion with auxiliary one-loop counterterms}
\label{se:TexpwithCT}

In this section we outline an alternative tadpole expansion
method that is widely used in 
multi-loop calculations of beta 
functions~\cite{Misiak:1994zw,beta_den_comp,Zoller:2014xoa}. This method 
makes it possible to isolate local
divergences without applying any auxiliary-mass expansion.
It can be understood by starting from the 
expansion~\refeq{eq:genoptexp} and 
disentangling the 
effects of the $\bfT_X$ and $\bfM^{(i)}_{X_i}$ operators. 
Applying only the $\bfT_X$ expansion to a 
generic $k$-loop integral results into combinations of tadpole integrals 
of type
\bea
\label{eq:tadexpintform}
\bfT_{X}\, 
\bigg(\prod_{i}
\bfM_{[0,X_i]}^{(i)}\bigg)\, 
\bar\calA_{k,\Gamma}
\,=\,
\bigg(\prod_{i}\bfM_{[0,X_i]}^{(i)}\bigg)\, 
\sum_{\vec P}
\int
\prod_{i}\rd\barq_i \frac{\calT_{\vec P}(\{\bar q_k, p_{ka}, m_{ka}\})}
{\prod_j(\barq_j^2-\hat M_j)^{P_i}}
\eea
where $\vec P=(P_1,\dots)$ describes the denominator 
powers of the various loop chains.
The numerators 
$\calT_{\vec P}(\{\bar q_k, p_{ka}, m_{ka}\})$, which result from
the $\bfT_X$ expansion of the original integral,
is a polynomials of homogeneous degree $X$ in 
$\{p_{ia}, m_{ia}\}$.
Moreover, in the case of amputated 1PI diagrams,
the superficial degree of divergence $X$ corresponds to the
mass dimension of the diagram. Therefore
the massive tadpole integrals on the rhs of~\refeq{eq:tadexpintform} 
must have vanishing global degree of divergence.
Let us now consider the effect of the auxiliary-mass expansions.
The leading contribution
\bea
\bfM_{0}\,=\,\prod_{i}\bfM_{0}^{(i)}\,,
\eea
which amounts to inserting by hand a 
mass term $\hat M_i=M$ in all denominators,
does not modify the 
superficial degree of divergence. Instead, the remnant part
\bea
\label{eq:homexp}
\Delta \bfM\,=\,\prod_{i}\bfM_{[0,X_i]}^{(i)} -\bfM_0\,,
\eea
is either vanishing (when $X_i<2$ for all $i$) or contains at least
one auxiliary-mass derivative, which results in a 
$M^2/\bar q_i^2$ suppression.
This implies that such terms have a negative degree of 
superficial divergence. 
Therefore they are only relevant 
for a correct description of all subdivergences, \ie
in order to
guarantee~\refeq{eq:proofsketchI}---\refeq{eq:proofsketchJ}, 
but they do not contribute to $\delta \calR_{2,\Gamma}$.

This implies that two-loop rational terms can be computed 
using the minimal expansion
\bea
\label{eq:minTexp}
\bar\calA_{k,\Gamma_\tad}
&=&
\bfT_{X}\, 
\bfM_{0}\, 
\bar\calA_{k,\Gamma}\,.
\eea
With this approach the formula for the calculation of 
$\delta \calR_2$ terms becomes 
\bea
\label{eq:r2masterformulaB}
\ratamp{2}{\Gamma}{}{}
&=&
\bfT_X
\int\rd\barq_1
\int \rd\barq_2
\left[
\bar\Gamma^{\bar\alpha_1\bar\alpha_2\bar\alpha_3}(\barq_1,\barq_2,\barq_3)\,\,
\bfM_{0} 
\prod_{i=1}^3 
\bar\calF^{(i)}_{\bar\alpha_i}(\barq_i) 
-\Gamma^{\alpha_1\alpha_2\alpha_3}(q_1,q_2,q_3)\,
\right.
\nonumber\\[3mm]
&&\left.{}\times
\bfM_{0} 
\prod_{i=1}^3 
\calF^{(i)}_{\alpha_i}(q_i)
\right]_{q_3=-q_1-q_2}
+\;
\sum_{i=1}^3
\,\bfT_X
\int\rd\barq_i\,
\Bigg[
\deltaZ{1}{\gamma_{\tad,i}}{\bar\alpha_i}{\barq_i,M}\,
\bfM_{0}\,
\bar\calF^{(i)}_{\bar\alpha_i}(\barq_i)
\nonumber\\[3mm]
&&
{}-
\left(\deltaZ{1}{\gamma_{\tad,i}}{\alpha_i}{q_i,M}
+\deltaZtilde{1}{\gamma_{\tad,i}}{\alpha_i}{\tilde q_i}
+\ratamp{1}{\gamma_{\tad,i}}{\alpha_i}{q_i,M}\right)
\bfM_{0}\,
\calF^{(i)}_{\alpha_i}(q_i) 
\Bigg]
\,,\qquad
\eea 
where it is  understood that, before applying $\bfT_{X}$,
all physical masses and momenta should be 
rescaled according to~\refeq{eq:globrescaling}.
Since the omission of the higher-order terms~\refeq{eq:homexp}
modifies all quadratic subdivergences,
the various UV and rational one-loop counterterms in~\refeq{eq:r2masterformulaB}
should be adapted to the modified one-loop subdiagrams
$\bfM_{0}\, 
\bar\calA_{1,\gamma_i}$.
For instance, the 
required
UV counterterm in $\numdim=D$ 
is 
\bea
\label{eq:modUVCT}
\delta Z_{1, \gamma_{\tad,i}}^{\bar\alpha_i}(\bar q_i, M)
&=&
- \bfK \,
\bfM_0\,
\bar\calA^{\bar\alpha_i}_{1\,\gamma_{i}}
(\bar q_i)\,.
\eea
Assuming that subdivergences are at most quadratic,
\refeq{eq:modUVCT} is related to the standard UV counterterm 
through 
\bea
\delta Z_{1, \gamma_{\tad,i}}^{\bar\alpha_i}(\bar q_i, M)
&=&
\delta Z_{1, \gamma_{i}}^{\bar\alpha_i}(\bar q_i)
+ \bfK \left[M^2\frac{\rd}{\rd M^2}
\left(\bfM_0\,\bar\calA^{\bar\alpha_i}_{1\,\gamma_{i}}({\barq_i})\right)\right]
\,.
\eea
The $M$-dependent term on the rhs contributes only 
in the presence of quadratic subdivergences and
can be treated as an extra auxiliary counterterm.
Similar $M$-dependent terms need to be included also in
the rational terms 
$\delta\calR_{1,\gamma_{\tad,i}}^{\alpha_i}(q_i,M)$,
while in renormalisable theories $\deltaZtilde{1}{\gamma_{\tad,i}}{\alpha_i}{\tilde q_i}$
is independent of $M$,
since the full $M$-dependence of the UV counterterm in 
$\numdim=4$ can be absorbed into 
$\deltaZ{1}{\gamma_{\tad,i}}{\alpha_i}{\tilde q_i,M}$.

\subsection{Invariance with respect to shifts of the loop momenta}
\label{app:paraminvariance}

As discussed in \refse{se:paramdep}, 
tadpole expansions can depend on the 
parametrisation of loop integrals, \ie shifts of the loop momentum can 
lead to a different result. 
For this reason, when computing $\delta \calR_2$ terms
care must be taken that parametrisation-dependent 
terms do cancel out as they should. 
To this end, when using the naive expansions of~\refses{se:naivetadpoledec}{se:paramdep},
two-loop diagrams and 
related one-loop counterterm insertions
should be parametrised in the same way.
On the contrary, as demonstrated in the following,
when using the optimised tadpole expansions~\refeq{eq:genoptexp} 
and~\refeq{eq:minTexp}
the parametrisations of all loop integrals can be 
chosen independently from one another.

As a starting point, let us consider the interplay of 
a loop-momentum shift $\bar q_i \to \bar q_i + \Delta p_i$ with the 
expansion~\refeq{eq:genoptexp} of a generic one-loop integral,
\bea
\label{eq:pioM}
\bfT_X\,
\bfM^{(i)}_{[0,X_i]}\, 
\bar\calA_1(\{p_{ia}+\Delta p_i, m_{ia}\}) 
&=&
\int \rd \bar q_i\, 
\tilde \bfT_X\,
\bfM^{(i)}_{[0,X_i]}\, 
\bar\calF(\{\hat \ell_{ia},\hat m_{ia}\}, \hat M_i^2)
\Bigg|_{\hat\ell_{ia} = \barq_i + \Delta \hat p_i + \hat p_{ia}}\,,
\nonumber\\
\eea
where $X$ is the superficial degree of divergence of the diagram at hand, 
while $X_i$ may have arbitrary values, and the case
$X_i=0$ corresponds to the 
expansion~\refeq{eq:minTexp}.
The momentum shift $\Delta p_i$ is a certain combination of the external momenta
$p_{ia}$. Thus $\Delta p_i$ should undergo the same 
$\lambda$-rescaling~\refeq{eq:globrescaling} and expansion~\refeq{eq:pioL}
as the original external momenta. 
However, for a transparent
bookkeeping of the dependence on $\Delta p_i$ we 
introduce an independent rescaling
\bea
\label{eq:pioN}
\Delta \hat p_i &=& \tilde \lambda\,\Delta p_i\,,
\eea
and on the rhs of~\refeq{eq:pioM} 
we carry out a simultaneous expansion in 
$\lambda$ and $\tilde \lambda$, which is 
embodied in the operator
\bea
\label{eq:pioP}
\tilde \bfT_{X}
&=&
\frac{1}{X!}
\left(\frac{\rd}{\rd \tilde\lambda}+\frac{\rd}{\rd \lambda}\right)^X 
\,\Bigg|_{\lambda=\tilde\lambda=0}
\,=\,
\sum_{K=0}^X
\frac{1}{K!(X-K)!}
\left(\frac{\rd}{\rd \tilde\lambda}\right)^K
\,\Bigg|_{\tilde\lambda=0}
\left(\frac{\rd}{\rd \lambda}\right)^{X-K} 
\,\Bigg|_{\lambda=0},
\nonumber\\
\eea
where 
\bea
\label{eq:pioQ}
\left(\frac{\rd}{\rd \tilde\lambda}\right)^K \,\Bigg|_{\tilde\lambda=0}\,
&=&
\left(\Delta p_i^\mu\,\sum_{a}
\frac{\partial}{\partial \hat \ell_{ia}^\mu}\right)^K
\,
 \Bigg|_{\tilde\lambda =0}\,
\,=\,
\left(\Delta p_i^\mu\, \frac{\partial}{\partial \bar q_i^\mu}\right)^K
\,
\Bigg|_{\tilde\lambda =0}\,
\,.
\eea
Since $\tilde \lambda$, $\lambda$ and $\omega_i$ are independent
expansion parameters, the corresponding derivatives commute. Thus 
combining~\refeq{eq:pioL}, \refeq{eq:pioM} and~\refeq{eq:pioP}--\refeq{eq:pioQ}
we can write
\bea
\label{eq:pioR}
&&
\bfT_X\, \bfM_{[0,X_i]}^{(i)}\,
\bar\calA_1(\{p_{ia}+\Delta p_i, m_{ia}\}) 
\,=\,
\nonumber\\&&{}\hspace{10mm}\,=\,
\sum_{K=0}^{X}
\int \rd \bar q_i\, 
\frac{1}{K!}
\left(\Delta p_i^\mu\, \frac{\partial}{\partial \bar q_i^\mu}\right)^K\,
\left\{
\bfT_{X-K}\,
\bfM_{[0,X_i]}^{(i)}\,
\bar\calF(\{\hat \ell_k,\hat m_k\}, \hat M^2)
\Bigg|_{\hat\ell_k = \bar q_i + \hat p_k}
\right\}\,.
\nonumber\\
\eea
Here all terms with $K>0$ on the rhs integrate to zero since the 
corresponding integrands have the form of a total $\partial/\partial \bar q_i^\mu$
derivative, while the remaining $K=0$ term corresponds to the expansion of
the  original integral with $\Delta p_i = 0$, \ie
\bea
\label{eq:pioS}
\bfT_X\, \bfM_{[0,X_i]}^{(i)}\,
\bar\calA_1(\{p_{ia}+\Delta p_i, m_{ia}\}) 
\,=\,
\bfT_X\, \bfM_{[0,X_i]}^{(i)}\,
\bar\calA_1(\{p_{ia}, m_{ia}\}). 
\eea
This demonstrates that, when applied to one-loop integrals,
the tadpole expansions~\refeq{eq:genoptexp} 
and~\refeq{eq:minTexp}
are invariant wrt shifts of the loop momentum.
Along similar lines one can show that this 
holds also beyond one loop.

\section{Renormalisation constants in the $\msbar$ scheme}
\label{app:msbarRCs}

For convenience of the reader, in this appendix  we list the explicit expressions of the
renormalisation constants that enter the Yang--Mills
Lagrangian~\refeq{eq:ymlag} for the case of the 
$\msbar$ scheme.
Similarly as in \refse{sec:qcdres} we adopt the Feynman gauge,
and we use the convention~\refeq{eq:qed2}  for the 
perturbative expansion of the various 
renormalisation constants.
In the the $\msbar$ scheme,
the rescaling factor $\msfact$ that enters 
$t= \msfact \mu_0^2/\mu_\rR^2$ in~\refeq{eq:qed2}
is defined through~\refeq{eq:msstG} and,
according to~\refeq{eq:genrsF},
the scale-independent parts 
of the renormalisation constants are the same as in the $\ms$ scheme,
\bea
\dcalZ^{(\ms)}_{k,\chi}\,=\,\dcalZ^{(\msbar)}_{k,\chi}\,=\,\dcalZ^{(\ms_0)}_{k,\chi}\,.
\eea
The gauge-fixing term does not receive any finite renormalisation
in the $\msbar$ scheme, \ie $\calZ_{\gpar}=1$, while the 
scale-independent parts of the other 
renormalisation constants read
\begin{align}
 \dcalZ^{(\ms)}_{1,\als} \, =  \; &  \lb -\f{11}{3} \, \CA + \f{4}{3} \, \TF \, \nf \rb \eps^{-1}    \,,  \nonumber \\
 \dcalZ^{(\ms)}_{2,\als}  \, = \; & \lb \f{121}{9} \, \CA^2 - \f{88}{9} \, \TF \, \nf \, \CA  + \f{16}{9} \, \TF^2 \, \nf^2  \rb \eps^{-2}   
- \bigg[\f{17}{3} \, \CA^2 
- \TF \, \nf\, \bigg( \f{10}{3} \, \CA
\nonumber
\qquad\qquad\quad
 \\& 
+ 2 \, \CF
\bigg)
\bigg] \eps^{-1}   \,, 
\qquad\qquad\qquad
\end{align}

\begin{align}
  \dcalZ^{(\ms)}_{1,\psif} \, =  \; & - \CF \, \eps^{-1}    \,,   \nonumber \\
  \dcalZ^{(\ms)}_{2,\psif}  \, = \; & \lb  \f{1}{2}\,\CF^2 + \CA \, \CF \rb \eps^{-2} 
+ \lb \f{3}{4} \, \CF^2 - \f{17}{4} \, \CA \, \CF + \TF \, \nf \, \CF   \rb \eps^{-1}  \,, 
\qquad\qquad
\end{align}

\begin{align}
    \delta \hat \calZ^{(\ms)}_{1,m_f} \, =  \; & -3 \, \CF \, \eps^{-1}    \,,   \nonumber \\
    \delta \hat \calZ^{(\ms)}_{2,m_f}  \, = \; & \CF 
\left[\lb \f{9}{2} \, \CF  + \f{11}{2} \, \CA  - 2 \, \TF\,\nf \rb \eps^{-2}  
- \lb    \f{3}{4} \, \CF  +\f{97}{12} \, \CA - \f{5}{3} \, \TF \, \nf \rb
\eps^{-1}\right] \,, \qquad
\end{align}

\begin{align}
    \delta \hat \calZ^{(\ms)}_{1,A} \, =  \; &  \lb \f{5}{3} \, \CA - \f{4}{3} \TF \, \nf \rb \eps^{-1}    \,,   \nonumber \\
    \delta \hat \calZ^{(\ms)}_{2,A}  \, = \; &  \lb -\frac{25}{12} \, \CA^2 + \frac{5}{3} \, \TF  \, \nf \, \CA \rb \eps^{-2} 
+ \left[\f{23}{8}\,  \CA^2 -  \TF \, \nf \lb \f{5}{2}\, \CA   + 2 \, \CF \rb
\right] \eps^{-1}\,, 
\end{align}

\begin{align}
    \delta \hat \calZ^{(\ms)}_{1,u} \, =  \; &  \f{\CA}{2} \, \eps^{-1}    \,,   \nonumber \\
    \delta \hat \calZ^{(\ms)}_{2,u}  \, = \; &  \lb - \CA^2 + \f{1}{2} \, \TF \, \nf \, \CA  \rb \eps^{-2} + \lb \f{49}{48} \, \CA^2 - \f{5}{12} \, \TF \, \nf \, \CA   \rb \eps^{-1}
\,. 
\quad\qquad\qquad
\end{align}

These renormalisation constants 
have been computed in the same framework as the rational terms and agree with those in the literature, which have been available for a long time
\cite{Jones1974531,PhysRevLett.33.244,Tarasov:1976ef}.
Specific results for SU(N) and U(1) gauge theories 
can be obtained by applying the substitutions in~\refta{tab:uonesun}.

\bibliographystyle{JHEP}
\bibliography{RT_from_tadpoles_QCD}

\end{document}